\documentclass[10pt,aps,fleqn,superscriptaddress,notitlepage,nofootinbib,preprintnumbers,nobalancelastpage]{revtex4-1}
\pdfoutput=1
\usepackage{amsmath,amssymb,graphicx,xspace,subfigure,tikz}
\newcommand{\eps}{\varepsilon}
\usepackage{hyperref}
\usepackage{bm}
\usepackage{ulem}

\hypersetup{
  pdfauthor={Stefan Hoeche,Andrew Long,Jonathan Kozaczuk,Jessica Turner,Yikun Wang},
  pdftitle={Towards an all orders calculation of the electroweak bubble wall velocity},
  pdfkeywords={Parton Shower,Electroweak Phase Transition}
}

\begin{document}
\preprint{FERMILAB-PUB-20-274-T, MCNET-20-14}
\title{Towards an all-orders calculation of the electroweak bubble wall velocity}
\author{Stefan~H{\"o}che}
\affiliation{Fermi National Accelerator Laboratory,Batavia, IL, 60510, USA}
\author{Jonathan Kozaczuk}
\affiliation{Department of Physics, University of California, San Diego, La Jolla, CA 92093, USA}
\author{Andrew~J.~Long}
\affiliation{Rice University, Houston, Texas 77005, USA}
\author{Jessica~Turner}
\affiliation{Fermi National Accelerator Laboratory,Batavia, IL, 60510, USA}
\author{Yikun~Wang}
\affiliation{Fermi National Accelerator Laboratory,Batavia, IL, 60510, USA}
\affiliation{Enrico Fermi Institute, University of Chicago, Chicago, Illinois, 60637, USA}
\begin{abstract}
We analyze Higgs condensate bubble expansion during a first-order
electroweak phase transition in the early Universe. The interaction of
particles with the bubble wall can be accompanied by the emission of
multiple soft gauge bosons. When computed at fixed order in
perturbation theory, this process exhibits large logarithmic
enhancements which must be resummed to all orders when the wall
velocity is large. We perform this resummation both analytically and
numerically at leading logarithmic accuracy. The numerical simulation
is achieved by means of a particle shower in the broken phase of the
electroweak theory.  The two approaches agree to the 10\% level. For
fast-moving walls, we find the scaling of the thermal pressure exerted
against the wall to be $P\sim \gamma^2T^4$, independent of the
particle masses, implying a significantly slower terminal velocity
than previously suggested.
\end{abstract}
\maketitle

\section{Introduction}
\label{sec:introduction}
A cosmological electroweak phase transition is expected to have
occurred in the early Universe when the thermal plasma cooled to the
weak scale.  The Standard Model predicts that this transition is a
continuous crossover~\cite{DOnofrio:2015mpa}, however minimal new
physics coupled to the Higgs can lead to a first order phase
transition~\cite{Anderson:1991zb,Quiros:1999jp,Grojean:2004xa}.  It is
important to understand the dynamics of Higgs-field bubbles during the
phase transition~\cite{Carrington:1993ng},
as they directly affect the production of various cosmological relics
including the matter-antimatter asymmetry~\cite{Cohen:1990it},
topological defects~\cite{Achucarro:1999it}, primordial magnetic
fields~\cite{Vachaspati:1991nm}, and especially a stochastic
background of gravitational wave radiation~\cite{Kamionkowski:1993fg}
that could be detected by the next generation of gravitational wave
experiments~\cite{Caprini:2019egz,Caprini:2015zlo, Kawamura:2011zz,Harry:2006fi}.

A first order electroweak phase transition proceeds through the
nucleation, growth, and percolation of bubbles.  Outside of the
bubbles, the expectation value of the Higgs field vanishes and
electroweak symmetry is restored.  Inside of the bubbles, the average
Higgs field has a nonzero value, giving mass to the quarks, charged
leptons, and weak gauge bosons.  A differential vacuum pressure across
the phase boundary drives the bubbles to expand and collide, on time
scales typically much less than one Hubble time, filling all of space with the broken-symmetry phase.
 
As the bubbles are expanding, their speed is controlled by a balance
of pressures.  A vacuum pressure, resulting from the underlying
symmetry-breaking Higgs potential, ``pushes'' the bubble walls
outward.  Meanwhile, a thermal pressure, resulting from the
interactions of the wall with the ambient plasma, retards the bubble's
expansion, and acts as a source of friction.  If the vacuum pressure
exceeds the thermal pressure, then the bubble wall will ``runaway''
with its velocity approaching the speed of
light~\cite{Bodeker:2009qy}. On the other hand, if the thermal
pressure balances the vacuum pressure then the wall reaches a
(possibly ultrarelativistic) terminal velocity~\cite{Bodeker:2017cim}.
Therefore, to understand the dynamics of Higgs-phase bubble walls
during a first-order electroweak phase transition, a key quantity of
interest is the thermal pressure induced by the plasma of Standard
Model particles.

Thermal pressure arises, in general, from the scattering of particles
whose masses or couplings vary across the bubble wall.  The authors of
Ref.~\cite{Bodeker:2017cim} argued that for fast-moving walls, the pressure is dominated by
the emission of soft vector bosons when particles cross the wall, a
phenomenon known as transition radiation. 
 In Ref.~\cite{Bodeker:2017cim} the authors
calculated the  thermal pressure on the bubble wall assuming a
\textit{single emission} of the soft vector boson.  They found that
this channel dominates the non-radiative process, due to its
enhancement in the infrared (IR) region. In this work, we compute
pressure exerted on the bubble wall from incoming 
particles, which can be put off-shell through their interaction with the Higgs condensate. We find that the pressure exerted on the wall from this process is IR-enhanced, featuring large logarithms when 
the wall velocity is large that invalidate
a fixed-order calculation of the pressure. Therefore, a resummation of the accompanying soft radiation is necessary to obtain an accurate estimate of the wall velocity in the ultrarelativistic regime.

In this work, we calculate the thermal pressure that results from the scattering of
Standard Model particles on an ultrarelativistic Higgs-phase bubble
wall while accounting for multiple soft emissions that collectively
comprise a particle shower.  We begin in Sec.~\ref{sec:motivation}
with the definition of thermal pressure and motivate the need for
resummation of soft radiation.  In Sec.~\ref{sec:formalism}
we establish the framework for our calculation and the relation to
Ref.~\cite{Bodeker:2017cim}. In Sec.~\ref{sec:framework} we compute
the logarithmically enhanced radiative corrections and perform an
analytic resummation.  In addition to the analytic result, we simulate
a particle shower in the broken Higgs phase in Sec.~\ref{sec:ps}. The
cosmological implications are presented in Sec.~\ref{sec:implications}
and we discuss and conclude in Sec.~\ref{sec:discussion}.

\section{Thermal pressure and why Higgs bubbles need a shower}
\label{sec:motivation}
In this section, we provide an intuitive understanding of the (net)
thermal pressure exerted against the wall, $P \equiv F/A$ (the 
retarding force per unit area) from particles interacting with 
the Higgs condensate at the phase interface.
We argue that a fixed-order calculation of $P$ must break down for
sufficiently large wall velocities $\gamma \gg 1$, and thus we
motivate an all-orders calculation which is carried out in the
following sections using well-known analytic resummation and numerical
techniques based on QCD parton showers.

Consider a Higgs-phase bubble in a plasma with weak-scale temperature
$T \sim 100 \ \mathrm{GeV}$.  On the length scales of interest, the
curvature of the bubble can be neglected, and the local bubble wall
can be treated as planar.  For concreteness let $\vec{v}_\mathrm{w} =
- v \, \vec{e}_z$ with $v > 0$ be the velocity of the wall in the rest
frame of the plasma, and let $\gamma = 1 / \sqrt{1 - v^2}$ be the
corresponding Lorentz factor.  We are interested in ultrarelativistic
walls for which $\gamma \gg 1$, and typically $\gamma \sim 10-1000$. 
In this regime, all SM particles are assumed to enter the broken phase 
with negligible reflection probability, the flux of particles passing 
the wall from inside the bubble is exponentially suppressed, and the 
distributions of the incoming particles are just the usual Bose-Einstein 
or Fermi-Dirac equilibrium distributions. 
Throughout this article, we work in a frame where the bubble wall is at
rest, and the plasma has an average velocity $\vec{v}_\mathrm{pl} = v
\, \vec{e}_z$.  In this frame, particles from the plasma are incident
on the wall with boosted energies $E \sim \gamma T$ and fluxes
$\mathcal{F} \sim \gamma T^3$.  As $\gamma T \gg 100 \ \mathrm{GeV}$,
there is no kinematic restriction toward producing a large number of
weak-scale particles when a particle interacts with the wall. The
thermal pressure $P$ can be written schematically as the product,
\begin{equation}
    P 
    \ = \ \mathcal{F} 
    \ \times \ \langle \Delta p_z \rangle \;, 
    \qquad \text{with} \qquad 
    \langle \Delta p_z \rangle 
    = \int \! {\rm d}\Delta p_{z}
    \ \Delta p_z \  \frac{d\mathbb{P}}{d\Delta p_z}
    \;,
\end{equation}
where $\mathcal{F}$ is the thermal flux of incident particles on the
wall (number per area per time) and $\langle \Delta p_z \rangle$ is the
average longitudinal momentum transferred to the wall by each incident particle.
For each particle that hits the wall, multiple scattering channels are
possible and $\mathbb{P}$ represents the probability for a given
scattering while $\Delta p_z$ is the longitudinal momentum transfer of
that scattering.  We are particularly interested in how the thermal
pressure scales with the Lorentz factor $\gamma$, as it will be
compared against the vacuum pressure that scales as $P_\mathrm{vac}
\sim \gamma^0$.  Since $\mathcal{F} \sim \gamma^1$ in general, we only
need to determine how $\langle \Delta p_z \rangle$ scales with $\gamma$.

Generally, $\langle \Delta p_z \rangle$ is given by a sum over all
possible scatterings with each weighted by its associated probability.
Ref.~\cite{Bodeker:2009qy} considered the pressure that results from a
particle entering the bubble and acquiring mass $m$, without any other
emission.  Based on simple kinematics, one can show $\Delta p_z \sim
m^2 / (2 E)$, which scales as $E^{-1} \sim \gamma^{-1}$.  Additionally
taking $\mathbb{P} = 1$, since nearly all particles are transmitted
into the bubble, Ref.~\cite{Bodeker:2009qy} found that the pressure
for such 1-to-1 transitions scales like $P \sim \gamma^0$ since
$\mathcal{F} \sim \gamma^1$.  The same authors revisited the
calculation of thermal pressure in Ref.~\cite{Bodeker:2017cim}, and
allowed the incident particle to emit an additional particle, which
they refer to as a 1-to-2 transition.  They argue that the pressure is
dominated by a region of phase space in which the emitted particle is
soft with transverse momentum close to its mass.  This leads to
$\Delta p_z \sim m_\mathrm{soft}$, where $m_\mathrm{soft}$ is the
on-shell mass of the emission, and $\mathbb{P} \sim \alpha \,
\gamma^0$, where $\alpha = g^2/4\pi$ is the appropriate three-particle
coupling.  As such, Ref.~\cite{Bodeker:2017cim} found $P \sim
\gamma^1$, implying that the thermal pressure grows as the wall speeds
up.  Since the vacuum pressure does not grow with increasing $\gamma$,
a balance of pressures is inevitable, and a terminal velocity will be
reached.

Such a possibility naturally leads to the question: what is the effect
of 1-to-many transitions?  To estimate whether these channels could be
relevant, we can calculate the probability for a 1-to-2 transition in
the soft region of phase space favored by Ref.~\cite{Bodeker:2017cim}. 
We utilize the  universal factorization properties of gauge theory
amplitudes in the soft gauge boson limit to factorize the 
1-to-2 transition probability into a 1-to-1 transition probability
and a logarithmic enhanced contribution. We find that, in the relevant
region of phase space, the emission probability for massless gauge
bosons is parametrically given by
\begin{equation}\label{eq:ll1}
    \mathbb{P}_{1\to2} \approx 
    \mathbb{P}_{1\to1} \times 
    \sum_i C_i \, \frac{\alpha_i}{2 \pi} \, 
    \log\frac{x\left(\gamma T\right)^2}{\vec{k}_\perp^{\,2}}\,,
\end{equation}
where $\vec{k}_{\perp}$ and $x$ are the transverse momentum and 
energy fraction of the soft emission respectively. The different 
emission channels are summed over with the appropriate couplings 
$\alpha_i = g_i^2 / 4\pi$ and charges $C_i$ (see
App.~\ref{sec:vertex_functions} for notation). 
The logarithm, even if cut off by a thermal mass, is in principle
unbounded, which makes it necessary to resum Eq.~\eqref{eq:ll1}
to all orders in perturbation theory. 
We find that the logarithmic enhancement produces a significant 
enhancement of the friction, which is then of order  $P \sim \gamma^2 T^4$, 
exhibiting the same scaling as found in  Ref.~\cite{Mancha:2020fzw}.

\section{Perturbative computation of the thermal pressure}
\label{sec:formalism}
In this section, we introduce the framework for the computation,
following the methods developed in Ref.~\cite{Bodeker:2017cim}.
We also establish the notation that will allow us to compute
radiative corrections in Sec.~\ref{sec:framework}.

\subsection{Kinematics and one-particle states}
We begin by defining the relevant kinematic variables.  We work in the
rest frame of the bubble wall, which is assumed to be planar and
located at $z = 0$.  In the rest frame of the plasma, the wall's
velocity is $\vec{v}_\mathrm{w} = - \vec{v} = - v \, \vec{e}_z$, and
in the rest frame of the wall, the plasma's velocity is
$\vec{v}_\mathrm{pl} = \vec{v} = v \, \vec{e}_z$; the associated
Lorentz factor is $\gamma = 1 / \sqrt{1-v^2}$.  We assume that
particles of type $a$ have a mass $m_{a,\mathrm{s}}$ in front of the
wall ($z < 0$, symmetric phase) and a mass $m_{a,\mathrm{h}}$ behind
the wall ($z > 0$, Higgs phase).  This mass-varying background breaks
spatial-translation invariance in the direction normal to the
wall. Noether's theorem implies that the $z$-component of momentum is
not conserved, and there is an ambiguity in the construction of a
complete Fock space because we cannot label single-particle states by
their $x$-, $y$-, and $z$-momentum, since the last one is not a good
quantum number.  To address this ambiguity we follow
Ref.~\cite{Bodeker:2017cim}.  One-particle states of flavor $a$ are
defined to be momentum eigenstates in the symmetric phase, and are
therefore labeled by the momentum $\vec{p}_{a,\rm s} =(p_{a,x}, \,
p_{a,y}, \, p_{a,z,\mathrm{s}})$ a particle has in the symmetric
phase, based on its energy and transverse momentum.  They are
normalized according to~\cite{Peskin:1995}
\begin{align}\label{eq:p_states}
  &\langle\vec{p}_{a,\rm s}^{\, \prime}|\vec{p}_{a,\rm s}\rangle
  = (2\pi)^3\,2E_{a} \,
  \delta^{(3)}(\vec{p}_{a,\rm s}^{\,\prime}-\vec{p}_{a,\rm s}) \;,
  &\int \! \! \frac{\mathrm{d}^3 \vec{p}_{a,\rm s}}{(2\pi)^3}
  \frac{1}{2E_{a}} \ |\vec{p}_{a,\rm s}\rangle \langle\vec{p}_{a,\rm s}| = 1 \;,
\end{align}
where $\vec{p}_{a,\perp}=(p_{a,x}, \, p_{a,y},0)$
and $p_{a,z,\rm s}^2=E_a^2-\vec{p}_{a,\perp}^{\,2}-m_{a,\rm s}^2$.
If the particle in question is located in front of the wall ($z < 0$),
and $\vec{p}_a$ is the particle's three momentum, then we have
the dispersion relation $E_{a}^2={\vec{p}_a^{\,2}+m_{a,\rm s}^2}$,
and the projection
\begin{equation}\label{eq:plane_wave}
  \langle\vec{x}\,|\vec{p}_{a,\rm s}\rangle
  =\sqrt{2E_a}\exp\{i\vec{p}_{a,\perp}\cdot\vec{x}_\perp\}\chi(z)\;,
\end{equation}
is a plane wave, i.e.\ $\chi(z)=\exp\{ip_{a,z,\rm s}z\}$, because the
label momentum, $\vec{p}_{a,\rm s}$, agrees with the kinematical
momentum.  If the particle is instead located behind the wall ($z>0$),
we have $E_{a}^2={\vec{p}_a^{\,2}+m_{a,\rm h}^2}$, and the free
particle states must be found by solving the associated evolution
equation in the presence of the bubble wall, because the label
momentum does not agree with the kinematical momentum. This can be
achieved by using the WKB approximation~\cite{Bodeker:2017cim} to
determine the mode functions $\chi(z)$, leading to the zeroth order
result $\chi(z)=\exp\{ip_{a,z,\rm h}z\}$, where
$p_{a,z,\rm h}^2=E_a^2-\vec{p}_{a,\perp}^{\,2}-m_{a,\rm h}^2$.
In summary, the functional form of $\chi(z)$ agrees with the vacuum case
for all $z$ if $\chi(z)$ is evaluated with the {\it kinematical}
$z$-momentum, $\chi(z)=\exp\{ip_{a,z}z\}$.%
\footnote{This result can be understood from a different perspective:
  The relative phase shift between the WKB solution and a plane wave
  amounts to the relative momentum transfer of a free particle to the
  bubble wall, which is given by
  $\Delta p_z/(\gamma T)\approx \Delta m^2/(\gamma T)^2$.
  In the high-energy limit $\gamma T\gg m_{a,\rm h}$
  the amplitude of the wave function is unaltered, because a
  small momentum change will not cause the particle to be reflected
  off the wall. Hence the free-particle momentum eigenstate at $z>0$
  is simply the plane wave solution in Eq.~\eqref{eq:plane_wave} with
  the appropriate dispersion relation for $z>0$.}  Note in particular
that the $\delta$ functions in Eq.~\eqref{eq:p_states} do not imply
$z$-momentum conservation. Written in terms of conserved kinematical
quantities they read instead
\begin{equation}\label{eq:delta_conversion}
  \delta^{(3)}(\vec{p}_{a,\rm s}^{\,\prime}-\vec{p}_{a,\rm s})=
  \frac{p_{a,z,\rm s}}{E_a}\,\delta(E_{a}'-E_{a})\,
  \delta^{(2)}(\vec{p}_{a,\perp}^{\,\prime}-\vec{p}_{a,\perp})\;.
\end{equation}

\subsection{Definition of thermal pressure}
We can now define the thermal pressure, $P$, and proceed to derive
a master formula that allows us to calculate this pressure from
scattering amplitudes. First we write
\begin{equation}\label{eq:pressure_sum}
    P = \sum_{a \in \mathcal{S}} \, P_a \, ,
\end{equation}
where $P_a$ is the pressure resulting from incident particles of
species $a$ specifically.  We sum $a$ over the set of massive 
Standard Model particle species, $\mathcal{S}$.  Next we can write
\begin{equation}\label{eq:Pa}
    P_a = \int \! \mathrm{d} \mathcal{F}_a \ \langle \Delta p^\nu
    \rangle_a N_\nu \,,
\end{equation}
where $N_\mu$ is a space-like four-vector normal to the wall,
${\rm d}\mathcal{F}_a=\mathrm{d} j_a^\mu N_\mu$ is the flux
of incident $a$-particles, $\mathrm{d} j_a^\mu$ is the differential
$a$-particle-number current density, and
$\langle \Delta p^\mu \rangle_a$ is the average four-momentum transfer
to the wall when a single $a$-particle is incident. To construct the
normal vector, $N_\mu$, suppose that the wall is parametrized by a
scalar field $m_a^2(x^\mu)$ that represents the inhomogeneous squared mass
of $a$-particles; then we define
$N_\mu = \partial_\mu m_a^2 / \sqrt{-(\partial m_a^2)^2}$,
which is evaluated at the wall, implying $N_\mu = \{ 0, 0, 0, 1 \}$
in the frame where the wall is at rest and $m^2(x^\mu)$ increases
from $m_{a,\mathrm{s}}^2$ at $z \to -\infty$ to
$m_{a,\mathrm{h}}^2$ at $z \to + \infty$.
The differential current density can be written as
\begin{equation}\label{eq:diff_flux}
    \mathrm{d} j_a^\mu = \nu_a \, \frac{\mathrm{d}^3
      \vec{p}_a}{(2\pi)^3} \, f_a \, \frac{p_a^\mu}{E_a} \,,
\end{equation}
where $\nu_a$ counts the redundant internal degrees of freedom (e.g.,
color and spin), and $f_a$ is the phase space distribution function of
$a$-particles. The plasma in front of the wall is assumed to be in
thermal equilibrium at temperature $T$.  If the wall moves
ultrarelativistically, as we assume throughout our study, the distribution
functions for bosons and fermions are the equilibrium Bose-Einstein and
Fermi-Dirac distributions, respectively, boosted from the plasma frame
to the wall frame:
\begin{equation}\label{eq:equilibrium_distributions}
  f_a = \frac{1}{e^{p^\mu u_\mu^\mathrm{pl} / T} \pm 1} =
  \frac{1}{e^{(\gamma E_a - \gamma {\vec{p}}_a \cdot
      \vec{v}_\mathrm{pl}) / T} \pm 1} \;,
\end{equation}
where $u^\mu_\mathrm{pl}$ is the four-velocity of the plasma that
equals $u^\mu_\mathrm{pl} = \{ \gamma, 0, 0, \gamma
\vec{v}_\mathrm{pl} \}$ in the rest frame of the wall.
Eq.~\eqref{eq:equilibrium_distributions} generates average
momenta of $p_{a,z} \sim \gamma T \gg p_{a,x} \sim p_{a,y} \sim T$.
We also assume $\gamma T \gg m_a$ such that $E_a \sim \gamma T$ and
parametrically the flux is Lorentz-boosted, $\mathrm{d} \mathcal{F}_a
\sim \gamma T^3$.

The average momentum transfer from incident $a$-particles,
$\langle \Delta p^\mu \rangle_a$, can be written as
\begin{equation}
\label{eq:Dp_avg}
  \langle \Delta p^\mu \rangle_a = \sum_{n=1}^{\infty} \sum_{\{ b \}
    \in \mathcal{S}} \int \! \mathrm{d} \mathbb{P}_{a \to b_1 b_2
    \cdots b_n} \ \Delta p^\mu\,,
\end{equation}
where $\mathrm{d} \mathbb{P}_{a \to b_1 b_2 \cdots b_n}$ is the
differential probability for a single $a$-particle with momentum
$\vec{p}_a$ to create a shower of $n$ particles of species $\{b\}$ and
label momenta $\{\vec{p}_{b,\rm s}\}$. The probability is weighted by
the four-momentum transferred to the wall, which we write as
\begin{equation}\label{eq:Dp}
    \Delta p^\mu = p_{a}^\mu - \sum_{i=1}^n p_{b_i}^\mu\;.
\end{equation}
The probability density $\mathrm{d} \mathbb{P}_{a \to b_1 \cdots b_n}$
enforces energy and transverse momentum conservation (see
Eq.~\eqref{eq:delta_conversion}), hence we find that $\Delta p^\mu
N_\mu = p_{a,z} - \sum_i p_{b_i,z}$.  We can write the
differential probability as~\cite{Peskin:1995}
\begin{equation}\label{eq:diff_probability}
  \mathrm{d} \mathbb{P}_{a \to b_1 \cdots b_n} = \Biggl[ \prod_{i=1}^n
    \frac{\mathrm{d}^3 \vec{p}_{b_i,\rm s}}{(2\pi)^32E_{b_i}} (1 \pm
    f_{b_i}) \Biggr] \ \bigl| \langle \vec{p}_{b_1,\rm s} \cdots
  \vec{p}_{b_n,\rm s} |i\hat{T}| \psi_a(\vec{p}_{a}) \rangle
  \bigr|^2\,,
\end{equation}
where $\mathrm{d}^3 \vec{p}_{b_i,\rm s} /((2\pi)^3 2 E_{b_i})$ is the
Lorentz-invariant differential phase space volume element for the
final-state particle $i$, $(1 \pm f_{b_i})$ accounts for Bose
enhancement (if $b_i$ is a boson) or Pauli blocking (if $b_i$ is a
fermion), and $\langle \vec{p}_{b_1,\rm s} \cdots \vec{p}_{b_n,\rm s}
|i\hat{T}| \psi_a(\vec{p}_{a}) \rangle$ is the transition matrix
element for particle $a$ represented by the state
$|\psi_a(\vec{p}_{a})\rangle$ to scatter into $n$ particles.  We will
assume that the occupation numbers are small and thus approximate $1
\pm f_{b_i} \approx 1$ in our calculation.  The incoming wave packet
is defined in terms of the wave function $\psi(\vec{p}_{a,\rm s}^{\,
  \prime}; \vec{p}_a)$ of particle $a$ with momentum $\vec{p}_a$ as
\begin{equation}\label{eq:wf_def}
  |\psi_a(\vec{p}_{a})\rangle = \int \! \!  \frac{\mathrm{d}^3
    \vec{p}_{a,\rm s}^{\, \prime}}{(2\pi)^3} \frac{1}{2E_a^\prime} \,
  \psi(\vec{p}_{a,\rm s}^{\, \prime}; \vec{p}_a) \, |\vec{p}_{a,\rm
    s}^{\, \prime}\rangle \;,
\end{equation}
which leads to the proper normalization
$\langle\psi_a(\vec{p}_{a})|\psi_a(\vec{p}_{a})\rangle = 1$.  The
transition amplitude can then be written as
\begin{equation}\label{eq:tmatrix_with_wavefunction}
  \langle \vec{p}_{b_1,\rm s} \cdots \vec{p}_{b_n,\rm s} |i\hat{T}|
  \psi_a(\vec{p}_{a}) \rangle = \int \!\! \frac{\mathrm{d}^3
    \vec{p}_{a,\rm s}^{\, \prime}}{(2\pi)^3 2E_a^\prime} \,
  \psi(\vec{p}_{a,\rm s}^{\, \prime}; \vec{p}_a) \, \langle
  \vec{p}_{b_1,\rm s} \cdots \vec{p}_{b_n,\rm s} |i\hat{T}|
  \vec{p}_{a,\rm s}^{\, \prime} \rangle\,.
\end{equation}
It can be expressed in terms of the corresponding scattering amplitude
$\mathcal{M}_{a \to b_1\cdots b_n}$ as follows%
\footnote{Our normalization convention for $\mathcal{M}$ differs from
  Ref.~\cite{Bodeker:2017cim} by a factor $p_{a,z,\rm s} / E_a$ due to
  Eq.~\eqref{eq:delta_conversion}.}
\begin{equation}\label{eq:tmatrix_element_in_terms_of_amplitude}
  \langle \vec{p}_{b_1,\rm s} \cdots \vec{p}_{b_n,\rm s} |i\hat{T}|
  \vec{p}_{a,\rm s} \rangle = (2\pi)^3 \,
  \delta^{(3)}({\vec{p}}_{a,\rm s} -\sum_{i=1}^n{\vec{p}}_{b_i,\rm s})
  \ i\mathcal{M}_{a \to b_1\cdots b_n}({\vec{p}}_{a,\rm s},
             {\vec{p}}_{b_1,\rm s}, \cdots {\vec{p}}_{b_n,\rm s})\,.
\end{equation}
Both the transition matrix element and the scattering amplitude depend
on the particles' spins, which we have suppressed to avoid unnecessary
notation.  Spins of final-state particles are summed, and the spin of
the initial-state particle is averaged.  Note that we only factor off
three Dirac delta functions in
Eq.~\eqref{eq:tmatrix_element_in_terms_of_amplitude}, rather than the
usual four~\cite{Peskin:1995}, because the $z$-component of momentum
is not conserved in a background with inhomogeneous particle masses.
The combination of Eqs.~\eqref{eq:tmatrix_with_wavefunction} and
\eqref{eq:tmatrix_element_in_terms_of_amplitude} leads to two
initial-state phase-space integrals, one of which can be evaluated
trivially using three of the $\delta$ functions. We obtain
\begin{equation}
  \begin{split}
    \bigl| \langle\vec{p}_{b_1,\rm s} \cdots \vec{p}_{b_n,\rm
      s}|i\hat{T}|\psi_a(\vec{p}_{a})\rangle \bigr|^2 =&\; \int \!
    \frac{\mathrm{d}^3 \vec{p}_{a,\rm s}^{\, \prime}}{(2\pi)^3}
    \frac{\bigl| \psi(\vec{p}_{a,\rm s}^{\, \prime}; \vec{p}_a)
      \bigr|^2}{(2E_a^{\prime})^2}\, (2\pi)^3
    \,\delta^{(3)}(\vec{p}_{a,\rm s}^{\, \prime} \!
    -\sum_{i=1}^n{\vec{p}}_{b_i,\rm s}) \, \bigl| \mathcal{M}_{a\to
      b_1 \ldots b_n} \bigr|^2 \;.
  \end{split}
\end{equation}
We now use the fact that the wave function $\psi(\vec{p}_{a,\rm s}^{\,
  \prime}; \vec{p}_a)$ is tightly peaked around $\vec{p}_a$ because
the incident particle has a well defined momentum. Formally, we can
write $|\psi(\vec{p}_{a,\rm s}^{\, \prime}; \vec{p}_a)|^2=
(2\pi)^3\,2E_a\,\delta^{(3)}(\vec{p}_{a,\rm
  s}^{\,\prime}-\vec{p}_{a})$.  The integral over $\vec{p}_{a,\rm
  s}^{\, \prime}$ is again trivial, and we obtain the differential
probability~\cite{Bodeker:2017cim}
\begin{align}\label{eq:dP_ab}
  \mathrm{d} \mathbb{P}_{a \to b_1 \cdots b_n} & = \frac{1}{2E_a}
  \Biggl[ \prod_{i=1}^n \frac{\mathrm{d}^3 \vec{p}_{b_i,\rm
        s}}{(2\pi)^32E_{b_i}} \Biggr] (2\pi)^3\,
  \delta^{(3)}({\vec{p}}_{a,\rm s} -\sum_{i=1}^n{\vec{p}}_{b_i,\rm s})
  \ |\mathcal{M}_{a \to b_1\cdots b_n}|^2 \;.
\end{align}
By combining Eqs.~\eqref{eq:pressure_sum}-
\eqref{eq:diff_flux}, \eqref{eq:Dp_avg}, \eqref{eq:Dp},
and~\eqref{eq:dP_ab}, and further assuming the high-energy limit, such
that $p_{a,z,\rm s}/E_a\approx 1$ in Eq.~\eqref{eq:diff_flux}, one
obtains a ``master formula'' for the thermal pressure, 
\begin{equation}\label{eq:master_formula}
\begin{split}
  P & = \sum_{a \in \mathcal{S}} \nu_a \int \!
  \frac{\mathrm{d}^3 \vec{p}_a}{(2\pi)^3 2E_a} \, f_a(\vec{p}_a)\,
  \sum_{n=1}^\infty \,
  \sum_{\{b\} \in \mathcal{S}} \Biggl[ \prod_{i=1}^n \int \! \!
    \frac{\mathrm{d}^3 \vec{p}_{b_i,\rm s}}{(2\pi)^32E_{b_i}}
    \Biggr] \\ & \qquad \times (2\pi)^3 \,
  \delta^{(3)}({\vec{p}}_{a,\rm s}
  -\sum_{i=1}^n{\vec{p}}_{b_i,\rm s}) \ |\mathcal{M}_{a \to
    b_1\cdots b_n}|^2 \bigl( p_{a,z,\mathrm{s}} - \sum_{i=1}^n p_{b_i,z,\mathrm{h}} \bigr) \,,
\end{split}
\end{equation}
that allows it to be calculated by specifying scattering processes and
calculating the associated scattering amplitudes.

\subsection{Transition radiation splitting}
\label{sec:pressure_at_lo}
As a first example, we will determine the $1\to 1$ transition matrix element to leading order in the perturbative expansion.  
This is given quite simply by comparing Eq.~\eqref{eq:tmatrix_element_in_terms_of_amplitude} with the normalization condition in Eq.~\eqref{eq:p_states} to find 
\begin{equation}\label{eq:lo_matrix_element}
    \langle \vec{p}_{b,\rm s} |\vec{p}_{a,\rm s}\rangle^{(0)} 
    = (2\pi)^3 \, \delta^{(3)}(\vec{p}_{a,\rm s}-\vec{p}_{b,\rm s}) \, \mathcal{M}_{a\to b}^{(0)} 
    \;, \qquad \text{with} \qquad 
    \mathcal{M}_{a\to b}^{(0)} = 2E_a 
    \;.
\end{equation}
The extension to fermion and vector fields is straightforward and yields identical results due to identical normalization of the one-particle eigenstates.  
Inserting the result into Eq.~\eqref{eq:master_formula} leads to the pressure formula for $1\to 1$ transitions which was derived in Ref.~\cite{Bodeker:2009qy}.

In what follows, the quantities we compute will be normalized to the
leading order matrix element $\mathcal{M}_{a\to b}^{(0)}$ for particle
$a$ interacting with the wall and producing particle $b$. This allows us 
to formulate the radiative corrections to the leading order $a \to b$ 
process probabilistically. The fact that $\mathcal{M}_{a\to b}^{(0)}=2E_{a}$
implies that all incoming particles, which couple to the Standard Model
Higgs condensate, will interact with the bubble wall with a probability 
of unity, independent of the size of their Yukawa coupling.
This is implicitly assumed in Refs.~\cite{Bodeker:2009qy, Bodeker:2017cim} and can be inferred from a VEV-insertion approximation where particle couplings to the Higgs VEV are treated as part of the interaction Hamiltonian. The quantity of interest is then the matrix element for $1\to 1$ transitions accompanied by an interaction with the wall. For massive SM species, this matrix element contains a factor of $m^2$ in the numerator (parametrizing the coupling to the wall) which is canceled by a factor of $1/\Delta p_z \approx 2E_a/m^2$ coming from the Fourier transform of the step function wall profile we assume, so that the matrix element is independent of the coupling to the Higgs field (as long as it is nonzero). The scaling of the pressure from $1 \to 1$ transitions, which is proportional to $\Delta m^2$, results from averaging the 
momentum transfer in the $z$-direction and not from  mass-dependence in the matrix element itself.

We emphasize, however, that in scenarios beyond the SM, this assumption may no longer apply. For example, a decoupled hidden sector particle will have zero probability of interacting with the wall and should not be counted as contributing to the pressure: even though naively one still has $\mathcal{M}_{a\to b}^{(0)}=2E_{a}$, the interpretation of $\mathcal{M}_{a\to b}^{(0)}$ as a matrix element for interacting with the wall no longer holds, since there is no coupling to the Higgs condensate in this case. We do not consider such scenarios further, and instead concentrate on the massive SM degrees of freedom. 

\begin{figure}[t]
\centering \includegraphics[scale=0.6]{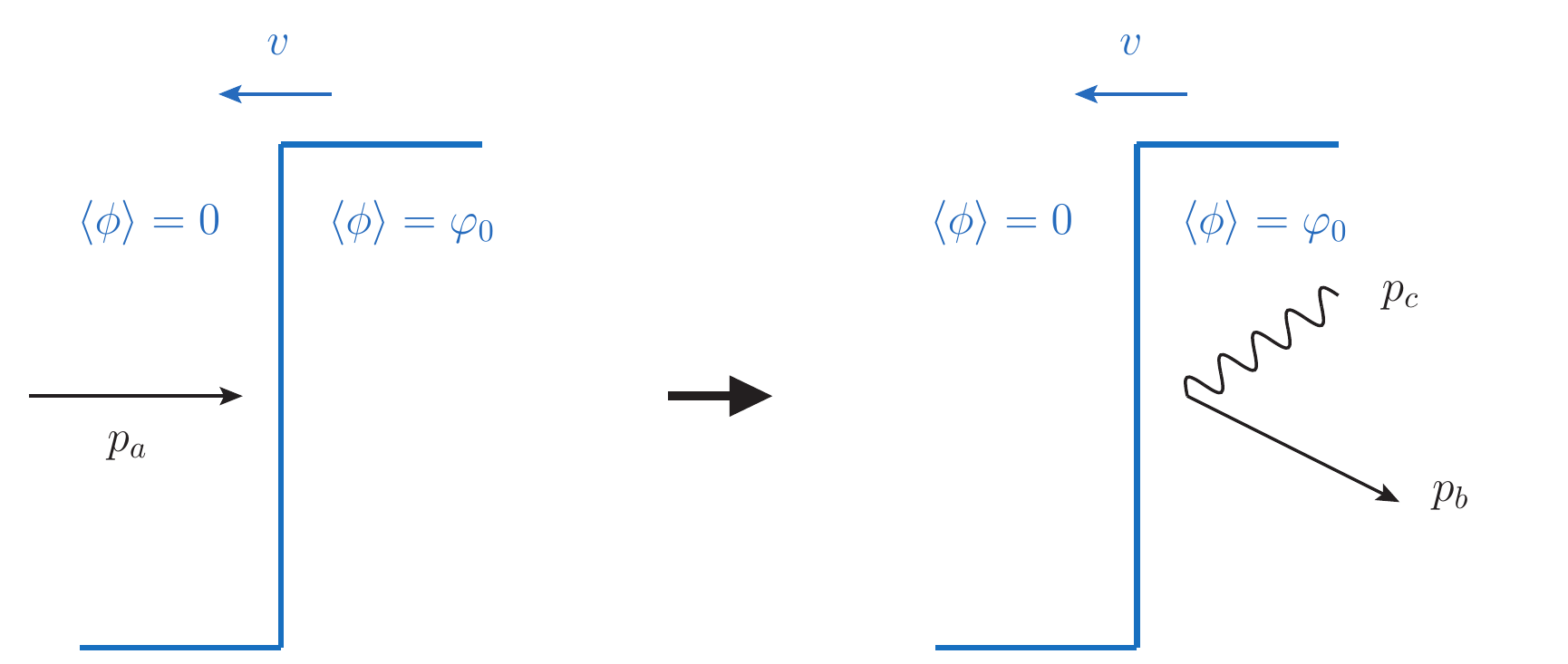}
\caption{Sketch of the real-emission kinematics.  The bubble wall is
  shown in blue and moves in the $-z$-direction with speed $v$ in the rest frame of the plasma in front of the wall.  The
  Higgs vacuum expectation value is denoted as $\langle \phi
  \rangle$. The incoming particle $a$ has a light-like four momentum
  $p_a$ in the wall's rest frame. The scattered particle $b$ and the soft emission, $c$, have
  four momenta $p_b$ and $p_c$.}
\label{fig:kinematics}
\end{figure}

From Eq.~(20) of Ref.~\cite{Bodeker:2017cim}, the leading order $1\to2$ particle scattering amplitude is given by 
\begin{equation}\label{eq:masterMSQ}
    \mathcal{M}_{a\to bc}^{(0)} 
    = 2iE_{a} \left( \frac{V_\mathrm{h}}{A_\mathrm{h}} - \frac{V_\mathrm{s}}{A_\mathrm{s}} \right) 
    \;
\end{equation}
where $V_\mathrm{h}$ and $V_\mathrm{s}$ are vertex functions, $A_\mathrm{h}$ and $A_\mathrm{s}$ are kinematical variables, and the subscripts indicate whether these quantities should be evaluated in the Higgs or the symmetric phase.  
Ref.~\cite{Bodeker:2017cim} writes\footnote{Here we denote the soft particle by $c$ and the limit of soft gauge boson emission
corresponds to $x \equiv E_c / E_a \ll 1$.  Ref.~\cite{Bodeker:2017cim} denotes the soft particle by $b$ instead and the limit
of soft emissions corresponds to $x \equiv E_b / E_a \ll 1$.}  
\begin{equation}\label{eq:As_Ah_from_BM}
\begin{split}
    A_\mathrm{s} 
    & = -2 E_a \bigl( p_{a,z,\mathrm{s}} - p_{b,z,\mathrm{s}} - p_{c,z,\mathrm{s}} \bigr) 
    \approx \vec{p}_{a,\perp}^{\,2} + m_{a,\mathrm{s}}^2 - \frac{\vec{p}_{b,\perp}^{\,2} + m_{b,\mathrm{s}}^2}{E_b/E_a} - \frac{\vec{p}_{c,\perp}^{\,2} + m_{c,\mathrm{s}}^2}{E_c/E_a} 
    \overset{x\ll 1}{\longrightarrow} - \frac{\vec{k}_\perp^{\,2} + m_{c,\mathrm{s}}^2}{x} \\
    A_\mathrm{h} 
    & = -2 E_a \bigl( p_{a,z,\mathrm{h}} - p_{b,z,\mathrm{h}} - p_{c,z,\mathrm{h}} \bigr) 
    \approx \vec{p}_{a,\perp}^{\,2} + m_{a,\mathrm{h}}^2 - \frac{\vec{p}_{b,\perp}^{\,2} + m_{b,\mathrm{h}}^2}{E_b/E_a} - \frac{\vec{p}_{c,\perp}^{\,2} + m_{c,\mathrm{h}}^2}{E_c/E_a} 
    \overset{x\ll 1}{\longrightarrow} - \frac{\vec{k}_\perp^{\,2} + m_{c,\mathrm{h}}^2}{x}
    \;.
\end{split}
\end{equation}
The first approximation is appropriate when all three particles are relativistic with large longitudinal momenta and energy is conserved.  
The second approximation is applicable in the frame where the incoming particle has zero transverse momentum, implying  $\vec{p}_{a,\perp} = 0$ and $\vec{p}_{c,\perp} = - \vec{p}_{b,\perp} \equiv \vec{k}_\perp$, and under the assumption that particle $c$ is soft; \textit{i.e.}, its energy fraction is $x=E_c/E_a \ll 1$ such that the $\mathcal{O}(1/x)$ term dominates.  
The vertex functions, $V_\mathrm{h}$ and $V_\mathrm{s}$, depend on the particles' spins and the nature of their three-body interaction.  
They are related to the Altarelli-Parisi splitting functions.
For example, in the case of soft vector boson emission, Table~I of Ref.~\cite{Bodeker:2017cim} gives the leading squared vertex function in the soft limit, $x \ll 1$, to be 
\begin{equation}\label{eq:Vs_vh_from_BM}
    |V_\mathrm{h}|^2 
    = |V_\mathrm{s}|^2 
    \approx 4 |g|^2 \frac{\vec{k}_\perp^{\,2}}{x^2} 
    \,,
\end{equation}
where $|g|^2$ is the coupling squared, which we define such as to include the appropriate quadratic Casimir operator $C_2[R]$.
In combination with the soft-limit expressions in Eq.~\eqref{eq:As_Ah_from_BM} the squared matrix element becomes 
\begin{equation}\label{eq:Masq_from_BM}
    |\mathcal{M}_{a\to bc}^{(0)}|^2 = 16 E_a^2|g|^2
    \frac{\vec{k}_\perp^{\,2} \ (m_{c,\mathrm{h}}^2 - m_{c,\mathrm{s}}^2)^2}{(\vec{k}_\perp^{\,2} + m_{c,\mathrm{h}}^2)^2 \ (\vec{k}_\perp^{\,2} + m_{c,\mathrm{s}}^2)^2} 
    \;.
\end{equation}
If the mass of the soft particle does not change across the wall, $m_{c,\mathrm{s}} = m_{c,\mathrm{h}}$, then the squared matrix element vanishes at $\mathcal{O}(x^0)$.  
If the masses are hierarchical, $m_{c,\mathrm{s}} \ll m_{c,\mathrm{h}}$, then the transverse momentum integral has a logarithmic enhancement, $\int \! {\rm d}^2\vec{k}_\perp |\mathcal{M}_{a\to bc}^{(0)}|^2 \propto \log(m_{c,\mathrm{h}}^2/m_{c,\mathrm{s}}^2)$~\cite{Vanvlasselaer:2020niz}.

In the analysis presented here, we do not use Eq.~\eqref{eq:As_Ah_from_BM} to evaluate the kinematical variables, $A_\mathrm{s}$ and $A_\mathrm{h}$, and we do not use Eq.~\eqref{eq:Vs_vh_from_BM} to evaluate the vertex factors, $V_\mathrm{s}$ and $V_\mathrm{h}$.  
Instead we write the kinematical variables as 
\begin{equation}\label{eq:As_Ah_def}
\begin{split}
    A_\mathrm{s} 
    & = -2p_{a,\mathrm{s}}p_{c,\mathrm{s}}
    \approx 2 \vec{p}_{a,\perp} \cdot \vec{p}_{c,\perp} - \frac{\vec{p}_{a,\perp}^{\,2} + m_{a,\mathrm{s}}^2}{E_a/E_c} - \frac{\vec{p}_{c,\perp}^{\,2} + m_{c,\mathrm{s}}^2}{E_c/E_a} 
    \overset{x\ll 1}{\longrightarrow} - \frac{\vec{k}_\perp^{\,2} + m_{c,\mathrm{s}}^2}{x} \\ 
    A_\mathrm{h} 
    & = -2p_{b,\mathrm{h}}p_{c,\mathrm{h}}
    \approx 2 \vec{p}_{b,\perp} \cdot \vec{p}_{c,\perp} - \frac{\vec{p}_{b,\perp}^{\,2} + m_{b,\mathrm{h}}^2}{E_b/E_c} - \frac{\vec{p}_{c,\perp}^{\,2} + m_{c,\mathrm{h}}^2}{E_c/E_b} 
    \overset{x\ll 1}{\longrightarrow} - \frac{\vec{k}_\perp^{\,2} + m_{c,\mathrm{h}}^2}{x}
\end{split}
\end{equation}
We choose Eq.~\eqref{eq:As_Ah_def} instead of Eq.~\eqref{eq:As_Ah_from_BM} since the former leads to 
a squared matrix element that matches the calculation in Sec.~\ref{sec:framework}, where particles 
hitting the bubble wall are treated semi-classically as decelerated charges emitting soft gauge field quanta.
Moreover it leads to a gauge invariant expression for the matrix element, as discussed both in Sec.~\ref{sec:framework} and App.~\ref{app:scalarQED}.
Note that the expressions for $A_\mathrm{s}$ and $A_\mathrm{h}$ in Eqs.~\eqref{eq:As_Ah_from_BM} and \eqref{eq:As_Ah_def} are equivalent in the limit, $x \ll 1$, which is the primary interest of Ref.~\cite{Bodeker:2017cim}. 
In our approach, the vertex factors in axial gauge with gauge vector $n^\mu=p_{b,\mathrm{h}}^\mu$ are given by 
(see App.~\ref{app:scalarQED} for a complete calculation)
\begin{equation}\label{eq:Vs_Vh_def}
\begin{split}
    |V_\mathrm{h}|^2 & = V_\mathrm{h}^\ast V_\mathrm{s} = V_\mathrm{s}^\ast V_\mathrm{h} = 0 \\
    |V_\mathrm{s}|^2 & = 4 \lvert g \rvert^2 \left(
    \frac{2(p_{a,\mathrm{s}}p_{b,\mathrm{h}})(p_{a,\mathrm{s}}p_{c})}{p_{b,\mathrm{h}}p_c}
    -\frac{p_{b,\mathrm{h}}^2 (p_{a,\mathrm{s}}p_{c})^2}{(p_{b,\mathrm{h}}p_c)^2} 
    -p^2_{a,\mathrm{s}}
    \right) \\
    &\approx\; 4|g|^2 \,\vec{k}_\perp^{\,2}\left(\frac{\vec{k}_\perp^{\,2} + x ( m_{b,\mathrm{h}}^2 - (1-x) m_{a,\mathrm{s}}^2)}{
    \vec{k}_\perp^{\,2} + x^2 m_{b,\mathrm{h}}^2}\right)^2
    \underset{\vec{k}_\perp^2\ll x^2m_{b,\mathrm{h}}^2}{\overset{m_{a,\mathrm{s}}^2\ll m_{b,\mathrm{h}}^2}{\longrightarrow}}
    4|g|^2 \,\frac{\vec{k}_\perp^{\,2}}{x^2}\;.
\end{split}
\end{equation}
We take the soft particle's mass to vanish so that $p_{c,\mathrm{s}} = p_{c,\mathrm{h}} \equiv p_c$.  
As a direct consequence, we find that the complete result for the matrix element of Eq.~\eqref{eq:masterMSQ} 
squared is given by
\begin{equation}
    |\mathcal{M}_{a\to bc}^{(0)}|^2=4E_a^2\frac{\left|V_s\right|^2}{A_s^2}\Big|_{n^\mu=p_{b,h}^\mu}\;.
\end{equation}
While this result has been derived in a particular gauge, for which it coincides with the form given
in~\cite{Bodeker:2017cim} in the ultra-collinear limit, $m_{a,\mathrm{s}}^2\ll m_{b,\mathrm{h}}^2$ and
$\vec{k}_\perp^2\ll x^2m_{b,\mathrm{h}}^2$, we emphasize that our calculation is gauge invariant and agrees 
with the semi-classical result in Sec.~\ref{sec:fixed_order}, which provides an important physical cross-check. 
The suppression factor $(m_{c,\mathrm{h}}^2-m_{c,\mathrm{s}}^2)^2$ in the numerator of Eq.~\eqref{eq:Masq_from_BM}, 
which was derived based on the assumption $V_s=V_h$ in Eq.~\eqref{eq:Vs_vh_from_BM} is therefore incorrect
and leads to a severe underestimation of the pressure on the bubble wall. 
This will be further discussed in Sec.~\ref{sec:analytic_pressure}, 
in particular Eq.~\eqref{eq:caesar_pressure_fc_fo_cut}.
The complete squared matrix element can be written as 
\begin{equation}\label{eq:bm_correct}
    \begin{split}
    |\mathcal{M}_{a\to bc}^{(0)}|^2
    =&\;4E_a^2 \lvert g \rvert^2 \Bigg(\frac{2p_{a,\mathrm{s}}p_{b,\mathrm{h}}}{ p_{a,\mathrm{s}}p_{c}\;p_{b,\mathrm{h}}p_{c}}-\frac{m_{a,\mathrm{s}}^2}{(p_{a,\mathrm{s}}p_{c})^2
    }-\frac{m_{b,\mathrm{h}}^2}{(p_{b,\mathrm{h}}p_{c})^2}\,\Bigg)\\
    \approx&\; 8E_a^2|g|^2 \,(2x^2) \frac{\vec{k}_\perp^{\,2} \bigl( \vec{k}_\perp^{\,2} + x ( m_{b,\mathrm{h}}^2 - (1-x) m_{a,\mathrm{s}}^2) \bigr)^2}{\bigl( \vec{k}_\perp^{\,2} + x^2 m_{a,\mathrm{s}}^2 \bigr)^2 \bigl( \vec{k}_\perp^{\,2} + x^2 m_{b,\mathrm{h}}^2 \bigr)^2} \;,
    \end{split}
\end{equation}
where the approximation makes use of the first (but not the second) approximation in Eq.~\eqref{eq:As_Ah_def}.
Note that this takes a particularly simple form if the mass of the incident particle is negligible, $m_{a,\mathrm{s}}\to 0$.
As discussed in Sec.~\ref{sec:fixed_order}, Eq.~\eqref{eq:bm_correct} is the matrix element squared of an accelerated 
semi-classical charge emitting soft gauge field quanta. A mass change across the bubble wall is generally necessary 
to generate a nonzero pressure on the wall. In our analysis, this change arises from the hard fermion, 
scalar or vector particle acquiring a mass, while the radiated vector boson can remain massless.

The most relevant limit of Eq.~\eqref{eq:bm_correct} is the small mass limit.  
Taking the incident particle's mass to be small gives
\begin{equation}\label{eq:bm_correct_approx}
    |\mathcal{M}_{a\to bc}^{(0)}|^2\overset{m_{a,\mathrm{s}}^2\ll\vec{k}_\perp^2,m_{b,\mathrm{h}}^2}{\longrightarrow}\;
    8E_a^2|g|^2\,\frac{2x^2}{\vec{k}_\perp^{\,2}}
        \left(\frac{\vec{k}_\perp^{\,2} + x\, m_{b,\mathrm{h}}^2}{\vec{k}_\perp^{\,2} + x^2 m_{b,\mathrm{h}}^2}\right)^2\;.
\end{equation}
In the limit of two small masses, with $m_{a,\mathrm{s}}\neq m_{b,\mathrm{h}}$, 
the result can be further simplified to 
\begin{equation}\label{eq:bm_correct_smallm}
    |\mathcal{M}_{a\to bc}^{(0)}|^2\overset{m_{a,\mathrm{s}}^2, m_{b,\mathrm{h}}^2\ll k_\perp^2}{\longrightarrow}\;8E_a^2|g|^2\,\frac{2x^2}{\vec{k}_\perp^{\,2}}\;.
\end{equation}
Note that Eq.~\eqref{eq:bm_correct_smallm} is independent of the mass difference
and is valid for any $m_{a,\mathrm{s}}\neq m_{b,\mathrm{h}}$. However, the result 
does vanish for $m_{a,\mathrm{s}}=m_{b,\mathrm{h}}$, due to overall momentum conservation 
and the constraint that the bubble wall does not absorb transverse momentum.
The small mass limit in Eq.~\eqref{eq:bm_correct_smallm} turns out to be the most relevant 
for the computation of the average pressure on the bubble wall. It will be analyzed 
in more detail in Sec.~\ref{sec:analytic_pressure} and~\ref{sec:ps}.
We conclude this section with a brief discussion of the singularity structure of
Eq.~\eqref{eq:bm_correct_approx}. Combining the matrix element squared with the 
differential phase space element for the emission of $p_c$ in the limit of large 
longitudinal momenta (cf.\ App.~\ref{sec:ps_kin}), and dividing by the leading-order result, 
Eq.~\eqref{eq:lo_matrix_element}, we obtain
\begin{equation}\label{eq:emission_probability_correct}
    \frac{1}{|\mathcal{M}_{a\to b}^{(0)}|^2}\frac{{\rm d}^3\vec{p}_c}{(2\pi)^3\,2E_c}
    |\mathcal{M}_{a\to bc}^{(0)}|^2=\frac{\alpha}{2\pi}\,
    \frac{{\rm d}\vec{k}_\perp^{\,2}}{\vec{k}_\perp^{\,2}}\,{\rm d}x\,2C_{abc}\,x
    \left(\frac{\vec{k}_\perp^{\,2} + x\, m_{b,\mathrm{h}}^2}{\vec{k}_\perp^{\,2} + x^2 m_{b,\mathrm{h}}^2}\right)^2\;,
\end{equation}
where $\alpha=|g|^2/(4\pi C_{abc})$ and $C_{abc}$ denotes the appropriate charge for the
transition $a\to bc$. It corresponds to the Casimir operator $C_2[R]$ in~\cite{Bodeker:2017cim}.
The $x$-integration is kinematically restricted by the transverse momentum and mass,
but in the limit of large incident energies these restrictions can be neglected,
and we obtain the approximate differential radiation probability
\begin{equation}\label{eq:emission_probability_correct_integrated}
    \frac{1}{|\mathcal{M}_{a\to b}^{(0)}|^2}\int\frac{{\rm d}^3\vec{p}_c}{(2\pi)^3\,2E_c}
    |\mathcal{M}_{a\to bc}^{(0)}|^2\approx\left\{\begin{array}{ccc}
    \displaystyle\frac{\alpha}{2\pi}\,\int\frac{{\rm d}\vec{k}_\perp^{\,2}}{\vec{k}_\perp^{\,2}}\,
    C_{abc}\log\frac{m_{b,\mathrm{h}}^2}{\vec{k}_\perp^{\,2}}
    & \quad\text{if}\quad& m_{a,\mathrm{s}}^2\ll k_\perp^2\ll m_{b,\mathrm{h}}^2\\[5mm]
    \displaystyle\frac{\alpha}{2\pi}\,\int\frac{{\rm d}\vec{k}_\perp^{\,2}}{\vec{k}_\perp^{\,2}}\,C_{abc} 
    & \quad\text{if}\quad& m_{a,\mathrm{s}}^2,m_{b,\mathrm{h}}^2\ll k_\perp^2
    \end{array}
    \right.\;,
\end{equation}
where terms subdominant in $\vec{k}_\perp^2$ have been neglected.
The double logarithmic behavior for soft vector boson emission in the region 
of small transverse momenta, and the single logarithmic behavior in the region 
of large transverse momenta necessitate the resummation of radiative corrections 
to all orders in perturbation theory. This will be discussed in more detail 
in the following section.

\section{Factorization and resummation of radiative corrections}
\label{sec:framework}
It was highlighted in Ref.~\cite{Bodeker:2017cim} that the transition
radiation effects discussed in Sec.~\ref{sec:pressure_at_lo}
significantly alter the pressure transfer in the $1\to 1$ transition.
The changes originate in logarithmically-enhanced radiative
corrections to the light-to-heavy current parametrizing a fast
particle that crosses the domain wall.  For massless emissions, these
logarithms will become infrared poles, which are canceled to all
orders by the virtual corrections to the light-to-heavy
transition. The appropriate treatment for such effects is Sudakov
resummation~\cite{Sudakov:1954sw}. Running coupling effects can easily
be incorporated in the calculation~\cite{Amati:1980ch}, and certain
higher-logarithmic corrections may be resummed for simple observables
as well~\cite{Catani:1990rr}.  In this section, we will develop the
formalism and compute an analytic estimate of the pressure at leading
logarithmic accuracy.

\subsection{Fixed-order perturbative computation}
\label{sec:fixed_order}
We derive the leading logarithmic approximation of the emission rate
using source theory. Due to the universal structure of the matrix
elements in the soft gauge boson limit, our eventual result will
resemble the treatment of infrared divergences in
QED~\cite{Yennie:1961ad}. The formal derivation here is strictly valid
only for non-flavor-changing reactions, but it can easily be extended
to flavor-changing processes. We will also find that non-flavor-changing reactions of QCD type provide the
largest contribution to the overall momentum transfer in the Standard
Model, which can be traced back to the relatively large number of
degrees of freedom, $\nu_q=2\times 3$ for quarks and the large number
of quarks (see Tab.~\ref{fig:pressure_comparison} for details).  This may seem somewhat surprising, given the small Yukawa couplings of the light quarks to the Higgs condensate, but follows from taking the probability for all massive SM species to interact with the bubble wall to be unity
as a consequence of the perturbative calculation in the VEV insertion approximation.

We
begin with a classical vector current associated with the moving
charge of the incoming particle $a$. This will allow us to treat the
radiative corrections for scalars, fermions and vector bosons in a
unified way.  The classical current is parametrized as
\begin{equation}
    j^\mu(x)=g\int{\rm d}s\,\frac{dy^\mu(s)}{ds}\,\delta^{(4)}(x-y(s))\, .
\end{equation}
In this context, $g$ is the coupling, $y^\mu(s)$ is the particle's
location, ${\rm d}s$ is the differential line element, and we
integrate along the particle's trajectory.  In momentum space, this
current reads
\begin{equation}
  j^\mu(k)=\int{\rm d}^{4}x\,e^{ikx}\,j^\mu(x)
  = g\int{\rm d}s\,\frac{dy^\mu(s)}{ds}\,e^{iky(s)}\,.
\end{equation}
For simplicity, throughout our analysis we will consider the thin wall limit, $mL_w \ll 1$, with $L_w$ the wall width, for which the bubble profile can be approximated as a step function at $z=0$.
We can then parametrize the line element as ${\rm d}s={\rm d}z$, and
write $y^\mu(s) = z\, p^\mu(z) / p_0(z)$ where we have $p(z) \to p_a$
if $z < 0$ and $p(z) \to p_b$ if $z > 0$.  This parametrization leads
to\footnote{ Note that Eq.~\eqref{eq:soft current_pre} is
  structurally equivalent to Eq.~(20) in Ref.~\cite{Bodeker:2017cim}
  in the soft limit, $x \to 0$.}
\begin{equation}\label{eq:soft current_pre}
  j^\mu(k)=g\int_{-\infty}^0{\rm d}z\,
  \frac{p_a^\mu}{p_{a,0}}\,
  \exp\left\{i\,\frac{p_ak}{p_{a,0}} z\right\}
  +g\int_0^{+\infty}{\rm d}z\,
  \frac{p_b^\mu}{p_{b,0}}\,
  \exp\left\{i\,\frac{p_bk}{p_{b,0}} z\right\}\,.
\end{equation}
Upon inserting a regulator, we obtain the classical current
\begin{equation}\label{eq:soft_current}
  \begin{split}
    j^\mu(k)=&\;g\int_{-\infty}^0{\rm d}z\,
    \frac{p_a^\mu}{p_{a,0}}\,
    \exp\left\{i\left(\frac{p_ak}{p_{a,0}}-i\eps\right) z\right\}
    +g\int_0^{+\infty}{\rm d}z\, \frac{p_b^\mu}{p_{b,0}}\,
    \exp\left\{i\left(\frac{p_bk}{p_{b,0}}+i\eps\right) z\right\}\\
    =&\;ig\left(\frac{p_b^\mu}{p_bk+i\eps}-
    \frac{p_a^\mu}{p_ak-i\eps}\right)\,.
  \end{split}
\end{equation}
Eq.~\eqref{eq:soft_current} could alternatively be obtained by
matching the result of the full QFT to the soft
limit~\cite{Becher:2014oda}, and by rewriting the factorized form of
the matrix element $\mathcal{M}_{a\to b}^{(1)}$ in
Sec.~\ref{sec:pressure_at_lo}~\cite{Bassetto:1984ik}.  We now proceed
to compute the radiation field $A_\mu(x)$ of this current.  Note that
only the cross section for radiation of massless vector boson fields
exhibits a double logarithmic enhancement in the soft region (a double
pole in dimensional regularization). If we work in the soft
approximation we therefore do not need to consider the radiation of
scalars and fermions. Let us consider the interaction Hamiltonian
density
\begin{equation}
  \mathcal{H}_{\rm int}(x)=j^\mu(x)A_\mu(x)\;,
\end{equation}
which we require to derive the vacuum persistence amplitude,
\begin{equation}\label{eq:vacuum_amplitude}
  W_{a\to b}=\langle 0|T\left[\exp
    \left\{i\int{\rm d}^{4}x\,j^\mu(x)A_\mu(x)\right\}\right]|0\rangle\,.
\end{equation}
The probability of no emission off the classical current is given by
$|W_{a\to b}|^2$. Note that the only dynamical degrees of freedom in
this calculation are the vector bosons radiated by the fast, classical
particle, hence the above notion of a ``vacuum'' persistence amplitude
is justified. In terms of the matrix elements used in
Eq.~\eqref{eq:master_formula}, we find, in the soft limit
\begin{equation}\label{eq:wab_relation_to_pab}
  |W_{a\to b}|^2=\frac{|\mathcal{M}_{a\to b}|^2}{|\mathcal{M}_{a\to b}^{(0)}|^2}\;,
\end{equation}
where $\mathcal{M}_{a\to b}^{(0)}$ is given by Eq.~\eqref{eq:lo_matrix_element}.
The vacuum persistence amplitude can be expanded into a power series 
in the coupling constant, $g$, as
$W_{a\to b}=\sum W_{a\to b}^{(n)}/n!$, with $W_{a\to b}^{(n)} \propto g^n$.
The zeroth order term is trivially $W_{a\to b}^{(0)}=1$. 
The first order term vanishes, as $\langle0|A_\mu(x)|0\rangle=0$.
The second-order term is the first non-trivial result and gives
\begin{equation}\label{eq:w2}
    \begin{split}
    W_{a\to b}^{(2)}=&\;-\int{\rm d}^4x\int{\rm d}^4y\,j^\mu(x)j^\nu(y)
    \langle0|T\left[A_\mu(x)A_\nu(y)\right]|0\rangle
    =\;-\int{\rm d}^4x\int{\rm d}^4y\,j^\mu(x)i\Delta_{F,\mu\nu}(x,y)j^\nu(y)\,.
    \end{split}
\end{equation}
This is a very intuitive result, as it describes the emission
and re-absorption of a soft field
quantum  by the same classical current after propagation
from $x$ to $y$. The propagation is described by the time-ordered 
Green's function $i\Delta_F(x,y)$. It can be written as
\begin{equation}\label{eq:greens_function}
\begin{split}
    i\Delta_F^{\mu\nu}(x,y)=&\;
    \Theta(y_0-x_0)\langle 0|A^\nu(y)A^\mu(x)|0\rangle
    +\Theta(x_0-y_0)\langle 0|A^\mu(x)A^\nu(y)|0\rangle\\
    =&\;\int\!\frac{{\rm d}^3\vec{k}}{(2\pi)^3\,2E_k}\left[
    \Theta(y_0-x_0)e^{-ik(y-x)}+\Theta(x_0-y_0)e^{ik(y-x)}\right]
    \sum_{\lambda=\pm}\eps_\lambda^\mu(k,l)\eps_\lambda^{\nu\,*}(k,l)\,,\\
    \end{split}
\end{equation}
where $\vec{k}$ is the soft particle's momentum, and $E_k$ is determined 
by the dispersion relation $E_k^2=\vec{k}^2+m(z)^2$, with $m(z)$ the
position dependent mass of the soft emission. We will discuss this 
position dependence further below.
The polarization vectors $\eps_\lambda^\mu$ can be constructed
for example by using the Weyl-van-der-Waerden spinor
formalism~\cite{Dixon:1996wi,Dittmaier:1998nn}.
For massless vector bosons they obey the relation
\begin{equation}\label{eq:polarisation_sum_wvdw}
  \sum_{\lambda=\pm}\,\varepsilon^\mu_\lambda(k,l)\,
  \varepsilon^{\nu\,*}_\lambda(k,l)\,
  =\;-g^{\mu\nu}+\frac{k^\mu l^\nu+ k^\nu l^\mu}{k l}\;,
\end{equation}
where $l$ represents a light-like auxiliary vector,
that must not be parallel to $k$.
Eq.~\eqref{eq:polarisation_sum_wvdw} will be sufficient
for computing helicity summed amplitudes in the remainder of
this section. We note that for massive bosons one instead
obtains the polarization sum
\begin{equation}\label{eq:polarisation_sum_wvdw_ms}
  \sum_{\lambda=\pm,0}\,\varepsilon^\mu_\lambda(k,l)\,
  \varepsilon^{\nu\,*}_\lambda(k,l)\,
  =\;-g^{\mu\nu}+\frac{k^\mu k^\nu}{k^2}\,,
\end{equation}
where the individual polarization vectors depend on $l$,
while their sum does not. Eq.~\eqref{eq:polarisation_sum_wvdw_ms} 
applies to all massive gauge bosons of the Standard Model
in the high-energy limit $E\gg m$ due to the Goldstone boson
equivalence theorem~\cite{Denner:2019vbn,Denner:2006xx}.
Both polarization sums lead to the same results when squared
matrix elements are computed using the soft current defined
in Eq.~\eqref{eq:soft_current}.

The on-shell mass $m(z)$ of the emitted vector boson may change
at the domain wall, which affects $E_k$ by means of the dispersion relation.
However, this is a dynamical effect, which must be described by (resummed) 
higher-order corrections in the VEV insertion approximation~\cite{
  Carena:2000id,Carena:2002ss,Lee:2004we}. We neglect these corrections here,
because they are suppressed by $\mathcal{O}(1/(\alpha\log(\gamma T/m)))$
and therefore irrelevant in the high-energy limit $\gamma\gg 1$.
In contrast, mass insertions scale like the leading-order terms in
Sec.~\ref{sec:pressure_at_lo}, and are therefore $\gamma$-independent.
In Sec.~\ref{sec:ps} we will nevertheless include all kinematic effects 
for $m_c\neq 0$ (cf.\ App.~\ref{sec:ps_kin}).

Using these approximations, we obtain the massless Feynman propagator
\begin{equation}\label{eq:greens_function_2}
\begin{split}
    \Delta_F^{\mu\nu}(x,y)=&\;
    \int\frac{{\rm d}^4k}{(2\pi)^4}
    \frac{e^{-ik(y-x)}}{k^2+i\eps}
    \sum_{\lambda=\pm}\eps_\lambda^\mu(k)\eps_\lambda^{\nu\,*}(k)\;.
    \end{split}
\end{equation}
Inserting Eq.~\eqref{eq:greens_function_2} into Eq.~\eqref{eq:w2} yields 
\begin{equation}
    \begin{split}
    W_{a\to b}^{(2)}=&\;-i\int{\rm d}^4x\int{\rm d}^4y\,
    \int\frac{{\rm d}^4k}{(2\pi)^4}
    \frac{e^{-ik(y-x)}}{k^2+i\eps}
    \sum_{\lambda=\pm}\big(j(x)\eps_\lambda(k)\big)\big(j(y)\eps_\lambda(k)\big)^*\\
    =&\;-i\int\frac{{\rm d}^4k}{(2\pi)^4}
    \frac{1}{k^2+i\eps}
    \sum_{\lambda=\pm}\big(j(k)\eps_\lambda(k)\big)\big(j(k)\eps_\lambda(k)\big)^*\;.
    \end{split}
\end{equation}
Inserting the soft current of Eq.~\eqref{eq:soft_current} and using the 
completeness relation, Eq.~\eqref{eq:polarisation_sum_wvdw} 
or~\eqref{eq:polarisation_sum_wvdw_ms} we obtain
\begin{equation}\label{eq:nlo_virtual_diff}
    \begin{split}
    W_{a\to b}^{(2)}=&\;-i\,|g|^2\int\frac{{\rm d}^4k}{(2\pi)^4}\frac{1}{k^2+i\eps}
    \left(\frac{2p_ap_b}{(p_ak)(p_bk)}
    -\frac{m_a^2}{(p_ak)^2}
    -\frac{m_b^2}{(p_bk)^2}
    \right)\;.
    \end{split}
\end{equation}
Note that while Eq.~\eqref{eq:polarisation_sum_wvdw} was derived in
a light-like axial gauge, the result is actually gauge independent.
The infrared divergent part of Eq.~\eqref{eq:nlo_virtual_diff}
can be computed using dimensional regularization in $D=4-2\eps$ dimensions with $\eps<0$.  
For one massive particle, $b$, we obtain, in the $\overline{\rm MS}$ scheme
(cf.\ App.~\ref{sec:virtual_integrals})
\begin{equation}\label{eq:onetoone_virtual}
    W^{(2)\,\rm IR}_{a\to b}=-\frac{\alpha}{\pi}\,C_{abc}
    \left(\frac{1}{2\eps^2}-\frac{1}{2\eps}
    \left(1+\log\frac{(2p_ap_b)^2}{\mu^2 p_b^2}\right)
    +\frac{1}{4}\log^2\frac{(2p_ap_b)^2}{\mu^2p_b^2}
    -\frac{1}{2}\log^2\frac{2p_ap_b}{p_b^2}+\ldots\right)\,.
\end{equation}
The quantity $4\pi\alpha=|g|^2$ is the coupling squared for the transition $a\to bc$,
and $C_{abc}$ is an associated charge factor in the collinear limit
(see App.~\ref{sec:vertex_functions} for details) and $\mu$ is the renormalization scale. 
We have only listed the poles and leading logarithmic terms, as the subleading logarithmic
and finite contributions are irrelevant for the resummation we intend to perform.
Note that, as $p_b^2\to 0$, Eq.~\eqref{eq:onetoone_virtual} develops an additional
infrared singularity. Comparing with Eq.~\eqref{eq:i1ir_zero}, one finds that the
leading pole is then doubled, which agrees with the intuitive notion that two
massless charged particles will radiate twice as many gauge bosons as a 
single particle. Note that the scenario of two massless particles serves only
as  a cross-check on our result. It is not relevant in practice because
the hard scale $2p_a p_b$ would tend to zero in this case. 

We now proceed to compute the real-emission corrections. 
The all-orders single emission amplitude squared is 
\begin{equation}\label{eq:Wabc}
    {\rm d}W_{a\to bc}^2(p_c)=
    \frac{{\rm d}^3\vec{p}_c}{(2\pi)^3\,2E_c}
    \left| \langle \vec{p}_c|T\left[\exp\left\{i\int{\rm d}^4x\,j^\mu(x)A_\mu(x)\right\}\right]|0\rangle\right|^2\;.
\end{equation}
It is related to the matrix elements used in Eq.~\eqref{eq:master_formula}
in the soft limit as
\begin{equation}
  {\rm d}W_{a\to bc}^2(p_c)
  =\frac{{\rm d}^3\vec{p}_c}{(2\pi)^3\,2E_c}\,
  \frac{|\mathcal{M}_{a\to bc}|^2}{|\mathcal{M}_{a\to b}^{(0)}|^2}\;,
\end{equation}
where $\mathcal{M}_{a\to b}^{(0)}$ is given by Eq.~\eqref{eq:lo_matrix_element}.
Eq.~\eqref{eq:Wabc} can be expanded into a power series in the 
coupling constant, $g$, as ${\rm d}W_{a\to bc}(p_c)=
\sum {\rm d}W_{a\to bc}^{(n)}(p_c)/n!$, with 
${\rm d}W_{a\to bc}^{(n)}(p_c) \propto g^n$.
The zeroth order term vanishes, as $\langle \vec{p}_c|0\rangle=0$.
The first-order term is the first non-trivial result and gives
\begin{equation}\label{eq:w1_real_diff}
    \begin{split}
    {\rm d}W_{a\to bc}^{2\,(1)}(p_c)=&\;
    \frac{{\rm d}^3\vec{p}_c}{(2\pi)^3\,2E_c}
    \bigg|\;i\int{\rm d}^4x\,j^\mu(x)\langle \vec{p}_c|A_\mu(x)|0\rangle\bigg|^2
    =\;-\frac{{\rm d}^3\vec{p}_c}{(2\pi)^3\,2E_c}
    \sum_{\lambda=\pm}\big(j(p_c)\eps_\lambda(p_c)\big)\big(j(p_c)\eps_\lambda(p_c)\big)^*\;,
    \end{split}
\end{equation}
which we integrate  over the full final-state phase space in order
to obtain the correction to the inclusive rate:
\begin{equation}\label{eq:w1_real}
    \begin{split}
      \int{\rm d}W_{a\to bc}^{2\,(1)}=&\;
    |g|^2\int \! \frac{{\rm d}^3\vec{p}_c}{(2\pi)^3\,2E_c}
    \left(\frac{2p_ap_b}{(p_ap_c)(p_bp_c)}
    -\frac{m_a^2}{(p_ap_c)^2}-\frac{m_b^2}{(p_bp_c)^2}
    \right)\;.
    \end{split}
\end{equation}
The infrared divergent part of Eq.~\eqref{eq:w1_real} 
can be extracted using dimensional regularization
in $D=4-2\eps$ dimensions. 
We obtain (cf.\ App.~\ref{sec:real_integrals})
\begin{equation}\label{eq:onetoone_real}
    \int{\rm d}W_{a\to bc}^{2\,(1)\,\rm IR}=+\frac{\alpha}{\pi}\,C_{abc}
    \left(\frac{1}{2\eps^2}-\frac{1}{2\eps}
    \left(1+\log\frac{(2p_ap_b)^2}{\mu^2 p_b^2}\right)
    +\frac{1}{4}\log^2\frac{(2p_ap_b)^2}{\mu^2p_b^2}
    -\frac{1}{2}\log^2\frac{2p_ap_b}{p_b^2}+\ldots\right)\;.
\end{equation}
Using the first-order expansion of $W_{a\to b}$ and $\int{\rm d}W_{a\to bc}^{2}$,
we find the no-emission and integrated one-emission probability are
\begin{equation}
\begin{split}
    \mathbb{P}_{a\to b}^{\rm IR\,(1)}=&\;\Big|\,1+
    \frac{W_{a\to b}^{(2)\,\rm IR}}{2!}+\mathcal{O}(\alpha^2)\,\Big|^2
    =1+W_{a\to b}^{(2)\,\rm IR}+\mathcal{O}(\alpha^2)\;,\\
    \int{\rm d} \mathbb{P}_{a\to bc}^{\rm IR\,(1)}
    =&\;\int{\rm d}W_{a\to bc}^{2\,(1)\,\rm IR}+\mathcal{O}(\alpha^2)
    =-W_{a\to b}^{(2)\,\rm IR}+\mathcal{O}(\alpha^2)\;.    
\end{split}
\end{equation}
The singular terms in the amplitudes cancel at first order in perturbation theory\footnote{Note that the cancellation of singularities can be derived
more elegantly. The loop integrand can be rewritten using
\begin{equation}
    \frac{1}{k^2+i\eps}={\rm PV}\,\frac{1}{k^2}-i\pi\delta(k^2)\;,
\end{equation}
where ${\rm PV}$ stands for the principal value.
This implies that Eq.~\eqref{eq:onetoone_real} can be obtained
from Eq.~\eqref{eq:onetoone_virtual} and vice 
versa~\cite{Yennie:1961ad,Eden:1966dnq,Bern:1994cg}.}.
Similar results will be obtained at all higher orders, 
but we will not proceed to compute these terms. Instead, 
we will use Eq.~\eqref{eq:w2} to construct 
the analytic resummation formalism in Sec.~\ref{sec:analytic_pressure}
and a numerical simulation in Sec.~\ref{sec:ps}.

We conclude this section by noting that it is not sufficient to compute
only the real-emission corrections to the $1\to 1$ transition. This is
apparent in the case where a massless particle becomes massive, as outlined
above. The situation is more subtle when both the incoming and the outgoing
particle is massive. However, we will show in Sec.~\ref{sec:all-orders} 
that in the limit $\gamma T\gg {\rm min}(m_a,m_b)$ the differential 
real-emission amplitude squared ${\rm d}W_{a\to bc}^{2}$
vanishes as the gauge boson transverse momentum tends to zero. This result is
qualitatively different from the behavior in~\cite{Bodeker:2017cim}, where the
amplitude tends to infinity instead. The difference is due to the fact
that we consider the transition radiation process to be a quantum correction
to the zeroth order light-to-heavy form factor, $|\mathcal{M}_{a\to b}^{(0)}|^2$,
while in~\cite{Bodeker:2017cim} it is considered to be a leading-order reaction
by itself.

\subsection{All-orders result}
\label{sec:all-orders}
To derive the all-orders result, we start from Eq.~\eqref{eq:vacuum_amplitude}.
Terms of order $2n+1$ in the expansion vanish, since 
$\langle 0|A(x_1)\ldots A(x_{2n+1})|0\rangle=0$.
The $2n$-th order term is given by
\begin{equation}\label{eq:vacuum_nth}
    W_{a\to b}^{(2n)}=\Big[\,\prod_{i=1}^{2n}i\int{\rm d}^4x_i j^{\mu_i}(x_i)\,\Big]\,
    \langle 0|T\Big[\prod_{i=1}^{2n}A_{\mu_i}(x_i)\Big]|0\rangle\;.
\end{equation}
One can use the decomposition of the time-ordered product into Feynman propagators
and the symmetry of the integrand in the currents to show that
\begin{equation}\label{eq:vacuum_nth_2}
\begin{split}
    \frac{W_{a\to b}^{(2n)}}{(2n)!}=&\;\frac{(2n-1)(2n-3)\ldots 3\cdot 1}{(2n)!}
    \Big[\,\prod_{i=1}^{2n}i\int{\rm d}^4x_i j^{\mu_i}(x_i)\,\Big]\;\prod_{i=1}^{n}\,
    \langle 0|T\left[A_{\mu_{2i}}(x_{2i})A_{\mu_{2i+1}}(x_{2i+1})\right]|0\rangle\\
    =&\;\frac{1}{2^n n!}
    \left(-\int{\rm d}^4x\int{\rm d}^4y\,j^\mu(x)i\Delta_{\mu\nu}(x,y)j^\nu(y)\right)^n
    =\frac{1}{n!}\bigg(\frac{W_{a\to b}^{(2)}}{2}\bigg)^n\;.
\end{split}
\end{equation}
Summing all orders in $\alpha$, we obtain the vacuum persistence amplitude squared
\begin{equation}\label{eq:resummed_noem}
  \mathbb{P}_{a\to b}=|W_{a\to b}|^2
  =\bigg|\sum_{n=0}^\infty\frac{1}{n!}
  \bigg(\frac{W_{a\to b}^{(2)}}{2}\bigg)^n\bigg|^{\,2}
   =\exp\Big\{W_{a\to b}^{(2)}\Big\}\;.
\end{equation}
Using Eqs.~\eqref{eq:onetoone_virtual} and~\eqref{eq:onetoone_real},
we obtain to leading logarithmic accuracy
\begin{equation}\label{eq:resummed_noem_2}
  \mathbb{P}_{a\to b}^{\rm IR}=\exp\Big\{W_{a\to b}^{(2)\,\rm IR}\Big\}
   =\exp\left\{-\int{\rm d}W_{a\to bc}^{2\,(1)\,\rm IR}\right\}\;.
\end{equation}
A similar calculation leads to
\begin{equation}\label{eq:resummed_oneem}
  {\rm d}\mathbb{P}_{a\to bc}^{\rm IR}
  ={\rm d}W_{a\to bc}^{2\,\rm IR}(k)
  ={\rm d}W_{a\to bc}^{2\,(1)\,\rm IR}(k)\,
  \exp\bigg\{-\int{\rm d}W_{a\to bc}^{2\,(1)\,\rm IR}\bigg\}\;.
\end{equation}
Note that these results still exhibit unphysical IR divergences, which are
canceled in the matching to the fragmentation function of the incoming and
outgoing particle. The matching can be interpreted as an experimental resolution, 
which requires a photon to be of sufficient energy and sufficiently separated 
in angle from the classical particle in order to be resolved as transition
radiation. In dimensional regularization, the fragmentation functions are 
pure IR divergences, hence for the light-to-massive transition we obtain 
the renormalized emission amplitude in the $\overline{\rm MS}$ scheme
\begin{equation}\label{eq:radiator_renormalized}
  \int{\rm d}W_{a\to b,\,r}^{2\,(1)}=\frac{\alpha}{2\pi}\,C_{abc}\left(
  \frac{1}{2}\log^2\frac{(2p_ap_b)^2}{\mu^2p_b^2}
  -\log^2\frac{2p_ap_b}{p_b^2}+\ldots\right)\;,
\end{equation}
where the dots stand for higher-logarithmic and finite contributions.
In this context, $\mu$ plays the role of the experimental resolution scale
which regularizes the above expression. 
In our case of interest, this implies the bubble wall 
is not sensitive to emissions which are arbitrarily soft. 
Eqs.~\eqref{eq:resummed_noem} and~\eqref{eq:resummed_oneem} are then related
to the zero and one-event probabilities according to a Poisson distribution
with mean value $\int{\rm d}W_{a\to b,\,r}^{2\,(1)}$. Using this result,
the all-orders computation can be performed by means of QCD-based
resummation techniques or parton showers.

\subsection{Momentum transfer at leading logarithmic accuracy}
\label{sec:analytic_pressure}
In this section we estimate the average momentum transfer to the wall
per incident particle. The structure of the perturbative result
in Sec.~\ref{sec:fixed_order} allows us to derive
a resummation formalism similar to the techniques employed
for the computation of collider observables~\cite{Banfi:2004yd}.
We work in the leading logarithmic approximation. We also assume that particle masses
are small compared to the particle's energies, and can therefore be treated according 
to Eq.~\eqref{eq:emission_probability_correct_integrated}.
Numerical studies will be carried out in Sec.~\ref{sec:ps} using the full
kinematical mass dependence. Here we focus instead on the qualitative
predictions at $\gamma T\gg \max(m_a,m_b)$ and for fixed coupling.
  
We begin by computing the so-called radiator function~\cite{Banfi:2004yd},
which corresponds to the sum of the integrated S-matrix elements
$\int{\rm d}W_{a\to bc}^{2\,(1)\,\rm IR}$ and $W_{a\to b}^{(2)\,\rm IR}$.
However, instead of being an inclusive quantity, the radiator function
implements the physical constraint that the momentum transfer in any
branching $a\to bc$ cannot be larger than the eventually observed value
of the momentum transfer for all emissions in the resummed theory.
This restriction is most easily implemented by working in the leading
logarithmic approximation and using the fact that the $1/\eps^2$ and
$1/\eps$ poles cancel between $\int{\rm d}W_{a\to bc}^{2\,(1)\,\rm IR}$ 
and $W_{a\to b}^{(2)\,\rm IR}$.
This implies that instead of computing the finite difference between two
individually IR divergent quantities, we may compute the finite remainder
directly by making use of the unitarity constraint. In practice, it is
achieved by placing a lower bound on the relative momentum transfer
per splitting 
\begin{equation}\label{eq:def_v}
  V(p_a,p_b,p_c)=\frac{\Delta p_z}{\gamma T}\approx\frac{\vec{k}_\perp^{\,2}/(2E_a^2)}{x(1-x)}\;,
\end{equation}
where the last equality holds for $m^2\ll \vec{k}_{\perp}^{\,2}$ (note that $E_a=\gamma T$).
In terms of $V(p_a,p_b,p_c)$, the radiator function $R(V)$ for any given value of $V$ is given by
\begin{equation}\label{eq:caesar_radiator}
 R_{abc}(V)=C_{abc} |g|^2\int\frac{{\rm d}^3\vec{p}_c}{(2\pi)^3\,2E_c}
 \left( \frac{2\;p_{b,\mathrm{h}}p_{a,\mathrm{s}}}{ p_{a,\mathrm{s}}p_c\;p_{b,\mathrm{h}}p_c}
 +\mathcal{O}\left(\frac{m_{a,\mathrm{s}}^2}{\vec{k}_\perp^{\;2}},
 \frac{m_{b,\mathrm{h}}^2}{\vec{k}_\perp^{\;2}}\right)\right) \Theta(p_{b,z,\mathrm{h}})\Theta(p_{c,z,\mathrm{h}}) \, \Theta(V(p_a,p_b,p_c)-V)
 \;,
 \end{equation}
where we have used the squared matrix element from
Eq.~\eqref{eq:w1_real}. The $\Theta$ function constraints arise
from the requirement that the emitted particles
must enter the Higgs condensate. Phase-space restrictions 
play a significant role for the agreement between
analytic and numeric resummation of event shape 
observables~\cite{Hoeche:2017jsi}. This problem is amplified here,
because we compute the average value of $V$, which is impacted
significantly by modest changes of the differential cross section
at large $V$.Using the parametrization of kinematics in
App.~\ref{sec:ps_kin}, we can write Eq.~\eqref{eq:caesar_radiator} as
\begin{equation}\label{eq:single_emission_nll}
  R_{abc}(V)=C_{abc}\,\frac{\alpha}{2\pi} \int_{V}^{\,1} \frac{{\rm d}V'}{V'} \; 
  \int_0^1 {\rm d}x\,2x
  \;\Theta\left(\frac{1}{1+V'}-x\right)\Theta\left(x-\frac{V'}{1+V'}\right)\;.
\end{equation}
Performing the integrals we obtain 
\begin{equation}\label{eq:caesar_rl}
  R_{abc}(V) = \frac{\alpha}{2\pi}\,C_{abc}\,\Big(L+2\log\big(1+e^{-L}\big)\Big),
  \qquad\text{where}\qquad
  L=\log\frac{1}{V}\;.
\end{equation}
While the techniques we employ are the same as for the computation
of the well-known radiator function for the thrust in $e^+e^-$
annihilation~\cite{Farhi:1977sg,Catani:1991kz}, the result shows
only a single logarithmic enhancement, as we focus on the region
$m^2\ll \vec{k}_\perp^2$.
To leading logarithmic accuracy, the resummed cumulative cross section
at $V$ is given by the vacuum persistence amplitude squared,
Eq.~\eqref{eq:resummed_noem}, with $W_{a\to b}^{(2)}$ replaced
by $R(V)$~\cite{Banfi:2004yd}.
This can be understood in the following intuitive way:
The radiator function $R_{abc}(V)$ corresponds to a probability
for the decay of particle $a$ into the two final states $b$ and $c$.
However, in order for particle $a$ to produce a relative momentum transfer
of $V$, it must not have produced a relative momentum transfer $V(p_a,p_b,k)>V$,
as otherwise, it would not exist anymore in its present state. This is
analogous to a nuclear decay process, where the nucleus can decay at
any given time only if it has not decayed at earlier times.  
This ``survival probability'' is encapsulated in an exponential
suppression factor conventionally called the Sudakov factor,
which can be read off Eq.~\eqref{eq:resummed_noem_2}
\begin{equation}\label{eq:sudakov_factor}
  \Delta_a(V)=\exp\Big\{-\sum_{b}R_{ab}(V)\Big\}\;,
  \qquad\text{where}\qquad
  R_{ab}(V)=\sum_c R_{abc}(V)\;.
\end{equation}
The rate at which particle $a$ branches into any particles
$b$ and $c$ is eventually given by
\begin{equation}\label{eq:caesar_dsigma}
  \frac{1}{N_a}\frac{{\rm d}N_{a}(V)}{{\rm d}V}
  = \sum_b\frac{{\rm d}R_{ab}(V)}{{\rm d}V}\,\Delta_a(V)
  = -\frac{{\rm d}\Delta_a(V)}{{\rm d}V}\;,
\end{equation}
which, at leading logarithmic accuracy, leads
to the normalized cumulative cross section~\cite{Banfi:2004yd}
\begin{equation}\label{eq:caesar_sigma}
  \frac{1}{\sigma}\int_V^{\,1}{\rm d}V^\prime\,
  \frac{{\rm d}\sigma(V^\prime)}{{\rm d}V^\prime}
  = \Delta_a(V)\;.
\end{equation}
The average relative momentum transfer from
all branchings is obtained by weighted summation over particle species,
where the weight is given by the incident flux, times the cross section
for interaction with the Higgs condensate. The leading-order cross section
in the high-energy limit is identical for all massive Standard Model particles, 
which we denote as the set $\mathcal{S}$ (cf.\ Eq.~\eqref{eq:Pa}). We can thus write 
\begin{equation}\label{eq:caesar_pressure}
  \Bigl<\frac{\Delta p_z}{\gamma T}\Bigr>=
  \int_0^1 {\rm d}V\,V\frac{{\rm d}}{{\rm d}V}
  \prod_{a\in\mathcal{S}}\Delta_a(V)\;.
\end{equation}
At fixed coupling, we obtain
\begin{equation}\label{eq:caesar_pressure_fc}
  \Bigl<\frac{\Delta p_z}{\gamma T}\Bigr>_{\,\rm FC}=
  \int_0^{\infty}{\rm d}L\,e^{-L}\,\frac{(\alpha C)_{\Sigma}}{2\pi}\,
  \frac{e^L-1}{e^L+1}\,
  \exp\bigg\{-\frac{(\alpha C)_{\Sigma}}{2\pi}\,\Big(L+2\log\big(1+e^{-L}\big)\Big)\bigg\}\;,
\end{equation}
where $(\alpha C)_{\Sigma}=\sum_{a,b\in\mathcal{S}}\alpha_{abc} C_{abc}$ and $\alpha_{abc}$ is the coupling associated to the splitting $a\to bc$. The relevant couplings are listed in App.~\ref{sec:vertex_functions}.
Eq.~\eqref{eq:caesar_pressure_fc} has the solution
\begin{equation}\label{eq:avg_pressure_ll_fc}
  \Bigl<\frac{\Delta p_z}{\gamma T}\Bigr>_{\,\rm FC}=
  \frac{4^{-\zeta}}{\zeta-1} \left(\zeta+1- \frac{\sqrt{\pi}\,\Gamma (\zeta+1)}{\Gamma \left(\zeta+1/2\right) }\right)
  \approx\zeta\big(\log 4-1\big)
  \qquad\text{where}\qquad
  \zeta=\frac{(\alpha C)_{\Sigma}}{2\pi}\;.
\end{equation}
The linear approximation is very simple and works up to relative pressures
of $\langle V\rangle_{\,\rm FC}\approx 1\%$. It can alternatively be obtained
from the fixed-order expansion of Eq.~\eqref{eq:caesar_pressure_fc} 
\begin{equation}\label{eq:caesar_pressure_fc_fo}
  \Bigl<\frac{\Delta p_z}{\gamma T}\Bigr>_{\,\rm FC,FO}=
 \int_0^\infty{\rm d}L\,e^{-L}\,\frac{(\alpha C)_{\Sigma}}{2\pi}
 \frac{e^L-1}{e^L+1}
 =\,\zeta\big(\log 4-1\big)\;.
\end{equation}
Running coupling effects should be included to obtain a more reliable
resummed result. They induce a mild change in the scaling behavior of
the radiator function~\cite{Banfi:2004yd}. For the qualitative discussion
in this section, it is sufficient, however, to consider
Eq.~\eqref{eq:avg_pressure_ll_fc}.

Note in particular that $\langle \Delta p_z\rangle\varpropto\gamma T$, 
independent of the particle masses, as long as $\gamma T\gg m_{b,\mathrm{h}}$. 
This is in  contrast to~\cite{Bodeker:2017cim}, where a 
$\gamma$-independent value was obtained for the average pressure transfer,
which implies $P \sim \gamma^1$ once the flux factor is taken into account.
This result was derived based on the assumption that the dominant contribution 
to the integral is obtained from the region $\vec{k}_\perp^2 \approx m_{b,\mathrm{h}}^2$, 
and by cutting off the divergent $\vec{k}_\perp$-integral at this value. 
We find a different result due to the single logarithmic growth of the radiation probability, 
Eq.~\eqref{eq:emission_probability_correct_integrated}, in the region $m_{b,\mathrm{h}}^2\ll \vec{k}_\perp^2$, 
independent of $\gamma T$, which necessitates the incorporation of virtual corrections 
to all orders. For details on the emergence of this behavior, see the discussion in 
Sec.~\ref{sec:formalism} and the explicit examples in App.~\ref{app:scalarQED}.
Cutting off the integrals in a procedure similar to~\cite{Bodeker:2017cim}
would instead turn Eq.~\eqref{eq:caesar_pressure_fc_fo} into  
\begin{equation}\label{eq:caesar_pressure_fc_fo_cut}
   \Bigl<\frac{\Delta p_z}{\gamma T}\Bigr>_{\,\rm FC,FO}^{\rm(cut)}\approx
   \zeta\,\frac{m_{b,\mathrm{h}}}{\gamma T}+
   \zeta\left(\log 4-1-2\log\left(1+\frac{m_{b,\mathrm{h}}}{\gamma T}\right)\right)
\end{equation}
The constant and logarithmic contributions are missing in~\cite{Bodeker:2017cim},
which indeed results in $\langle\Delta p_z\rangle\varpropto m_{b,\mathrm{h}}$.
The origin of the discrepancy is the fundamentally different behavior of
the real radiative corrections in the regime $m_{b,\mathrm{h}}^2\ll \vec{k}_\perp^2\ll(\gamma T)^2$.

\section{Numerical simulation}
\label{sec:ps}
In the following, we establish the connection of the above formalism to parton showers,
which can be used at leading logarithmic accuracy to simulate the physics encapsulated
in Eqs.~\eqref{eq:resummed_noem},~\eqref{eq:resummed_oneem}, and the corresponding
equations for higher particle multiplicity. Again, it is important to note that
the renormalized counterparts of these equations describe the distribution of
emissions according to a Poissonian with average value 
$\int{\rm d}W_{a\to bc,r}^{2\,(1)\,\rm IR}$.
The regularization of ${\rm d}W_{a\to bc,r}^{2\,(1)\,\rm IR}$ in the 
parton shower follows the procedure outlined in Sec.~\ref{sec:analytic_pressure}
and may be performed in any way that allows for an infrared and collinear safe
simulation, such as using a transverse momentum cutoff at scales $k_{\perp,0}\ll \gamma T$.
We can define the transverse momentum-dependent radiator function for the parton shower as
(cf.~Eq.\eqref{eq:w1_real})
\begin{equation}\label{eq:ps_radiator}
  R_{abc}^{\rm PS}\bigg(\frac{2k_{\perp,0}}{\gamma T}\bigg)
  =|g|^2\int\frac{{\rm d}^3\vec{p}_c}{(2\pi)^3\,2E_c}
  \left(\frac{2p_ap_b}{(p_ap_c)(p_bp_c)}
  -\frac{m_a^2}{(p_ap_c)^2}-\frac{m_b^2}{(p_bp_c)^2}\right)\,
  \Theta(p_{c,\perp}-k_{\perp,0})\;.
\end{equation}
As long as the cutoff is small compared to the average transverse momentum generated
by Eq.~\eqref{eq:resummed_oneem}, the results of infrared and collinear safe observables
such as the relative $z$-momentum transfer will be independent of $k_{\perp,0}$. We can approximate Eq.~\eqref{eq:ps_radiator} as
\begin{equation}\label{eq:ps_split}
  R_{abc}^{\rm PS}(t,Q^2)=
  \int_{t}^{Q^2/4}\frac{{\rm d}\bar{t}}{\bar{t}}
  \int_{x_-}^{x_+}{\rm d}x\,
  \sum_{a,b}\frac{\alpha}{2\pi}\,2C_{abc}\left\{
  \begin{array}{ll}
  x^{-1}&\qquad\text{if}\qquad k_\perp^2\ll m_b^2\\
  x^{+1}&\qquad\text{if}\qquad m_b^2\ll k_\perp^2
  \end{array}\right.\;,
\end{equation}
where $t\varpropto \vec{k}_\perp^2$ is called the parton-shower evolution variable,
and $Q/2=\gamma T$ is the kinematical boundary.
The quantity $4\pi\alpha=|g|^2$ is the coupling squared for the transition $a\to bc$,
and $C_{abc}$ is an associated charge factor in the collinear limit
(see App.~\ref{sec:vertex_functions} for details). The integration boundaries
$x_{\pm}$ are determined by the constraint $k_{\perp}>k_{\perp,0}$.
We can then generate emissions by setting the Sudakov factor $\Delta_a(t,Q^2)$
equal to a random number and solving for the evolution variable $t$, where
\begin{equation}\label{eq:sudakov}
  \Delta_a(t,Q^2)=\exp\bigg\{-\sum_{b,c}R_{abc}^{\rm PS}(t,Q^2)\bigg\}\;.
\end{equation}
In addition, we select the splitting variable $x$
according to $2/x$ and sample the azimuthal angle
from a uniform distribution. 
For QCD partons, we also choose a color configuration.
The kinematics mapping is described in App.~\ref{sec:ps_kin},
and more details of the algorithm are given in~\cite{Webber:1986mc,
  Sjostrand:1995iq,Buckley:2011ms}.
Note that the splitting variable $x$ is chosen differently 
for initial- and final-state showering, as detailed in Sec.~\ref{sec:ps_kin}.
Our numerical implementation is based on the QCD parton shower
published in~\cite{Hoche:2014rga}. The evolution variable $t$
is chosen to be the relative transverse momentum in the collinear limit
(cf.~App~\ref{sec:ps_kin}), and we solve the soft double-counting
problem~\cite{Marchesini:1987cf} by means of phase-space partitioning
in the dipole rest frame~\cite{Banfi:2004yd}. We have checked that this
gives similar results as angular ordered evolution, using the formalism
of~\cite{Platzer:2009jq}.

\begin{table}[t]
  \centering
  \begin{tabular}{lccccccc}
    & & & & & \multicolumn{2}{c}{$\langle \Delta p_z/(\gamma T)\rangle_{\rm FC}$}\\
    Particle & & $n_f$ & & $\nu_a$ & ~analytic~ & ~numeric~ \\\hline
    $l^\pm$ & & $2\times 3$ & & 2 & 0.17\% & 0.17\% \\
    $u$ & & $2\times 3$ & & $2\times 3$ & 0.46\% & 0.44\% \\
    $d$ & & $2\times 3$ & & $2\times 3$ & 0.46\% & 0.44\% \\
    $W^\pm$ & & 2 & & 2 & 0.52\% & 0.52\% \\
    $Z$ & & 1 & & 2 & 0.41\% & 0.40\% \\
    $h$ & & 1 & & 1 & 0.22\% & 0.21\% \\
    $G_{W^\pm}$ & & 2 & & 1 & 0.22\% & 0.21\% \\
    $G_Z$ & & 1 & & 1 & 0.22\% & 0.21\%
  \end{tabular}
  \caption{Average relative momentum transfer per degree of freedom,
    $\langle \Delta p_z/(\gamma T)\rangle_{\rm FC}$, assuming that
    a particle of the given species is incident on the wall, and
    allowed to shower into the full SM.
    We compare analytic results from Eq.~\eqref{eq:caesar_pressure}
    and the numerical simulation described in Sec.~\ref{sec:ps}.
    We have chosen $\gamma=10^6$, $\alpha_s=0.04$ and $\alpha=0.01$,
    and we have fixed the couplings in order to satisfy the assumptions
    leading to Eq.~\eqref{eq:caesar_pressure}.
    Differences are due to
    flavor-changing effects, which are not taken into account in
    Eq.~\eqref{eq:caesar_pressure}, and due to the definition of
    the momentum transfer in terms of the initial- and final-state momenta.
    This is computed approximately in~\ref{sec:analytic_pressure}
    and treated exactly in the numerical simulation~\cite{Hoeche:2017jsi}.
    We also list the number of flavors of this type, $n_f$, and the
    number of the corresponding degrees of freedom, $\nu_a$,
    per flavor in Eq.~\eqref{eq:master_formula}, assuming that
    the flavors are active, i.e. that they contribute to the pressure 
    due to a mass change at the bubble wall.}
  \label{fig:pressure_comparison}
\end{table}
We have implemented the above described algorithm in a numerical program
based on the QCD simulation published in~\cite{Hoche:2014rga}, which was
validated against the public event generator Sherpa~\cite{Bothmann:2019yzt}.
We employ a 2-loop running strong coupling with threshold matching up to
$n_f=6$ and $\alpha_s(M_Z)=0.118$. The electroweak input parameters are
$\alpha(0)=1/137$, $m_W=80.385~{\rm GeV}$, $m_Z=91.1876~{\rm GeV}$
and $m_H=125~{\rm GeV}$, leading to $\sin^2\theta_W=1-(m_W/m_Z)^2=0.223$.
The relative flux factors for incident particles are given simply by the
number of degrees of freedom of the particle. This can be understood by
computing the (trivial) leading-order transition amplitudes 
in Sec.~\ref{sec:pressure_at_lo} in a VEV-insertion
approximation~\cite{Carena:2000id,Carena:2002ss,Lee:2004we}.

Due to the exact kinematics in the numerical simulation, the results
display significant threshold effects at small momentum transfer
($\Delta p_z\approx T$), which become irrelevant as $\gamma\to\infty$
due to the parametric behavior derived in Eq.~\eqref{eq:avg_pressure_ll_fc}.
This leads to a slight distortion of the scaling behavior, which goes
beyond running coupling effects. In addition, the computation of the
observable with exact kinematics induces a shift in the average pressure,
but does not change the qualitative behavior.
Table~\ref{fig:pressure_comparison} shows a comparison between the
analytic results from Eq.~\eqref{eq:caesar_pressure} and the numerical
simulation for the most relevant particle species. Given the simplicity
of the analytical estimate, the two results agree very well.

\begin{figure}[t]
  \centerline{
  \includegraphics[width=0.45\textwidth]{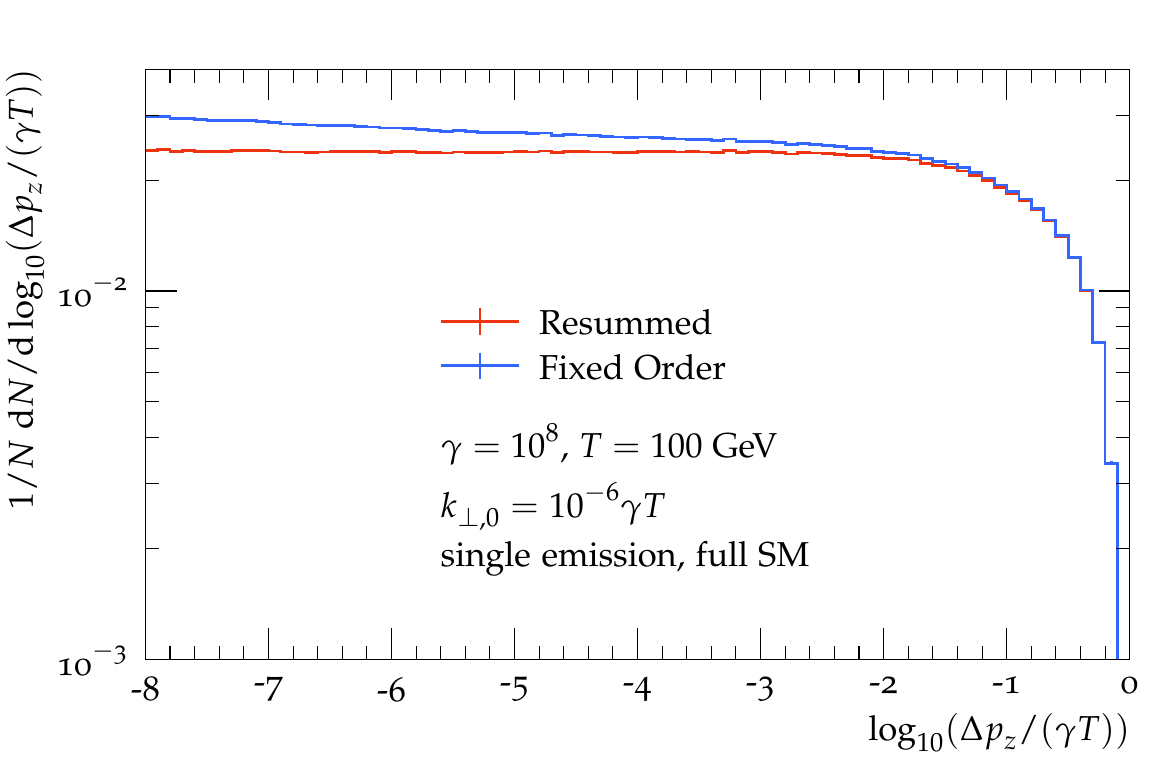}}
  \caption{Comparison between fixed-order and resummed result
    for the relative momentum transfer distribution in the full
    Standard Model using the parton-shower approximation.
    The number of emissions is limited to one. Note that the
    fixed-order result is not normalized to the total rate,
    as the rate tends to infinity in this case.
    \label{fig:sudakov}}
\end{figure}
Fig.~\ref{fig:sudakov} exemplifies the structural difference
between the fixed-order and all-orders resummed predictions
for the relative momentum transfer distribution in the
parton-shower approximation. While the fixed-order result
diverges as $\Delta p_z/(\gamma T)\to 0$
(cf.\ Eqs.~\eqref{eq:onetoone_real} and~\eqref{eq:caesar_rl}),
the resummed result remains finite and slowly approaches zero
(see Sec.~\ref{sec:all-orders}). This leads to a cutoff-dependence
in the average relative momentum transfer at fixed order in the
method of Ref.~\cite{Bodeker:2017cim} which changes the scaling
behavior with $\gamma$ (cf.\ Eq.~\eqref{eq:caesar_pressure_fc_fo_cut}).
In resummed perturbation theory the average relative momentum transfer
is cutoff-independent, and is primarily determined by the coupling
strength, cf.\ Eq.~\eqref{eq:avg_pressure_ll_fc}.
\begin{figure}[t]
  \centerline{
    \includegraphics[width=0.45\textwidth]{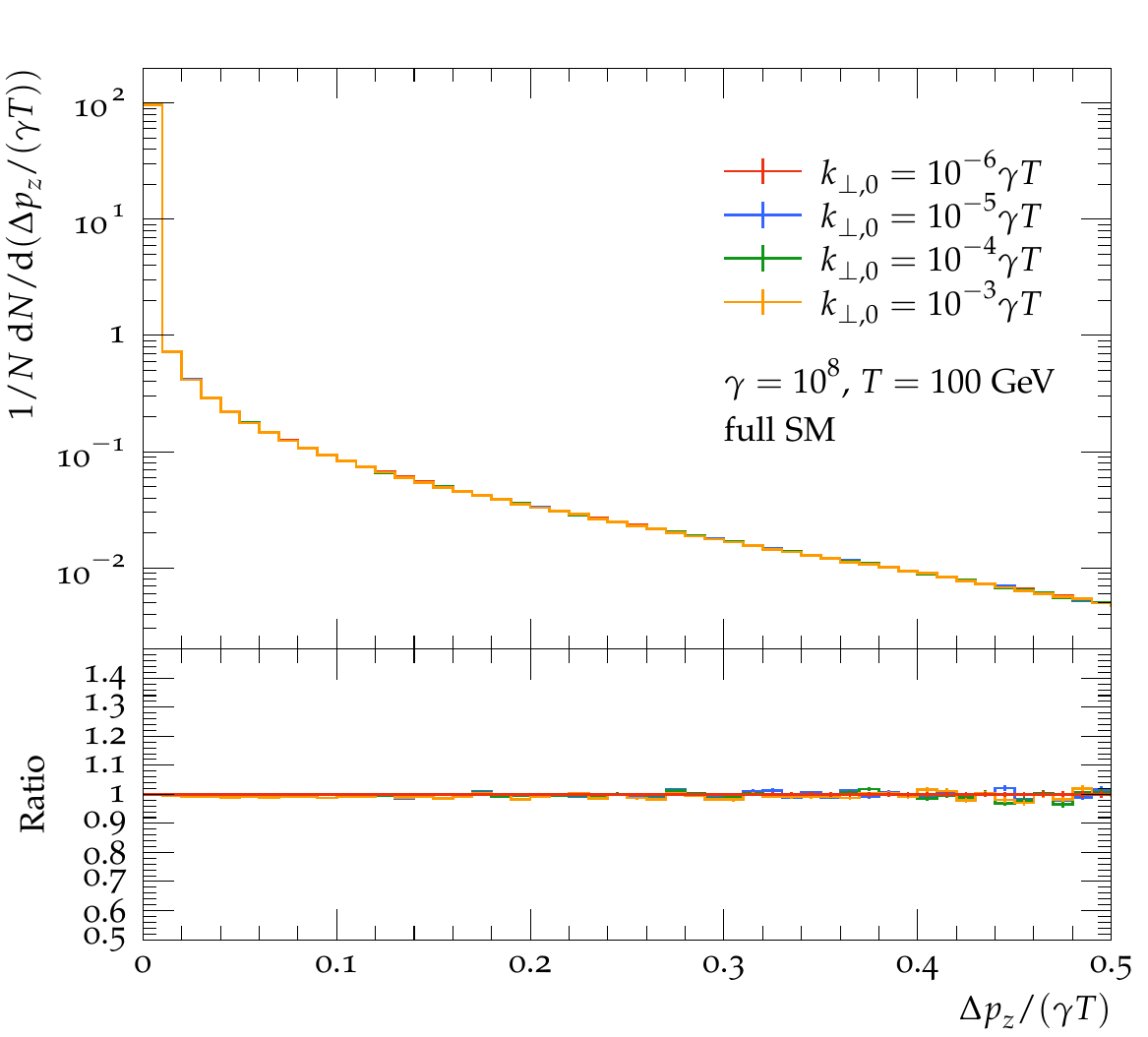}\hskip 3mm
    \includegraphics[width=0.45\textwidth]{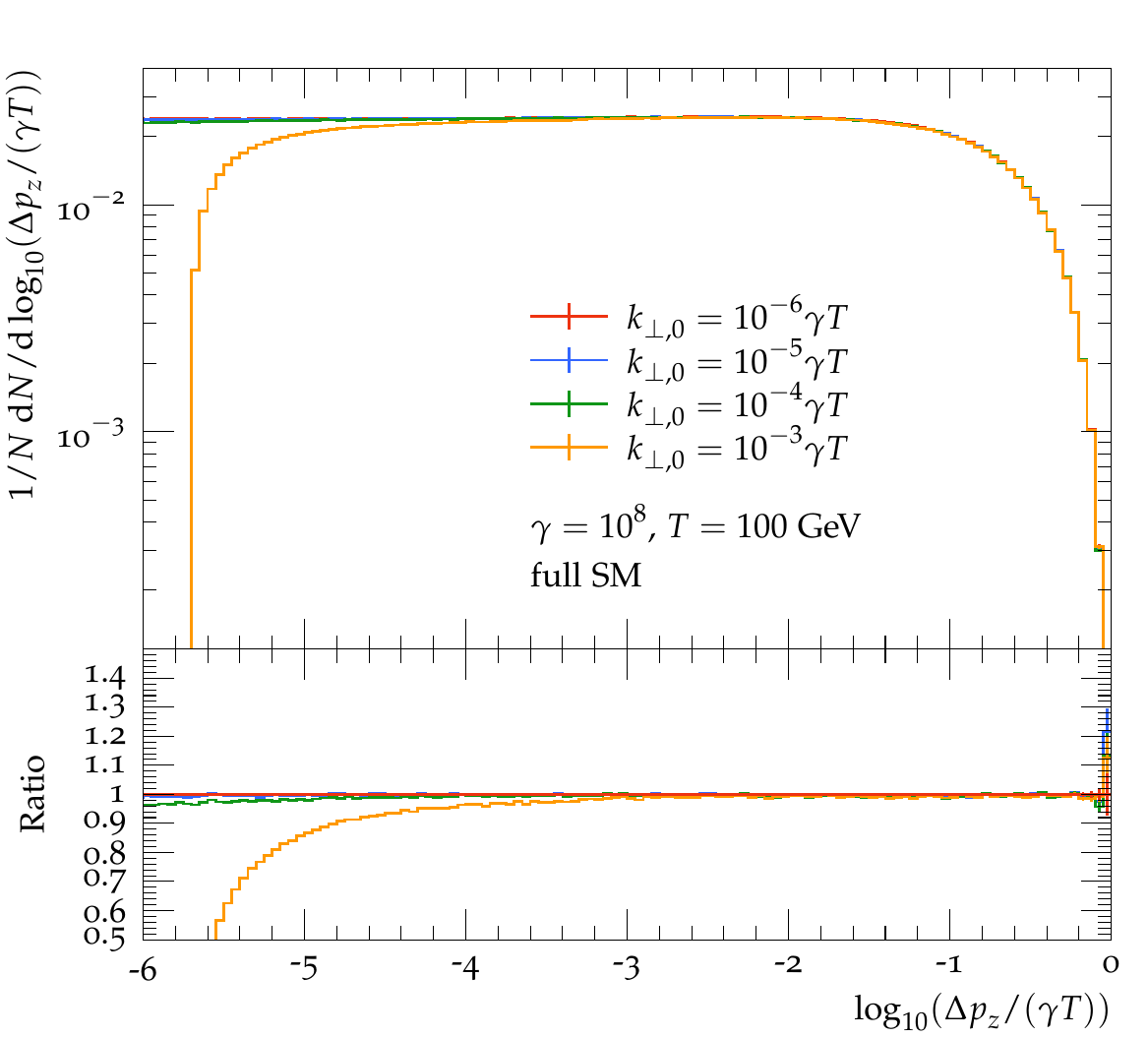}}
  \caption{Cutoff-dependence of the relative momentum transfer
    in the parton-shower approximation.
    \label{fig:cutoff}}
\end{figure}
This is tested in detail in Fig.~\ref{fig:cutoff}, which shows
the cutoff dependence of the relative momentum transfer spectrum
in the parton-shower approximation at $\gamma=10^{8}$.
Note that the cutoff is varied over three orders of magnitude
with no significant effect on the spectrum on a linear scale
(Fig.~\ref{fig:cutoff} left). The corresponding values for the
average relative momentum transfer are given by
$\langle\Delta p_z/(\gamma T)\rangle=0.401\%-0.409\%$.
Similar results are obtained for all values of $\gamma$.

\begin{figure}[t]
  \centerline{
    \includegraphics[width=0.45\textwidth]{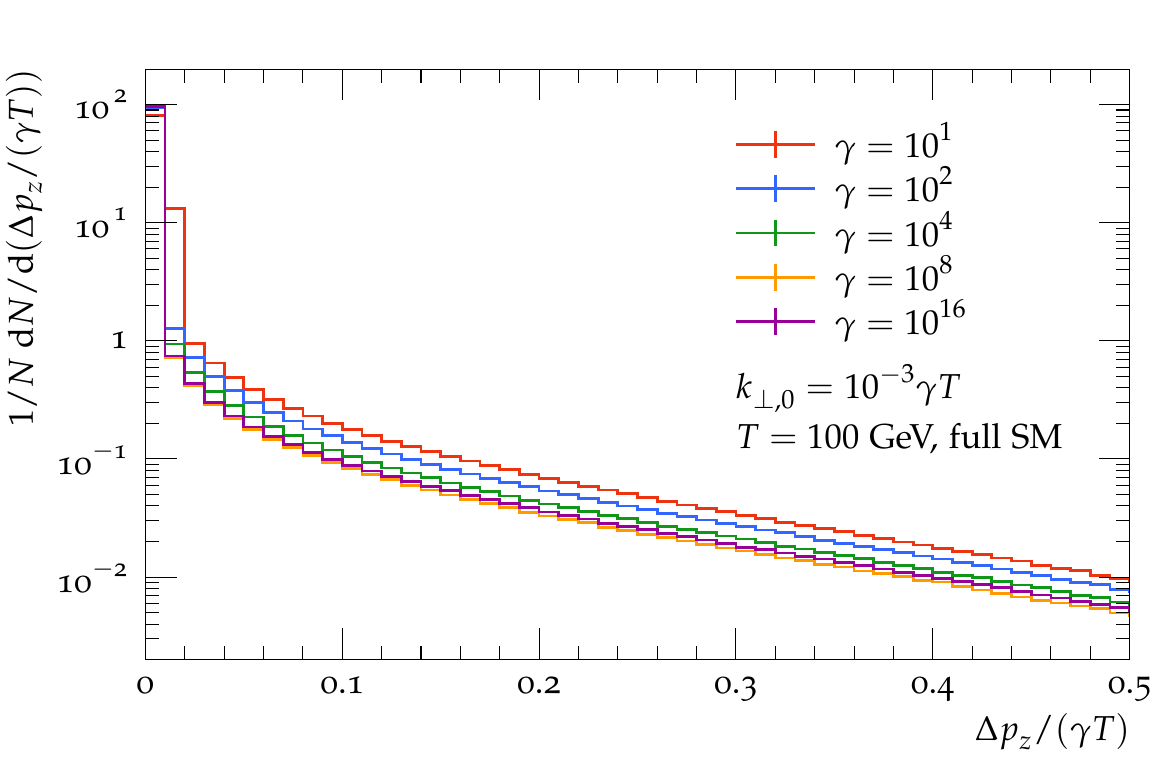}\hskip 3mm
    \includegraphics[width=0.45\textwidth]{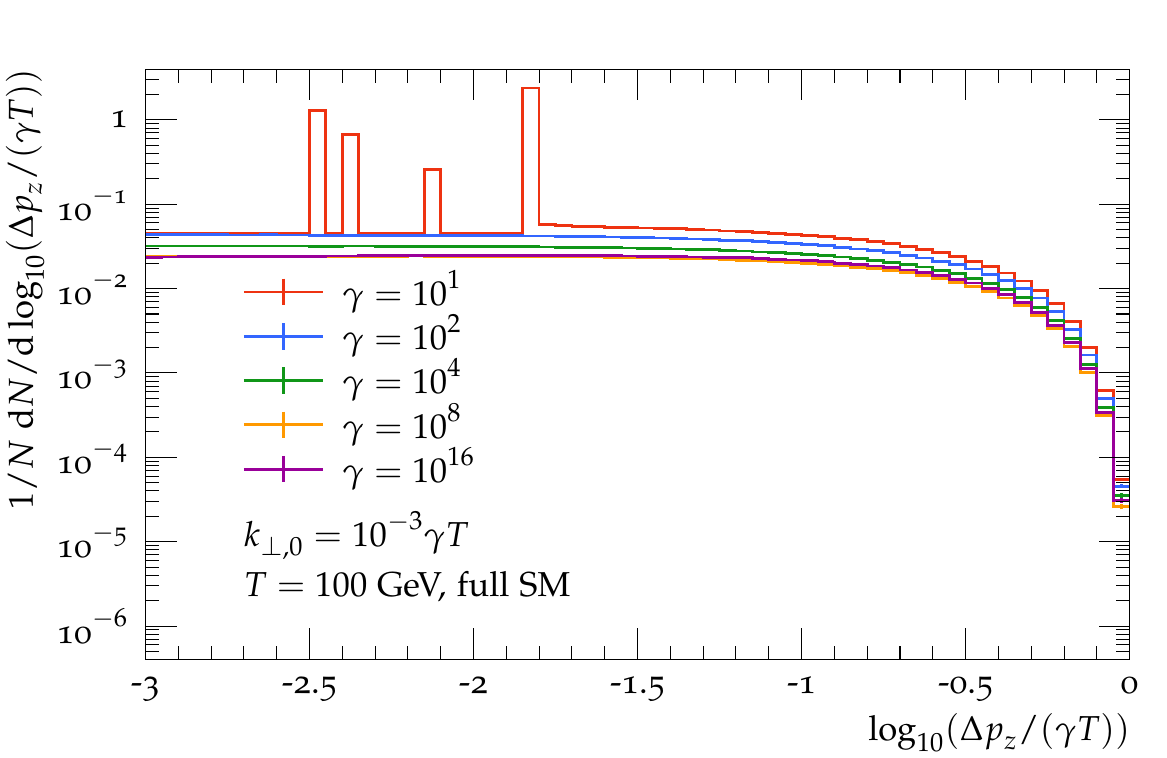}}
  \caption{Boost factor dependence of the relative momentum transfer
    in the parton-shower approximation.
    \label{fig:gamma}}
\end{figure}
Fig.~\ref{fig:gamma} shows the change in the relative momentum transfer
with changing $\gamma$, which is relatively mild. Note in particular the
features in the spectrum at $\gamma=10$, which originate in kinematic
effects when top-quarks, Higgs bosons, $Z$ bosons and $W^\pm$ bosons
are put on-shell or produced in $1\to 2$ splittings at threshold.
The net result is monochromatic lines and discontinuities in the
momentum transfer spectrum, accounting for a larger average
momentum transfer at small $\gamma$. However, in this region the soft
approximation breaks down, and a more precise calculation should
be performed. Such a calculation is not needed in order to determine
the scaling behavior of the relative momentum transfer with $\gamma$.
In the ultrarelativistic  scenario, for $10^{2}\lesssim \gamma\lesssim 10^{16}$,
we can fit the numerical results for the average momentum transfer
in the full Standard Model at $T=100$~GeV to the following form
\begin{equation}\label{eq:mom_transfer_fit}
    \Bigl<\frac{\Delta p_z}{\gamma T}\Bigr> =
    0.89(17)\%-0.14(3)\%\,\log_{10}\gamma+0.015(3)\%\,\log_{10}^2\gamma
    -0.0008(1)\%\log_{10}^3\gamma+0.000017(3)\%\log_{10}^4\gamma\;,
\end{equation}
where the numbers in parentheses are uncertainties on the last digit,
determined from cutoff-, ordering parameter and kinematics variations.
The above fit is still accurate to within a factor of 2 down to 
$\gamma \sim 10$, where several of our other approximations 
begin to break down. Our numerical simulations qualitatively
confirm the analytic result of Sec.~\ref{sec:analytic_pressure}.
Modifications that are due to running coupling effects
are captured by Eq.~\eqref{eq:mom_transfer_fit}, and the threshold
effects only increase the average relative momentum transfer.

Note that our numerical simulations assumed particles in the 
broken phase to have masses set by their $T=0$ values. In a 
realistic phase transition, $\langle \phi \rangle \neq 246$ GeV, 
and so the broken phase masses depend non-trivially on the details
of the Higgs finite-temperature effective potential. 
However, for sufficiently large velocities such that $\gamma T \gg m$
(required by several of the approximations we have made), 
the precise values of the broken phase masses are not important 
and the relative momentum transfer is still expected to be 
$\sim 0.5\% \times  \gamma T$, with the scaling determined 
by Eq.~\eqref{eq:avg_pressure_ll_fc}.

\section{The bubble wall velocity and its implications for cosmology}
\label{sec:implications}

Using the above results, we can obtain an estimate for the terminal wall velocity as a function of the phase transition strength. The bubble wall reaches a terminal velocity when the pressure difference between the interior and exterior vanishes. The total pressure difference, $\Delta P_{\rm tot}$ is given by 
\begin{equation} \label{eq:pressure_balance}
\Delta P_{\rm tot} = -\Delta V + P\,,
\end{equation}
where $P$ is the thermal pressure exerted against the wall arising from particles impinging on it and showering as computed in Secs.~\ref{sec:framework}-\ref{sec:ps}, and $\Delta V$ is the difference in vacuum energy between the two phases. From Eqs.~\eqref{eq:Dp_avg}, \eqref{eq:dP_ab}, and~\eqref{eq:master_formula}, we can relate the thermal pressure to the average momentum transfer obtained from either the analytic or numerical treatments above. We find:
\begin{equation} \label{eq:P_est}
P= \sum_{a \in \mathcal{S}}\nu_a \int \frac{\mathrm{d}^3\vec{p}_a}{(2\pi)^3 E_a}f_a(\vec{p}_a)\,p^2_{a,z}\,\Bigl<\frac{\Delta p_z}{\gamma T}\Bigr>\,,
\end{equation}
which, up to $\log \gamma$ effects stemming from thresholds and the running of the couplings, is simply a constant times the total $z$-direction wall-frame plasma pressure in the symmetric phase from particles coupled to the Higgs. In other words, we have found that in the limit of large wall velocities, the net thermal pressure experienced by the bubble wall in its rest frame is simply a constant fraction ($\sim 0.5\%$) of the total pressure of the gas of symmetric-phase particles that couple directly to the wall.  We can approximate 
\begin{equation}
P\simeq P_{1\rightarrow 1}+\gamma^2 \times \lim_{\gamma \rightarrow \infty}\left(\frac{P_{\rm FC}}{\gamma^2}\right)\,,
\end{equation}
where 
\begin{equation}
P_{1 \rightarrow 1}\equiv \sum_{a \in \mathcal{S}} \int \frac{\mathrm{d}^3\vec{p}_a}{(2\pi)^3 2 E_a}f_a(\vec{p}_a)\Delta m^2_a\,,
\end{equation}
is the $1\rightarrow 1$ pressure and the `FC' subscript denotes the fixed-coupling approximation of the relative momentum transfer, $\langle\Delta p_z/(\gamma T)\rangle\approx \langle\Delta p_z/(\gamma T)\rangle_{\rm FC}$ in Eq.~\eqref{eq:P_est} (the results will not be very sensitive to the precise scale chosen).

To obtain a parametric estimate of the terminal wall velocity, Eq.~\eqref{eq:pressure_balance} indicates that $P$ should be compared against the quantity $\Delta V$, which is model-dependent. However, in many cases of interest, it is set roughly by the energy density of the radiation bath during the transition. We can define the following parameters 
\begin{equation} \label{eq:alpha_def}
\alpha_\theta \equiv \frac{\Delta V}{\rho_{\rm rad}}, \qquad \alpha_{\infty} \equiv \frac{P_{1\rightarrow 1 }}{\rho_{\rm rad}} \approx 5 \times 10^{-3} \, \frac{\langle \phi \rangle ^2}{T^2}, \qquad \alpha_{\rm eq} \equiv  \lim_{\gamma \rightarrow \infty}\left(\frac{ P_{\rm FC}}{\gamma^2}\right)\frac{1}{\rho_{\rm rad}} \approx 5 \times 10^{-3}\,,
\end{equation}
where $\rho_{\rm rad}= \pi^2/30 g_*(T)T^4$ with $g_*(T)\approx 100$, and the numerical values in Eq.~\eqref{eq:alpha_def} assume SM-like plasma content with $\langle\Delta p_z/(\gamma T)\rangle\approx \langle\Delta p_z/(\gamma T)\rangle_{\rm FC} \approx 0.5\%$ evaluated from Table.~\ref{fig:pressure_comparison}. The quantity $\alpha_\theta$ parametrizes the strength of the phase transition, and appears often in the gravitational wave literature (see e.g.~Ref.~\cite{Caprini:2019egz} for an in-depth discussion).  Models for which $\alpha_\theta>\alpha_{\infty}$ would satisfy the original runaway wall condition of Ref.~\cite{Bodeker:2009qy} if transition radiation was not taken into account. Inserting Eqs.~\eqref{eq:P_est}, \eqref{eq:alpha_def} into Eq.~\eqref{eq:pressure_balance} and requiring $\Delta P_{\rm tot}=0$, we find the terminal velocity, expressed in terms of the equilibrium value of $\gamma$ is
\begin{equation} \label{eq:gameq_1n}
\gamma_{\rm eq}=\left(\frac{\alpha_{\theta}-\alpha_{\infty}}{\alpha_{\rm eq}} \right)^{1/2}\approx 14 \, \sqrt{\alpha_{\theta}}\,,
\end{equation}
where the final approximation applies for $\alpha_{\theta}\gg \alpha_{\infty}$ as is the case when $\gamma T \gg m$ (i.e.~where our analysis remains self-consistent). Note that, for strong phase transitions, the terminal velocity depends on the strength of the phase transition and the gauge boson couplings, but not the particle masses. This is a strikingly different result than that implied by the fixed-order calculation of Ref.~\cite{Bodeker:2017cim}.  

Throughout our analysis we have assumed $\gamma T \gg m$ for all massive SM particles in the plasma\footnote{We also worked in the thin-wall limit, $m L_w \ll 1$, which can be violated by the heaviest SM degrees of freedom if $L_w$ is sufficiently thick. However, $L_w$ depends on the underlying effective potential and implicitly on the terminal velocity itself, and so this criterion should be checked on a model-by-model basis.}. This allowed us to neglect the reflection of particles back into the symmetric phase, the transmission of particles from the broken into the symmetric phase, and the plasma interactions which drive the distributions back to equilibrium, which substantially complicates the calculation (see Refs.~\cite{Moore:1995ua, Moore:1995si, John:2000zq, Konstandin:2014zta, Kozaczuk:2015owa, Dorsch:2018pat, Mancha:2020fzw} for calculations in this slow-wall regime). For an electroweak-scale transition with $T \gtrsim \mathcal{O}(10)$ GeV and a SM-like plasma where the largest masses are $\mathcal{O}(100)$ GeV, the approximations we have made break down for $\gamma \lesssim \mathcal{O}(10)$. Meanwhile, for many models with SM-like plasma content, one finds $\alpha_{\theta} \lesssim 1$ (see e.g.~\cite{Caprini:2019egz} and the corresponding benchmark points compiled at \href{http://ptplot.org}{ptplot.org}), implying that $\gamma_{\rm eq} \lesssim 10$ in these conventional cases. Although our approximations break down for these relatively small values of $\gamma$, we can still interpret our result as an upper bound on $\gamma_{\rm eq}$: if $\gamma_{\rm eq}$ were in fact larger than $\mathcal{O}(10)$, then our approximations would be justified and we would find no self-consistent solution for the terminal velocity, implying that the true value of $\gamma_{\rm eq}$ must fall within the regime where our analysis breaks down. For BSM scenarios with large values of $\alpha_{\theta}$, our predicted value of $\gamma_{\rm eq}$ will become increasingly accurate. 

It is illuminating to compare our results for the terminal velocity to those obtained from the earlier fixed-order results of Ref.~\cite{Bodeker:2017cim}. To do so, we  follow the approach of Ref.~\cite{Ellis:2019oqb}, which amounts to using $\langle\Delta p_z/(\gamma T)\rangle_{\rm FC, FO}^{(\rm cut)}$ in place of $\langle\Delta p_z/(\gamma T)\rangle$ and neglecting the logarithmic term.  Eq.~\eqref{eq:pressure_balance} then becomes
\begin{equation}
P_{\rm tot} \simeq -\Delta V + P_{1 \rightarrow 1} + P_{1 \rightarrow 2}^{\rm FO, \, NL}\,,
\end{equation}
where 
\begin{equation}
\begin{aligned}
P_{1 \rightarrow 2}^{\rm FO, \, NL} &\equiv \sum_{a \in \mathcal{S}} \int \frac{\mathrm{d}^3\vec{p}_a}{(2\pi)^3}f_a(\vec{p}_a)\,\gamma T\,\Bigl<\frac{\Delta p_z}{\gamma T}\Bigr>_{\rm FC, FO}^{(\rm cut)}\,,
\end{aligned}
\end{equation}
with the logarithmic term dropped in $\langle\Delta p_z/(\gamma T)\rangle_{\rm FC, FO}^{(\rm cut)}$. In analogy with Eq.~\eqref{eq:alpha_def}, we can define 
\begin{equation}
 \alpha_{\rm eq}^{{\rm FO, \,NL}} \equiv \frac{P_{1 \rightarrow 2}^{\rm FO, \, NL}}{\gamma \, \rho_{\rm rad}} \approx 7 \times 10^{-5} \, \frac{\langle \phi \rangle}{T}\,,
\end{equation} 
where the numerical value again assumes SM-like plasma content and only accounts for the electroweak gauge boson contributions, as in Ref.~\cite{Ellis:2019oqb}.  The analog of Eq.~\eqref{eq:gameq_1n} is
\begin{equation}
\gamma_{\rm eq} = \frac{\alpha_\theta-\alpha_{\infty}}{\alpha_{\rm eq}^{{\rm FO, \,NL}}} \approx 1 \times 10^4  \, \left(\frac{\langle \phi \rangle}{T}\right)^{-1} \times \alpha_{\theta}, \qquad ({\rm fixed \, order, \, no \, logarithm})\,,
\end{equation}
with the last approximation again holding for $\alpha_{\theta} \gg \alpha_{\infty}$. In this approximation, the terminal velocity in the large-$\alpha_{\theta}$ limit scales as $\alpha_{\theta}$ instead of $\sqrt{\alpha_{\theta}}$, and further depends on the order parameter of the phase transition due to the mass cutoff in the integration.

\begin{figure}[t]
\centering
\includegraphics[scale=0.6]{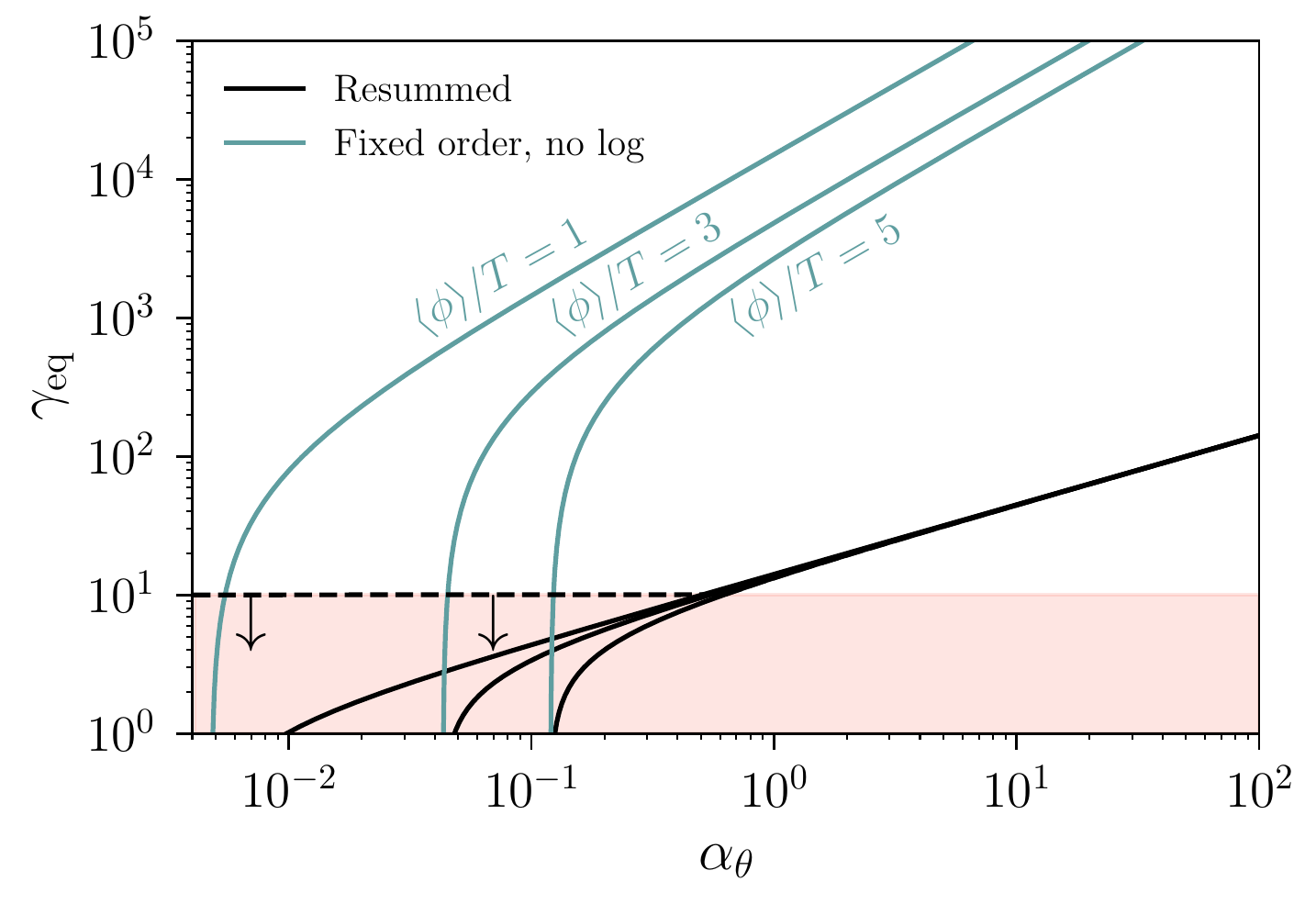}
\caption{Comparison of the terminal Lorentz factor from our analysis (black) compared to those inferred from a fixed order calculation (blue). We show predictions for different values of the order parameter $\langle\phi\rangle/T$, with $\langle\phi\rangle/T=1$, 3, 5 for the set of black curves from top to bottom, and similarly for the blue curves. For large $\alpha_{\theta}$ (strong transitions), our results predict significantly slower walls than implied by previous analyses, with the terminal velocity independent of the particle masses. For an electroweak-scale phase transition in a SM-like plasma,  several of the approximations made break down for $\gamma_{\rm eq} \lesssim 10$, rendering the predictions in the shaded pink region unreliable. There, our result should be interpreted as an upper bound on the terminal Lorentz factor, indicated by the dashed black line and the arrows.}  
\label{fig:gamma_eq}
\end{figure}

We compare our result for the terminal Lorentz factor to the fixed order prediction in Fig.~\ref{fig:gamma_eq}.  We show the corresponding results for $\langle\phi\rangle/T=1$, 3, 5. For strong transitions, our resummed result implies significantly slower walls than suggested by the fixed order calculation. One should bear in mind, however, that for $\gamma_{\rm eq} \lesssim 10$ or so (for a SM-like plasma) several of the approximations we have made break down - this is indicated by the shaded pink region of Fig.~\ref{fig:gamma_eq}. In this regime, we instead interpret our results as an upper bound on the wall velocity, indicated by the black dashed line and the arrows: applying our analysis for larger $\gamma$ would be self-consistent and predict a thermal pressure overwhelming the vacuum energy difference, indicating that such velocities are never reached.  We discuss the implications of these results for cosmology in the remainder of this section.

\subsection{Cosmological implications}

If the cosmological electroweak phase transition was  first order in nature, the dynamics of the Higgs-phase bubble walls affect the production of various cosmological relics.  
In this subsection, we briefly discuss the implications of our work for several possible relics.  

\subsubsection{Gravitational wave radiation}
The departure from thermal equilibrium during a first-order electroweak phase transition provides a suitable environment for gravitational wave radiation production.
Gravitational waves arise predominantly from three sources~\cite{Caprini:2019egz,Caprini:2015zlo,Hogan:1983zz,Kamionkowski:1993fg}: the collisions of bubble walls, sound waves that are produced when bubble walls push through the plasma, and the decay of magnetohydrodynamic turbulence that is produced when bubbles collide.
If the bubble walls were able to runaway, such that $\gamma \to \infty$ as $v \to 1$ without bound before colliding with other bubbles, then the latent heat of the phase transition would be transferred predominantly to the kinetic energy of the bubble walls and their collisions would provide the dominant source of gravitational wave energy.  
However, when the bubble walls reach a terminal velocity before they collide with each other, then the kinetic energy of the bubble walls saturates and most of the energy of the phase transition is transferred into the kinetic motion of the plasma~\cite{Hindmarsh:2014aa}.  
In this case, the sound waves and turbulence are the dominant sources of gravitational wave radiation. The scalar field, sound wave, and turbulence contributions to the GW background differ in terms of their spectral shapes and dependence on the phase transition parameters. This implies that the detection prospects of the corresponding signal can depend quite strongly on which source(s) dominates. 

To assess the impact of our results on GW predictions, note that at early stages of expansion when the friction is negligible, the bubble expands as in vacuum, for which the radius grows as 
\begin{equation} \label{eq:R}
R\sim \gamma R_0\,,
\end{equation}
where $R_0$ is the initial radius, which itself is usually close to the critical radius $R_c$. The average size reached by the bubbles at collision, $R_*$, is set by the typical separation between bubble centers~\cite{Caprini:2019egz}, 
\begin{equation}\label{eq:Rstar}
R_* \simeq (8\pi)^{1/3} \beta^{-1}\,,
\end{equation}
for fast walls, where $\beta$ parametrizes the duration of the PT and is typically $\sim \mathcal{O}(10-1000) H_*$ for models predicting a sizable GW signal with SM-like field content, with $H_*$ the Hubble parameter at the transition (see Ref.~\cite{Caprini:2019egz} for a detailed discussion and examples). Eqs.~\eqref{eq:R} and~\eqref{eq:Rstar} can be combined to define a characteristic Lorentz factor at collision in the absence of significant friction~\cite{Ellis:2019oqb}:
\begin{equation}
\gamma_* \equiv \frac{(8\pi)^{1/3}}{R_c \beta} \gtrsim \frac{10^{-3}}{R_c H_*}\,,
\end{equation}
for models with $\beta/H_* \lesssim 10^3$ as typically required for detection at LISA~\cite{Caprini:2019egz}. This Lorentz factor should be compared with our prediction for $\gamma_{\rm eq}$: if $\gamma_* \lesssim \gamma_{\rm eq}$, the bubbles effectively run away and the GW signal is dominated by the scalar field contributions. If $\gamma_* \gtrsim \gamma_{\rm eq}$, then the fluid sources (sound waves and turbulence) dominate. For thermal transitions, one typically expects $R_c \sim 1/T$, $H_* \sim T^2/M_{\rm Pl}$ with $M_{\rm Pl}$ the reduced Planck mass. Then
\begin{equation}
\gamma_* \gtrsim 10^{-3} \frac{M_{\rm Pl}}{T} \sim 10^{13}\,,
\end{equation}
where the last approximate equality holds for EW-scale transitions with $T\sim 100$ GeV.

From these considerations, we find that the GW signal will be completely dominated by contributions from the fluid for transitions with $\gamma_{\rm eq} \ll 10^{13}$ in models of interest. From Fig.~\ref{fig:gamma_eq}, we observe that for electroweak phase transition scenarios with SM-like potentials and $\alpha \sim \mathcal{O}(1)$ or so, both our result and the earlier $1\rightarrow 2$ estimates predict $\gamma_{\rm eq} \ll \gamma_*$, so the qualitative picture does not change once higher-order effects are accounted for. However, there exist models with phase transitions arising from non-polynomial potentials, which can feature extremely large values of $\alpha$ (see ~\cite{Caprini:2019egz,Ellis:2019oqb,Randall:2006py,Konstandin:2011dr,Konstandin:2011ds,Jinno:2016knw,Iso:2017uuu,vonHarling:2017yew,Kobakhidze:2017mru,Marzola:2017jzl,Prokopec:2018tnq,Hambye:2018qjv,Marzo:2018nov,Baratella:2018pxi,Bruggisser:2018mrt,Aoki:2019mlt,DelleRose:2019pgi} for examples). In these cases, our results will likely yield substantially different predictions for the GW signal. It could be the case that due to the $P \propto \gamma^2$ scaling, models which were previously thought to exhibit $\gamma_{\rm eq} > \gamma_*$, and hence a large scalar field contribution to the GW signal, do not once higher order effects are accounted for. We intend to apply our methods to such BSM scenarios in future work. 

\subsubsection{Matter-antimatter asymmetry}
The cosmological excess of matter over antimatter may have arisen at the electroweak phase transition through the physics of electroweak baryogenesis~\cite{Kuzmin:1985mm,Shaposhnikov:1986jp}~(see Ref.~\cite{Morrissey:2012db} for a review).  
There are various implementations of electroweak baryogenesis.  

Theories of \textit{non-local} electroweak baryogenesis~\cite{Cohen:1990it} rely on the transport of charge (particle number asymmetries) from the Higgs-phase bubble wall, where \textsf{CP} is violated, to the symmetric phase in front of the wall, where \textsf{B} is violated.  
These theories require the non-relativistic wall velocities, as the \textsf{CP}-violating source becomes suppressed for walls moving significantly faster than the  speed of sound in the plasma\footnote{Ref.~\cite{Cline:2020jre} recently showed that the resulting baryon asymmetry does not necessarily vanish for velocities exceeding $c_s$, however it is still generally suppressed relative to the subsonic case and vanishes as $v\rightarrow 1$.}, $c_s \approx 1 / \sqrt{3} \simeq 0.577$.  
Otherwise, charge simply enters the bubble without diffusing into the symmetric phase, and there is insufficient time for \textsf{B}-violation to act.  
Our results imply that the friction from Standard Model particles in the plasma will lead to a relativistic wall with $\gamma = \mathcal{O}(10)$ for SM-like electroweak transitions with $\alpha_{\theta}\sim 1$, which does not provide the necessary environment for non-local electroweak baryogenesis~\cite{Joyce:1994zn,Joyce:1994zt}.  
However, a viable scenario can be obtained for smaller values of $\alpha_{\theta}$ or by lowering the wall's terminal velocity by introducing new species of particles in the plasma to raise the thermal pressure beyond the Standard Model prediction.  

Alternatively, theories of \textit{local} electroweak baryogenesis~\cite{Trodden:1998ym} implement both \textsf{CP} and \textsf{B} violation at the Higgs-phase bubble wall.  
For instance, the passage of bubble walls through regions of plasma with nonzero gauge-Higgs field winding will trigger these configurations to unwind and generate an anomalous \textsf{B} number.
A related idea was proposed recently in Ref.~\cite{Katz:2016adq} where the collisions of ultrarelativistic bubble walls creates heavy particles that decay out of equilibrium to generate the baryon asymmetry.  
In these theories, there is no upper limit on the wall's speed, and baryogenesis may be viable even for ultrarelativistic walls.  However, our results preclude the possibility of runaway bubbles at SM-like electroweak transitions, which could impact the viability of mechanisms like that proposed in Ref.~\cite{Katz:2016adq}, depending on the details of the plasma content and potential assumed.

\subsubsection{Primordial magnetic fields}

The collisions of Higgs-phase bubble walls at the electroweak phase transition are expected to generate a primordial magnetic field~\cite{Grasso:2000wj}.  
For instance, bubble collisions stir up the charged constituents of the plasma~\cite{Hindmarsh:2017gnf}, and the associated magnetohydrodynamic turbulence leads to a magnetic field with coherence on the scale of the bubble radius.  

The subsequent evolution of this primordial magnetic field from the electroweak phase transition until today can be studied with the theory of magnetohydrodynamics, for instance using numerical lattice simulations~\cite{Kahniashvili:2012uj,Brandenburg:2017neh}.  
This cosmological magnetic field is expected to survive in the Universe today where it may play an important role in the generation of galactic magnetic fields and may be probed by various cosmological and astrophysical observations; see Ref.~\cite{Durrer:2013pga} for a general review, and see Ref.~\cite{Ellis:2019tjf} for a discussion of the electroweak phase transition, in particular.  
We are not aware of any studies that specifically address how the primordial magnetic field's strength depends on the bubble wall's speed in the regime where $\gamma \gg 1$, and this would be interesting to investigate further.  
Nevertheless, it is reasonable to expect some level of magnetic field creation for any $\gamma$, since electromagnetic radiation can arise even for a vacuum electroweak phase transition~\cite{Zhang:2019vsb}, which corresponds to the regime $\gamma \to \infty$ with a runaway bubble wall.

\section{Summary and Discussion}\label{sec:discussion}
We have presented an all-orders calculation of the pressure
exerted by Standard Model particles on fast-moving bubble walls produced during a first-order electroweak phase transition in the early Universe.
We built on and extended the pioneering works of Refs.~\cite{Bodeker:2009qy} and
\cite{Bodeker:2017cim} which calculated the pressure induced from $1\to 1$ and $1\to2$ processes respectively. These fixed order calculations receive large corrections in the limit of large wall velocities, motivating 
an all-orders calculation in the leading-logarithmic approximation, which we have performed for the first time. 

We carried out a fixed order calculation where we parametrized the radiating particle in terms of a classical current. From this, the vacuum persistence amplitude was calculated, at leading logarithmic accuracy, for both the real emissions and virtual corrections. This is necessary as infrared divergences cancel once the virtual corrections and real emissions are combined at the same order in perturbation theory. The vacuum persistence amplitude squared is exponentiated to calculate the resummed average momentum transfer, $\langle \Delta p_z\rangle$, to the wall. This calculation  closely follows the resummation of the thrust observable at colliders.  As seen in Eq.~(\ref{eq:avg_pressure_ll_fc}), we found $\langle \Delta p_z\rangle \sim \gamma T$, where the coefficient depends on the coupling of the incident particle species to the radiated particle.  The corresponding pressure is given by Eqs.~(\ref{eq:pressure_sum})~and~(\ref{eq:Pa}).
In addition to the analytic resummation we numerically simulated a particle shower inside the bubble and extracted the average momentum transfer to the wall. We found the numerical and analytical resummation results to be consistent with each other at the $10\%$ level and exhibit the same parametric dependence on the boost factor; see Table.~\ref{fig:pressure_comparison}.

For a wall with Lorentz factor $\gamma \gg 1$, both approaches indicate an average momentum transfer of $\sim 0.5\% \times \gamma T$. The results in Table.~\ref{fig:pressure_comparison} show that the pressure is dominated by both incoming  vector bosons and quarks. Interestingly, gluon emission from light quarks presents a non-negligible contribution to the pressure due to the number of degrees of freedom for quarks and the magnitude of the strong coupling. This contrasts with previous work, which found that the pressure is always dominated by the showered particles receiving the largest mass at the transition. As the pressure is a product of the incident particle flux and the average momentum transfer, our results indicate   
that the net thermal pressure experienced by the wall is parametrically $P \sim  \gamma^2 T^4$ in its rest frame, and is simply a constant fraction ($\sim 0.5\%$) of the total pressure of the ideal gas of symmetric-phase particles that couple directly to the wall. This result in fact matches the scaling of the pressure in a different class of scenarios for which local thermal equilibrium is maintained across the wall, as shown recently in~\cite{Mancha:2020fzw}. 
However, $P\sim \gamma^2 T^4$ is in contrast with Ref.~\cite{Bodeker:2017cim}, which instead found $P\sim\gamma \Delta m T^3$ with $\Delta m$ the change in mass of the emitted gauge bosons. We trace this difference back to an ad-hoc cutoff of the momentum integrals, the different behavior of emission matrix elements, and (as a consequence) the neglect of large logarithms in Ref.~\cite{Bodeker:2017cim}, which must be resummed. As such, we conclude that Higgs-phase bubble walls at strong electroweak phase transitions reach significantly slower terminal velocities than previously estimated.

 
Our results have implications for various cosmological observables.
Of particular interest is their impact on the scaling of the terminal bubble wall velocity with $\gamma$ and the strength of the phase transition, as this affects the associated gravitational
wave (GW) spectrum. In Fig.~\ref{fig:gamma}, we show the terminal Lorentz boost factor ($\gamma_{\rm eq}$) of the bubble wall as a function of the 
vacuum energy difference normalized to the thermal energy of the bath ($\alpha_\theta$). The scaling we calculated is shown in black while the scaling of \cite{Bodeker:2017cim} is shown in blue. This figure demonstrates that for a given value of $\alpha_\theta$, incorporation of off-shell effects significantly reduces the bubble wall velocity and results in $\gamma_{\rm eq} \propto \sqrt{\alpha_{\theta}}$ as opposed to $\propto \alpha_{\theta}$.  This is important as slower moving bubble walls imply the predominant source of gravitational waves stems from the fluid. As a result, the GW spectrum in models with large $\alpha_{\theta}$ can have a markedly difference shape than in the case where the contribution from collisions is assumed to dominate.

Our work has focused on first order electroweak phase transitions where the plasma content is dominated by SM particles. This led to gluon and electroweak gauge boson emission being the dominant source of momentum transfer at the wall. This is not necessarily the case in other models of interest from the standpoint of gravitational wave production. It would be interesting to extend our analysis to these scenarios in the future.

\section*{Acknowledgments}
We thank John Campbell and Marek Sch{\"o}nherr for numerous discussions on the interpretation of transition radiation and on the YFS formalism.  We are grateful to Nikita Blinov, Dietrich B\"odeker, Kimmo Kainulainen, Heather McAslan, and the members of the LISA Cosmology Working Group for useful discussions.  We are grateful to Aleksandr Azatov and Miguel Vanvlasselaer for constructive comments on the initial version of this manuscript.  This research was supported by the Fermi National Accelerator Laboratory (Fermilab), a U.S. Department of Energy, Office of Science, HEP User Facility. Fermilab is managed by Fermi Research Alliance, LLC (FRA), acting under Contract No. DE--AC02--07CH11359.  This work was initiated and performed in part at the Aspen Center for Physics, which is supported by National Science Foundation grant PHY-1607611. The work of JK was supported by Department of Energy (DOE) grant DE-SC0019195.

\appendix
\section{Vertex functions}
\label{sec:vertex_functions}
In this appendix we compute the charge factors needed to evaluate
Eq.~\eqref{eq:avg_pressure_ll_fc} and to perform the numerical simulations
in Sec.~\ref{sec:ps}.
For strong interactions, we extract the strong coupling $g_3^2=4\pi\alpha_s$.
Ignoring subleading $N_c$ contributions, the charge factors are then given
by the color Casimir operators
\begin{equation}\label{eq:QCD_charge}
  C_{qqg}=C_F=\frac{N_c^2-1}{2N_c}\;,\qquad
  C_{ggg}=C_A=N_c\;.
\end{equation}
The relevant electroweak couplings can be obtained from existing approaches
to electroweak showers~\cite{Chen:2016wkt,Bauer:2017bnh,Denner:2019vbn,Kleiss:2020rcg}.
We give the helicity averaged results, which are sufficient for the target accuracy
in our calculation. We denote $\cos\theta_W=m_W/m_Z$ as $c_W$ and $\sin\theta_W$ as $s_W$.
We extract the electromagnetic coupling as $g_1^2=4\pi\alpha$.
This leaves the following charge factors for bosonic interactions~\cite{Denner:2019vbn}
\begin{equation}\label{eq:tgcs}
  \begin{split}
    C_{W^\pm W^\mp\gamma}=&\;C_{G_{W^\pm}G_{W^\mp}\gamma}=\;1\;,\quad
    &C_{W^\pm W^\mp Z}=&\;\frac{c_W^2}{s_W^2}\;,\quad
    &C_{G_{W^\pm}G_{W^\mp}Z}=&\;\left(\frac{c_W^2-s_W^2}{2c_Ws_W}\right)^2\;,\\
    C_{hG_{W^\pm}W^\mp} =&\;C_{G_ZG_{W^\pm}W^\mp}=\frac{1}{4s_W^2}\;,\quad
    &C_{hG_ZZ} =&\;\frac{1}{4c_W^2s_W^2}\;.
  \end{split}
\end{equation}
For radiation off of fermions we obtain
\begin{equation}\label{eq:fermioncpls}
  \begin{split}
    C_{f_if_j\gamma}=&\;Q_f^2\,\delta_{ij}\;,\quad
    &C_{f_if_jZ}=&\;\Big(Q_f\frac{s_W}{c_W}\Big)^2\delta_{ij}+
    \Big(Q_f\frac{s_W}{c_W}-\frac{I_f^3}{c_Ws_W}\Big)^2\delta_{ij}\;,\\
    C_{\bar{u}_id_jW^+}=&\;\frac{1}{4s_W^2}\,|V_{ij}|^2\;,\quad
    &C_{\bar{\nu}_il_jW^+}=&\;\frac{1}{4s_W^2}\,\delta_{ij}\;,
  \end{split}
\end{equation}
where $Q_f$ and $I_f^3$ are the electric charge and third component of
weak isospin for the fermion $f$, and where $V$ is the CKM matrix.

\section{Scalar QED as an example}
\label{app:scalarQED}
In this appendix we illustrate the computation of the full matrix element 
squared for scalar QED. The theory has simple Feynman rules and provides
the same solution as more realistic theories involving fermions, both in 
the soft gauge boson limit, and in the limit $m^2\ll k_\perp^2$. This
can be attributed to the universal behavior of soft gauge boson radiation
which was introduced from the semi-classical perspective 
in Sec.~\ref{sec:fixed_order}.

In scalar QED, the vertex functions are simply the matrix elements
in the symmetric and broken phase:
\begin{equation}
\begin{split}
        V_\mathrm{s} &  = ig\,(p_{a,\mathrm{s}} + p_{b,\mathrm{s}})_{\mu}\, \varepsilon_{\lambda}^{\mu}\;,
        \qquad\qquad
        &V_\mathrm{h} & = ig\,(p_{a,\mathrm{h}} + p_{b,\mathrm{h}})_{\mu}\, \varepsilon_{\lambda}^{\mu}\;.
\end{split}
\end{equation}
We use energy-momentum conservation at the radiative vertex to define 
$p_{b,\mathrm{s}} = p_{a,\mathrm{s}} - p_{c,\mathrm{s}}$ and 
$p_{a,\mathrm{h}} = p_{b,\mathrm{h}} + p_{c,\mathrm{h}}$.
The interaction with the wall and the associated momentum transfer is included
in these expressions in terms of a featureless vertex as it would be obtained 
in the VEV insertion approximation.
Assuming that the soft emission has a negligible (thermal) mass
in both phases lets us take  $p_{c,\mathrm{s}}=p_{c,\mathrm{h}}\equiv p_{c}$.
Importantly, we do not apply the approximation that $V_\mathrm{s}\sim V_\mathrm{h}\sim V$
(cf.\ the discussion in Sec.~\ref{sec:formalism}).
In axial gauge with arbitrary gauge vector, $n^\mu$, 
\begin{equation}\label{eq:axial_gauge_sqed}
  \sum_{\lambda=\pm}\eps^\mu_\lambda(p_c,n)\eps^{\nu,*}_\lambda(p_c,n)=
  -g^{\mu\nu}+\frac{p_c^\mu n^\nu+p_c^\nu n^\mu}{p_cn}-n^2\frac{p_c^\mu p_c^\nu}{(p_cn)^2}\;,
\end{equation}
the squared vertex functions summed over polarization states are given by
\begin{equation}\label{eq:scalarqed_vertices_squared}
    \begin{split}
        V_\mathrm{s}V^*_\mathrm{s} & = |g|^2\left[ -4p^2_{a,\mathrm{s}}
        +8p_{a,\mathrm{s}}n\,\frac{p_{a,\mathrm{s}}p_{c}}{p_cn}
        -4n^2\frac{(p_{a,\mathrm{s}}p_{c})^2}{(p_cn)^2}
        \right]\,,\\
        V_\mathrm{h}V^*_\mathrm{h} & = |g|^2\left[ -4p^2_{b,\mathrm{h}}
        +8p_{b,\mathrm{h}}n\,\frac{p_{b,\mathrm{h}}p_{c}}{p_cn}
        -4n^2\frac{(p_{b,\mathrm{h}}p_{c})^2}{(p_cn)^2}\right]\,,\\
        V_\mathrm{s}V^*_\mathrm{h} = V_\mathrm{h}V^*_\mathrm{s} & = |g|^2\left[ 
        -4 \left(p_{a,\mathrm{s}} p_{b,\mathrm{h}}\right)
        +4p_{b,\mathrm{h}}n\,\frac{p_{a,\mathrm{s}}p_{c}}{p_cn}
        +4p_{a,\mathrm{s}}n\,\frac{p_{b,\mathrm{h}}p_{c}}{p_cn}
        -4n^2\frac{p_{a,\mathrm{s}}p_{c}\,p_{b,\mathrm{h}}p_{c}}{(p_cn)^2}
        \right]\,,\\
\end{split}
\end{equation}
where we have also taken $p_c^2 = 0$.
The squared matrix element is calculated using Eq.~\eqref{eq:masterMSQ}, 
where the propagator terms $A_\mathrm{s}$ and $A_\mathrm{h}$ are given by
$A_\mathrm{s}=-2p_{a,\mathrm{s}}p_{c}$ and $A_\mathrm{h}=-2p_{b,\mathrm{h}}p_{c}$.
Putting this together we find 
\begin{equation}\label{eq:bm_correct2}
    \begin{split}
        |\mathcal{M}_{a\to bc}^{(0)}|^2=&\;4E_a^2 \lvert g \rvert^2 \Bigg(\frac{2\,p_{a,\mathrm{s}}p_{b,\mathrm{h}}}{ p_{a,\mathrm{s}}p_{c}\;p_{b,\mathrm{h}}p_{c}}-\frac{p_{a,\mathrm{s}}^2}{(p_{a,\mathrm{s}}p_{c})^2
        }-\frac{p_{b,\mathrm{h}}^2}{(p_{b,\mathrm{h}}p_{c})^2}\Bigg)
        \;.
    \end{split}
\end{equation}
Note that the result agrees with the integrand
in Eq.~\eqref{eq:w1_real} up to normalization factors, and that it is gauge independent,
i.e.\ there is no residual dependence on the auxiliary vector $n^\mu$. This is not
the case if we work with the approximate expressions for $A_s$ and $A_h$ 
in Eq.~\eqref{eq:As_Ah_from_BM}, or if we assume the momentum transfer to the wall 
to be associated with the radiative vertex itself, such that four-momentum conservation
would be violated locally during the emission of the gauge boson. For the Standard Model, 
the relevant interactions that exert pressure on the bubble wall are predominantly from fermions
and vector bosons emitting gauge bosons. The derivation of the squared matrix element 
is slightly more complicated in these cases due to the spinor algebra, but the result 
is identical to Eq.~\eqref{eq:bm_correct2} in the limits relevant to our computation. 
For the emission of a massive gauge boson, there is a change in polarization of the gauge bosons 
across the wall, and a complete treatment would require the matching of gauge fields at the boundary.
However, the resulting effects are suppressed by $m^2/k_\perp^2$ and therefore
irrelevant at large $\gamma T$, where the dominant contribution to the pressure
comes from the region $m^2\ll k_\perp^2$. 

In order to better understand the correspondence with the calculation in~\cite{Bodeker:2017cim}
we analyze Eq.~\eqref{eq:scalarqed_vertices_squared} for the special choice of
$n^\mu=p_{b,\mathrm{h}}^\mu$. We obtain
\begin{equation}\label{eq:scalarqed_vertices_squared_llaxial}
    \begin{split}
        V_\mathrm{s}V^*_\mathrm{s} & = 4|g|^2\left[ -p^2_{a,\mathrm{s}}
        +2p_{b,\mathrm{h}}p_{a,\mathrm{s}}\frac{p_{a,\mathrm{s}}p_{c}}{p_{b,\mathrm{h}}p_c}
        -p_{b,\mathrm{h}}^2\frac{(p_{a,\mathrm{s}}p_{c})^2}{(p_{b,\mathrm{h}}p_c)^2}\right]\;,\\
        V_\mathrm{h}V^*_\mathrm{h} & = V_\mathrm{s}V^*_\mathrm{h} = V_\mathrm{h}V^*_\mathrm{s} = 0\,.\\
\end{split}
\end{equation}
The remarkable result is that the complete matrix element 
is given by a single Feynman diagram, if the auxiliary vector
is chosen to be the momentum of one of the particles forming the soft current.
The squared matrix element still coincides with Eq.~\eqref{eq:bm_correct2},
as it must, because the complete result is gauge invariant.
This is a well-known result that holds in any gauge theory with massless
vector bosons~\cite{Field:1989uq}. It can be derived more generally based 
on spinor algebra~\cite{Dixon:1996wi} and is the basis for both 
analytic resummation techniques as well as parton showers~\cite{Ellis:1991qj}.

\section{Kinematics mapping and phase-space factorization}
\label{sec:ps_kin}
In this section we describe the algorithm to generate the kinematics 
in the numerical simulation. For splittings involving the incident parton,
$p_a$, we use the simple parametrization
\begin{equation}
    \begin{split}
        p_a^\mu=&\;\left(E_a,\vec{0},\sqrt{E_a^2-m_a^2}\right)
        \approx\Big(E_a,\vec{0},E_a\left(1-\frac{m_a^2}{2E_a^2}\right)\Big)\\
        p_b^\mu=&\;\left((1-x)E_a,-\vec{k}_\perp,\sqrt{(1-x)^2E_a^2-\vec{k}_\perp^{\,2}-m_b^2}\right)
        \approx\Big((1-x)E_a,-\vec{k}_\perp,(1-x)E_a\left(1-\frac{\vec{k}_\perp^{\,2}+m_b^2}{2(1-x)^2E_a^2}\right)\Big)\\
        p_c^\mu=&\;\left(x\,E_a,\vec{k}_\perp,\sqrt{x^2E_a^2-\vec{k}_\perp^{\,2}-m_c^2}\right)
        \approx\Big(x\,E_a,\vec{k}_\perp,x\,E_a\left(1-\frac{\vec{k}_\perp^{\,2}+m_c^2}{2x^2E_a^2}\right)\Big)
    \end{split}
\end{equation}
where the numerical simulation is in fact generated without any approximations.
The differential phase space element for the emission of $p_c$ can be written as
\begin{equation}
    \frac{{\rm d}^3\vec{p}_c}{(2\pi)^3\,2E_c}=
    \frac{1}{16\pi^2}\,d\vec{k}_{\perp}^{\,2}
    \frac{{\rm d}x}{\sqrt{x^2-(\vec{k}_{\perp}^{\,2}+m_c^2)/E_a^2}}
    \frac{{\rm d}\phi}{2\pi}\;
    \overset{\vec{k}_{\perp}^{\,2}+m_c^2\ll x^2E_a^2}{\longrightarrow}\;
    \frac{1}{16\pi^2}\,d\vec{k}_{\perp}^{\,2}
    \frac{{\rm d}x}{x}\frac{{\rm d}\phi}{2\pi}
\end{equation}

The algorithm for constructing the splitting kinematics in
final-state radiation is modeled on Ref.~\cite{Catani:2002hc}.%
\footnote{By suitable crossing of momenta to the initial state, 
  this algorithm can also be employed to simulate radiation off 
  the incident particle, although it presents the slight disadvantage 
  that it leads to energy transfer into the bubble, which corresponds to
  plasma heating. While the results for the pressure transfer are 
  unaffected by this, we prefer to work with the simpler initial-state
  radiation formalism in order to accurately reflect the symmetries
  of the problem.}
We use the following variables for a dipole splitting
$\{\widetilde{ij},\tilde{k}\}\to\{i,j,k\}$ with momentum
configuration $\tilde{p}_{ij}+\tilde{p}_k\to p_i+p_j+p_k$:
\begin{equation}\label{eq:cs_variables}
  \begin{split}
    \tilde{z}_i=&\;\frac{p_ip_k}{p_ip_k+p_jp_k}\;,
    \qquad
    &s_{ij}=&\;(p_i+p_j)^2\;,
    \qquad\text{and}\qquad
    Q=p_i+p_j+p_k\;.
  \end{split}
\end{equation}
The variable $\tilde{z}_i$ corresponds to the splitting variable $1-x$
of the parton shower, while the invariant mass $s_{ij}$ is computed
from the evolution and splitting variable as (cf.~Sec.~\ref{sec:ps})
\begin{equation}
  s_{ij}=t\,\times
  \left\{\begin{array}{lll}
  \big(\tilde{z}_i(1-\tilde{z}_i)\big)^{-1} & \text{if $t=|\vec{k}_\perp|^2$} &\text{cf.~\cite{Schumann:2007mg}}\\[1mm]
  \;\,\tilde{z}_i(1-\tilde{z}_i) & \text{if $t=\tilde{q}_T^2$} &\text{cf.~\cite{Platzer:2009jq}}
  \end{array}
  \right.\;.
\end{equation}
The spectator particle $k$ serves as a source of anti-collinear
momentum, but is otherwise unaffected by the splitting of the mother
particle $\widetilde{ij}$ into the daughter particles $i$ and $j$.
For primary branchings, i.e.\ those where the particle $\tilde{ij}$
radiates coherently with the incoming particle $p_a$, we choose
$p_k^\mu=(M,\vec{0})$, where $M\gg\gamma T$. This corresponds to
the reaction with the wall being modeled as a fixed-target collision,
with an energy transfer that is suppressed by $(\Delta p)^2/2M^2$
compared to the momentum transfer $\Delta p$, such that it can be
neglected. For secondary branchings we use the standard parton shower
assignment of $p_k$~\cite{Hoche:2014rga}. 
The kinematics mapping proceeds as follows
\begin{enumerate}
\item Determine the new momentum of the spectator parton as
  \begin{equation}\label{eq:def_ff_pk}
    \begin{split}
      p_k^{\,\mu}=&\;\left(\tilde{p}_k^{\,\mu}-
        \frac{Q\cdot\tilde{p}_k}{Q^2}\,Q^\mu\right)\,
      \sqrt{\frac{\lambda(Q^2,s_{ij},m_k^2)}{\lambda(Q^2,m_{ij}^2,m_k^2)}}
      +\frac{Q^2+m_k^2-s_{ij}}{2\,Q^2}\,Q^\mu\;,
    \end{split}
  \end{equation}
  with $\lambda$ denoting the K{\"a}llen function 
  $\lambda(a,b,c)=(a-b-c)^2-4\,bc$\\ and
  $s_{ij}\,=\;y_{ij,k}\,(q^2-m_k^2)+(1-y_{ij,k})\,(m_i^2+m_j^2)$.
\item Construct the new momentum of the emitter parton, $p_i$, as
  \begin{align}\label{eq:def_ff_pi_pj}
    p_i^\mu\,=&\;\bar{z}_i\,\frac{\gamma(Q^2,s_{ij},m_k^2)\,p_{ij}^\mu
      -s_{ij}\,p_k^\mu}{\beta(Q^2,s_{ij},m_k^2)}
    +\frac{m_i^2+{\rm k}_\perp^2}{\bar{z}_i}\,
    \frac{p_k^\mu-m_k^2/\gamma(Q^2,s_{ij},m_k^2)\,p_{ij}^\mu}{
      \beta(Q^2,s_{ij},m_k^2)}+k_\perp^\mu\;,
  \end{align}
  where $\beta(a,b,c)={\rm sgn}(a-b-c)\sqrt{\lambda(a,b,c)}$,
  $2\,\gamma(a,b,c)=(a-b-c)+\beta(a,b,c)$ and $p_{ij}^\mu=Q^\mu-p_k^\mu$.\\
  The parameters $\bar{z}_i$ and ${\rm k}_\perp^2=-k_\perp^2$ 
  of this decomposition are given by
  \begin{equation}\label{eq:def_ff_zi_kt}
    \begin{split}
      \bar{z}_i\,=&\;\frac{Q^2-s_{ij}-m_k^2}{\beta(Q^2,s_{ij},m_k^2)}\,
      \left[\;\tilde{z}_i\,-\,\frac{m_k^2}{\gamma(Q^2,s_{ij},m_k^2)}
          \frac{s_{ij}+m_i^2-m_j^2}{Q^2-s_{ij}-m_k^2}\right]\;,\\
      {\rm k}_\perp^2\,=&\;\bar{z}_i\,(1-\bar{z}_i)\,s_{ij}-
        (1-\bar{z}_i)\, m_i^2-\bar{z}_i\, m_j^2\;,
    \end{split}
  \end{equation}
\item The transverse momentum is constructed using an azimuthal angle, $\phi_{ai}$
  \begin{equation}\label{eq:likt}
    k_\perp^\mu={\rm k}_\perp\left(\cos\phi_{ai} \frac{n_\perp^\mu}{|n_\perp|}
    +\sin\phi_{ai} \frac{l_\perp^{\,\mu}}{|l_\perp|}\right)\;,
    \quad\text{where}\quad
    n_\perp^\mu=\eps^{0\mu}_{\;\;\;\nu\rho}\,
    \tilde{p}_{ij}^{\,\nu}\,\tilde{p}_k^{\,\rho}\;,
    \qquad
    l_\perp^{\,\mu}=\eps^\mu_{\;\nu\rho\sigma}\,
    \tilde{p}_{ij}^{\,\nu}\,\tilde{p}_k^{\,\rho}\,n_\perp^\sigma\;.
  \end{equation}
  In kinematical configurations where $\vec{\tilde{p}}_{aij}=\pm\vec{\tilde{p}}_k$,
  $n_\perp$ in the definition of Eq.~\eqref{eq:likt} vanishes. It can then be computed
  as $n_\perp^\mu=\eps^{0\,i\mu}_{\;\;\;\;\;\nu}\,\tilde{p}_{aij}^{\,\nu}$,
  where $i$ may be any Lorentz index that yields a nonzero result.
\end{enumerate}
The phase-space factorization was derived in~\cite{Dittmaier:1999mb}, App.~B.
Standard $s$-channel factorization over $p_{ij}$ gives~\cite{Byckling:1969sx,Byckling:1971vca}
\begin{equation}\label{eq:psfactorization_1}
  \begin{split}
    \int{\rm d}\Phi(p_i,p_j,p_k|\,Q)
    =&\int\frac{{\rm d}s_{ij}}{2\pi}\,
    \int{\rm d}\Phi(p_{ij},p_k|\,Q)\,
    \int{\rm d}\Phi(p_i,p_j|\,p_{ij})\\
    =&\int\frac{{\rm d}s_{ij}}{2\pi}\,
    \sqrt{\frac{\lambda(Q^2,s_{ij},m_k^2)}{
        \lambda(Q^2,m_{ij}^2,m_k^2)}}
    \int{\rm d}\Phi(\tilde{p}_{ij},\tilde{p}_k|\,Q)\,
    \int{\rm d}\Phi(p_i,p_j|\,p_{aij})\\
    =&\int{\rm d}\Phi(\tilde{p}_{ij},\tilde{p}_k|\,Q)\,
    \int\Big[{\rm d}\Phi(p_i,p_j|\,\tilde{p}_{ij},\tilde{p}_k)\Big]
  \end{split}    
\end{equation}
where ${\rm d}\Phi(p_{i_1},\ldots,p_{i_n}|Q)$ is given by the $n$-particle
final-state phase space integral in Eq.~\eqref{eq:diff_probability},
times a 4-momentum conservation constraint in the form $(2\pi)^4 \delta^{(4)}(Q-p_{i_1}-\ldots-p_{i_n})$.
We can rewrite Eq.~\eqref{eq:psfactorization_1} as
\begin{equation}\label{eq:ff_ps}
  \begin{split}
    \int\Big[{\rm d}\Phi(p_i,p_j|\,\tilde{p}_{ij},\tilde{p}_k)\Big]
    =&\int\frac{{\rm d}s_{ij}}{2\pi}\,
    \sqrt{\frac{\lambda(Q^2,s_{ij},m_k^2)}{
        \lambda(Q^2,m_{ij}^2,m_k^2)}}
    \int{\rm d}\Phi(p_i,p_j|p_{ij})\\
    =&\int\frac{{\rm d}s_{ij}}{2\pi}\,
    \frac{1}{\sqrt{\lambda(q^2,m_{ij}^2,m_k^2)}}
    \int\frac{{\rm d}s_{ik}\,{\rm d}\phi_i}{4(2\pi)^2}
    =\frac{J_{\rm FF}}{16\pi^2}\int{\rm d}s_{ij}
    \int{\rm d}\tilde{z}_i\int\frac{{\rm d}\phi_{i}}{2\pi}\;,
  \end{split}
\end{equation}
where $s_{ik}=(p_i+p_k)^2$, and where we have defined the Jacobian factor
\begin{equation}
  J_{\rm FF}=\frac{Q^2-s_{ij}-m_k^2}{
    \sqrt{\lambda(Q^2,m_{ij}^2,m_k^2)}}\;.
\end{equation}
Note that for massless partons, $J_{\rm FF}$ takes the simple form
$J_{\rm FF}=1-s_{ij}/Q^2$~\cite{Catani:1996vz}. In the soft limit,
we have $J_{\rm FF}=1$.

\section{Soft virtual integrals}
\label{sec:virtual_integrals}
In $D=4-2\eps$ dimensions with $\eps < 0$, Eq.~\eqref{eq:nlo_virtual_diff} reads
\begin{equation}\label{eq:nlo_virtual_diff_d}
    \begin{split}
    W_{a\to b}^{(2)}=&\;-4i\,|g|^2\mu^{2\eps}\Big(
    I_1(p_a,p_b)
    -I_2(p_a)
    -I_2(p_b)\Big)\;,
    \end{split}
\end{equation}
where we have defined the basic integrals
\begin{equation}
    \begin{split}
    I_1(p,q)=&\;\int\frac{{\rm d}^D k}{(2\pi)^{D}}
    \frac{1}{k^2}\frac{2pq}{(2pk)(2qk)}\;,
    \qquad\qquad
    &I_2(p)=&\;\Theta(p^2)
    \displaystyle\int\frac{{\rm d}^D k}{(2\pi)^{D}}
    \frac{1}{k^2}\frac{p^2}{(2pk)^2}\;.
    \end{split}
\end{equation}
Note that these integrals are both IR and UV divergent. Naive evaluation
using dimensional regularization would yield an ill-defined result.
However, we are interested only in the cancellation of the IR singularities,
which can be separated out at the integrand level. Using the decomposition
$I_{1/2}(p,q)=I_{1/2,\rm UV}(p,q)+I_{1/2,\rm IR}(p,q)$, we define
\begin{equation}
    \begin{split}
    I_{1,\rm IR}(p,q)=&\;\int\frac{{\rm d}^D k}{(2\pi)^{D}}
    \frac{1}{k^2}\frac{2pq}{(2pk+k^2)(2qk+k^2)}\;,
    \qquad
    &I_{2,\rm IR}(p)=&\;
    \displaystyle\int\frac{{\rm d}^D k}{(2\pi)^{D}}
    \frac{1}{k^2}\frac{p^2}{(2pk+k^2)^2}\\
    \end{split}
\end{equation}
We use Feynman parameters to write the integrand of $I_{1,\rm IR}$ as
\begin{equation}
\begin{split}
    \frac{1}{k^2(2pk+k^2)(2qk+k^2)}=&\;
    \int_0^1{\rm d}x\int_0^1{\rm d}y\,
    \frac{2x}{[K^2-C^2]^3}\;,
    \qquad&&\text{where}\qquad
    \begin{array}{l}
      K=k+C\\ C=x(yp+(1-y)q)
    \end{array}\\
\end{split}
\end{equation}
with the obvious change $C\to xp$ in the case of $I_{2,\rm IR}$.
Using the basic integral
\begin{equation}
\begin{split}
    \int\frac{{\rm d}^DK}{(2\pi)^D}\frac{1}{[K^2-C^2]^M}
    =&\;\frac{i(-1)^M}{(16\pi^2)^{D/4}}
    \frac{{\rm B}(D/2,M-D/2)}{\Gamma(D/2)(C^2)^{M-D/2}}
    \end{split}
\end{equation}
we obtain, in the massless case
\begin{equation}
\begin{split}\label{eq:i1ir_zero}
    I_{1,\rm IR}(p,q)=&\;-\frac{i}{16\pi^2}\,\frac{(4\pi)^\eps}{\Gamma(1-\eps)}
    \frac{(2pq)^{-\eps}}{\eps^2}\frac{\Gamma(1-\eps)^3\Gamma(1+\eps)}{\Gamma(1-2\eps)}
\end{split}
\end{equation}
For one massive leg, $p$, we obtain
\begin{equation}\label{eq:i1ir_massive}
\begin{split}
    I_{1,\rm IR}(p,q)=&\;-\frac{i}{16\pi^2}\,\frac{(4\pi)^\eps}{\Gamma(1-\eps)}\,
    (2pq)^{-\eps}\left(\frac{1}{2\eps^2}
    +\frac{\log\mu_p^2}{2\eps}-\frac{\log^2\mu_p^2}{4}
    -{\rm Li}_2(1-\mu_p^2)+\frac{\pi^2}{12}+\mathcal{O}(\eps)
    \right)\\
    I_{2,\rm IR}(p)=&\;\frac{i}{16\pi^2}\,\frac{(4\pi)^\eps}{\Gamma(1-\eps)}
    \frac{(p^2)^{-\eps}}{2\eps}\,\Gamma(1-\eps)\Gamma(1+\eps)
\end{split}
\end{equation}
where we have defined $\mu_p^2=p^2/(2pq)$.

\section{Soft real-emission integrals}
\label{sec:real_integrals}
In $D=4-2\eps$ dimensions with $\eps < 0$, Eq.~\eqref{eq:w1_real} reads
\begin{equation}\label{eq:nlo_real_diff_d}
    \begin{split}
    \int{\rm d}W_{a\to bc}^{2\,(1)}=&\;-4|g|^2\mu^{2\eps}\Big(
    \tilde{I}_1(p_a,p_b)
    -\tilde{I}_2(p_a,p_b)
    -\tilde{I}_2(p_b,p_a)\Big)\;,
    \end{split}
\end{equation}
where we have defined the basic integrals
\begin{equation}
    \begin{split}
    \tilde{I}_1(p,q)=&\;\int\frac{{\rm d}^D k}{(2\pi)^{D-1}}
    \frac{2pq}{(2pk)(2qk)}\,\delta^{+}(k^2)\;,
    \qquad
    &\tilde{I}_2(p,q)=&\;\Theta(p^2)
    \displaystyle\int\frac{{\rm d}^D k}{(2\pi)^{D-1}}
    \frac{p^2}{(2pk)^2}\,\delta^{+}(k^2)\;.
    \end{split}
\end{equation}
We use a Sudakov parametrization of the emission in terms of two
light-like hard momenta $n$ and $\bar{n}$
\begin{equation}
  {\rm d}^Dk
  =n\bar{n}\,{\rm d}\alpha\, {\rm d}\beta\,{\rm d}^{D-2}\vec{k}_\perp\;,
  \qquad\text{where}\qquad
  \alpha=\frac{\bar{n}k}{n\bar{n}}\;,\qquad
  \beta=\frac{nk}{n\bar{n}}\;.
\end{equation}
The hard momenta are the solutions of
\begin{equation}
\begin{split}
    p=&\;n+\frac{p^2}{2n\bar{n}}\,\bar{n}\;,
    \qquad
   & q=&\;\bar{n}+\frac{q^2}{2n\bar{n}}\,n\;,
   \qquad\text{where}\qquad
   2n\bar{n}=pq\left(1+v_{p,q}\right)\;,
   \qquad v_{p,q}=\sqrt{1-\frac{p^2q^2}{(pq)^2}}\;.
\end{split}
\end{equation}
The transverse momentum integral can be solved with the help 
of the on-shell condition
\begin{equation}\label{eq:pt_int_1p}
  \int {\rm d}^{D-2}\vec{k}_\perp\,\delta^+(k^2)=
  \frac{2\pi^{1-\eps}}{\Gamma(1-\eps)}\,\frac{1}{2}\,
  (2n\bar{n})^{-\eps}\big(\alpha\beta\big)^{-\eps}\;.
\end{equation}
This leads to the following result for the massless case
\begin{equation}
  \begin{split}
    \tilde{I}_1(p,q)=&\;-\frac{1}{16\pi^2}\frac{(4\pi)^\eps}{\Gamma(1-\eps)}\,
    \frac{(2pq)^{-\eps}}{\eps^2}\;.
  \end{split}
\end{equation}
In the case of one massive leg, $p^2>0$, we obtain instead
\begin{equation}
  \begin{split}
    \tilde{I}_1(p,q)=&\;-\frac{1}{16\pi^2}\frac{(4\pi)^\eps}{\Gamma(1-\eps)}\,
    (2pq)^{-\eps}\,\left(\frac{1}{2\eps^2}
    +\frac{\log\mu_p^2}{2\eps}-\frac{\log^2\mu_p^2}{4}
    -{\rm Li}_2(-\mu_p^2)-\frac{\pi^2}{12}
    +\mathcal{O}(\eps)\right)\\
    \tilde{I}_2(p,q)=&\;\frac{1}{16\pi^2}\frac{(4\pi)^\eps}{\Gamma(1-\eps)}\,
    (2pq)^{-\eps}\,\left(\frac{1}{2\eps}
    -\frac{\log\mu_p^2}{2}+\log(1+\mu_p^2)
    +\mathcal{O}(\eps)\right) \, .\\
  \end{split}
\end{equation}

\bibliographystyle{JHEP}

\bibliography{journal}

\providecommand{\href}[2]{#2}\begingroup\raggedright\begin{thebibliography}{10}

\bibitem{DOnofrio:2015mpa}
M.~D'Onofrio and K.~Rummukainen, \emph{{Standard model cross-over on the
  lattice}}, \href{https://doi.org/10.1103/PhysRevD.93.025003}{\emph{Phys.
  Rev.} {\bfseries D93} (2016) 025003},
  [\href{https://arxiv.org/abs/1508.07161}{{\ttfamily 1508.07161}}].

\bibitem{Anderson:1991zb}
G.~W. Anderson and L.~J. Hall, \emph{{The Electroweak phase transition and
  baryogenesis}}, \href{https://doi.org/10.1103/PhysRevD.45.2685}{\emph{Phys.
  Rev.} {\bfseries D45} (1992) 2685--2698}.

\bibitem{Quiros:1999jp}
M.~Quiros, \emph{{Finite temperature field theory and phase transitions}},  in
  \emph{{ICTP Summer School in High-Energy Physics and Cosmology}},
  pp.~187--259, 1, 1999, \href{https://arxiv.org/abs/hep-ph/9901312}{{\ttfamily
  hep-ph/9901312}}.

\bibitem{Grojean:2004xa}
C.~Grojean, G.~Servant and J.~D. Wells, \emph{{First-order electroweak phase
  transition in the standard model with a low cutoff}},
  \href{https://doi.org/10.1103/PhysRevD.71.036001}{\emph{Phys. Rev.}
  {\bfseries D71} (2005) 036001},
  [\href{https://arxiv.org/abs/hep-ph/0407019}{{\ttfamily hep-ph/0407019}}].

\bibitem{Carrington:1993ng}
M.~E. Carrington and J.~I. Kapusta, \emph{{Dynamics of the electroweak phase
  transition}}, \href{https://doi.org/10.1103/PhysRevD.47.5304}{\emph{Phys.
  Rev.} {\bfseries D47} (1993) 5304--5315}.

\bibitem{Cohen:1990it}
A.~G. Cohen, D.~B. Kaplan and A.~E. Nelson, \emph{{Baryogenesis at the weak
  phase transition}},
  \href{https://doi.org/10.1016/0550-3213(91)90395-E}{\emph{Nucl.Phys.}
  {\bfseries B349} (1991) 727--742}.

\bibitem{Achucarro:1999it}
A.~Achucarro and T.~Vachaspati, \emph{{Semilocal and electroweak strings}},
  \href{https://doi.org/10.1016/S0370-1573(99)00103-9}{\emph{Phys.Rept.}
  {\bfseries 327} (2000) 347--426},
  [\href{https://arxiv.org/abs/hep-ph/9904229}{{\ttfamily hep-ph/9904229}}].

\bibitem{Vachaspati:1991nm}
T.~Vachaspati, \emph{{Magnetic fields from cosmological phase transitions}},
  \href{https://doi.org/10.1016/0370-2693(91)90051-Q}{\emph{Phys.Lett.}
  {\bfseries B265} (1991) 258--261}.

\bibitem{Kamionkowski:1993fg}
M.~Kamionkowski, A.~Kosowsky and M.~S. Turner, \emph{{Gravitational radiation
  from first order phase transitions}},
  \href{https://doi.org/10.1103/PhysRevD.49.2837}{\emph{Phys. Rev.} {\bfseries
  D49} (1994) 2837--2851},
  [\href{https://arxiv.org/abs/astro-ph/9310044}{{\ttfamily
  astro-ph/9310044}}].

\bibitem{Caprini:2019egz}
C.~Caprini et~al., \emph{{Detecting gravitational waves from cosmological phase
  transitions with LISA: an update}},
  \href{https://doi.org/10.1088/1475-7516/2020/03/024}{\emph{JCAP} {\bfseries
  03} (2020) 024}, [\href{https://arxiv.org/abs/1910.13125}{{\ttfamily
  1910.13125}}].

\bibitem{Caprini:2015zlo}
C.~Caprini et~al., \emph{{Science with the space-based interferometer eLISA.
  II: Gravitational waves from cosmological phase transitions}},
  \href{https://doi.org/10.1088/1475-7516/2016/04/001}{\emph{JCAP} {\bfseries
  04} (2016) 001}, [\href{https://arxiv.org/abs/1512.06239}{{\ttfamily
  1512.06239}}].

\bibitem{Kawamura:2011zz}
S.~Kawamura et~al., \emph{{The Japanese space gravitational wave antenna:
  DECIGO}}, \href{https://doi.org/10.1088/0264-9381/28/9/094011}{\emph{Class.
  Quant. Grav.} {\bfseries 28} (2011) 094011}.

\bibitem{Harry:2006fi}
G.~M. Harry, P.~Fritschel, D.~A. Shaddock, W.~Folkner and E.~S. Phinney,
  \emph{{Laser interferometry for the big bang observer}},
  \href{https://doi.org/10.1088/0264-9381/23/15/008}{\emph{Class. Quant. Grav.}
  {\bfseries 23} (2006) 4887--4894}.

\bibitem{Bodeker:2009qy}
D.~Bodeker and G.~D. Moore, \emph{{Can electroweak bubble walls run away?}},
  \href{https://doi.org/10.1088/1475-7516/2009/05/009}{\emph{JCAP} {\bfseries
  05} (2009) 009}, [\href{https://arxiv.org/abs/0903.4099}{{\ttfamily
  0903.4099}}].

\bibitem{Bodeker:2017cim}
D.~Bodeker and G.~D. Moore, \emph{{Electroweak Bubble Wall Speed Limit}},
  \href{https://doi.org/10.1088/1475-7516/2017/05/025}{\emph{JCAP} {\bfseries
  1705} (2017) 025}, [\href{https://arxiv.org/abs/1703.08215}{{\ttfamily
  1703.08215}}].

\bibitem{Mancha:2020fzw}
M.~Barroso~Mancha, T.~Prokopec and B.~Swiezewska, \emph{{Field theoretic
  derivation of bubble wall force}},
  \href{https://arxiv.org/abs/2005.10875}{{\ttfamily 2005.10875}}.

\bibitem{Peskin:1995}
M.~E. Peskin and D.~V. Schroeder, \emph{{A}n {I}ntroduction to {Q}uantum
  {F}ield {T}heory, Sec.~4.5}.
\newblock {W}estview {P}ress, {B}oulder, {C}olorado, 1995.

\bibitem{Vanvlasselaer:2020niz}
A.~Azatov and M.~Vanvlasselaer, \emph{{Bubble wall velocity: heavy physics
  effects}},  \href{https://arxiv.org/abs/2010.02590}{{\ttfamily 2010.02590}}.

\bibitem{Sudakov:1954sw}
V.~V. Sudakov, \emph{{Vertex parts at very high-energies in quantum
  electrodynamics}}, {\emph{Sov. Phys. JETP} {\bfseries 3} (1956) 65--71}.

\bibitem{Amati:1980ch}
D.~Amati, A.~Bassetto, M.~Ciafaloni, G.~Marchesini and G.~Veneziano, \emph{{A
  treatment of hard processes sensitive to the infrared structure of QCD}},
  \href{https://doi.org/10.1016/0550-3213(80)90012-7}{\emph{Nucl. Phys.}
  {\bfseries B173} (1980) 429}.

\bibitem{Catani:1990rr}
S.~Catani, B.~R. Webber and G.~Marchesini, \emph{{QCD coherent branching and
  semiinclusive processes at large $x$}},
  \href{https://doi.org/10.1016/0550-3213(91)90390-J}{\emph{Nucl. Phys.}
  {\bfseries B349} (1991) 635--654}.

\bibitem{Yennie:1961ad}
D.~R. Yennie, S.~C. Frautschi and H.~Suura, \emph{{The Infrared Divergence
  Phenomena and High-Energy Processes}}, {\emph{Ann. Phys.} {\bfseries 13}
  (1961) 379--452}.

\bibitem{Becher:2014oda}
T.~Becher, A.~Broggio and A.~Ferroglia, \emph{{Introduction to Soft-Collinear
  Effective Theory}}, vol.~896.
\newblock Springer, 2015,
  \href{https://doi.org/10.1007/978-3-319-14848-9}{10.1007/978-3-319-14848-9}.

\bibitem{Bassetto:1984ik}
A.~Bassetto, M.~Ciafaloni and G.~Marchesini, \emph{{Jet structure and infrared
  sensitive quantities in perturbative QCD}}, {\emph{Phys. Rept.} {\bfseries
  100} (1983) 201--272}.

\bibitem{Dixon:1996wi}
L.~J. Dixon, \emph{{Calculating scattering amplitudes efficiently}},  in
  \emph{{Theoretical Advanced Study Institute in Elementary Particle Physics
  (TASI 95): QCD and Beyond}}, pp.~539--584, 1, 1996,
  \href{https://arxiv.org/abs/hep-ph/9601359}{{\ttfamily hep-ph/9601359}}.

\bibitem{Dittmaier:1998nn}
S.~Dittmaier, \emph{{Weyl-van der Waerden formalism for helicity amplitudes of
  massive particles}}, {\emph{Phys. Rev.} {\bfseries D59} (1999) 016007},
  [\href{https://arxiv.org/abs/hep-ph/9805445}{{\ttfamily hep-ph/9805445}}].

\bibitem{Denner:2019vbn}
A.~Denner and S.~Dittmaier, \emph{{Electroweak Radiative Corrections for
  Collider Physics}},
  \href{https://doi.org/10.1016/j.physrep.2020.04.001}{\emph{Phys. Rept.}
  {\bfseries 864} (2020) 1--163},
  [\href{https://arxiv.org/abs/1912.06823}{{\ttfamily 1912.06823}}].

\bibitem{Denner:2006xx}
A.~Denner.

\bibitem{Carena:2000id}
M.~Carena, J.~Moreno, M.~Quiros, M.~Seco and C.~Wagner, \emph{{Supersymmetric
  CP violating currents and electroweak baryogenesis}},
  \href{https://doi.org/10.1016/S0550-3213(01)00032-3}{\emph{Nucl. Phys. B}
  {\bfseries 599} (2001) 158--184},
  [\href{https://arxiv.org/abs/hep-ph/0011055}{{\ttfamily hep-ph/0011055}}].

\bibitem{Carena:2002ss}
M.~Carena, M.~Quiros, M.~Seco and C.~Wagner, \emph{{Improved Results in
  Supersymmetric Electroweak Baryogenesis}},
  \href{https://doi.org/10.1016/S0550-3213(02)01065-9}{\emph{Nucl. Phys. B}
  {\bfseries 650} (2003) 24--42},
  [\href{https://arxiv.org/abs/hep-ph/0208043}{{\ttfamily hep-ph/0208043}}].

\bibitem{Lee:2004we}
C.~Lee, V.~Cirigliano and M.~J. Ramsey-Musolf, \emph{{Resonant relaxation in
  electroweak baryogenesis}},
  \href{https://doi.org/10.1103/PhysRevD.71.075010}{\emph{Phys. Rev. D}
  {\bfseries 71} (2005) 075010},
  [\href{https://arxiv.org/abs/hep-ph/0412354}{{\ttfamily hep-ph/0412354}}].

\bibitem{Eden:1966dnq}
R.~J. Eden, P.~V. Landshoff, D.~I. Olive and J.~C. Polkinghorne, \emph{{The
  analytic S-matrix}}.
\newblock Cambridge Univ. Press, Cambridge, 1966.

\bibitem{Bern:1994cg}
Z.~Bern, L.~J. Dixon, D.~C. Dunbar and D.~A. Kosower, \emph{{Fusing gauge
  theory tree amplitudes into loop amplitudes}},
  \href{https://doi.org/10.1016/0550-3213(94)00488-Z}{\emph{Nucl. Phys.}
  {\bfseries B435} (1995) 59--101},
  [\href{https://arxiv.org/abs/hep-ph/9409265}{{\ttfamily hep-ph/9409265}}].

\bibitem{Banfi:2004yd}
A.~Banfi, G.~P. Salam and G.~Zanderighi, \emph{{Principles of general
  final-state resummation and automated implementation}},
  \href{https://doi.org/10.1088/1126-6708/2005/03/073}{\emph{JHEP} {\bfseries
  03} (2005) 073}, [\href{https://arxiv.org/abs/hep-ph/0407286}{{\ttfamily
  hep-ph/0407286}}].

\bibitem{Hoeche:2017jsi}
S.~H{\"o}che, D.~Reichelt and F.~Siegert, \emph{{Momentum conservation and
  unitarity in parton showers and NLL resummation}},
  \href{https://doi.org/10.1007/JHEP01(2018)118}{\emph{JHEP} {\bfseries 01}
  (2018) 118}, [\href{https://arxiv.org/abs/1711.03497}{{\ttfamily
  1711.03497}}].

\bibitem{Farhi:1977sg}
E.~Farhi, \emph{{A QCD Test for Jets}},
  \href{https://doi.org/10.1103/PhysRevLett.39.1587}{\emph{Phys. Rev. Lett.}
  {\bfseries 39} (1977) 1587--1588}.

\bibitem{Catani:1991kz}
S.~Catani, G.~Turnock, B.~Webber and L.~Trentadue, \emph{{Thrust distribution
  in e+ e- annihilation}},
  \href{https://doi.org/10.1016/0370-2693(91)90494-B}{\emph{Phys.Lett.}
  {\bfseries B263} (1991) 491--497}.

\bibitem{Webber:1986mc}
B.~Webber, \emph{{Monte Carlo Simulation of Hard Hadronic Processes}},
  {\emph{Ann. Rev. Nucl. Part. Sci.} {\bfseries 36} (1986) 253--286}.

\bibitem{Sjostrand:1995iq}
T.~Sjostrand, \emph{{PYTHIA 5.7 and JETSET 7.4: Physics and manual}},
  \href{https://arxiv.org/abs/hep-ph/9508391}{{\ttfamily hep-ph/9508391}}.

\bibitem{Buckley:2011ms}
A.~Buckley et~al., \emph{{General-purpose event generators for LHC physics}},
  \href{https://doi.org/http://dx.doi.org/10.1016/j.physrep.2011.03.005}{\emph{Phys.
  Rept.} {\bfseries 504} (2011) 145--233},
  [\href{https://arxiv.org/abs/1101.2599}{{\ttfamily 1101.2599}}].

\bibitem{Hoche:2014rga}
S.~H{\"o}che, \emph{{Introduction to parton-shower event generators}},  in
  \emph{{Proceedings of TASI 2014}}, pp.~235--295, 2015,
  \href{https://arxiv.org/abs/1411.4085}{{\ttfamily 1411.4085}},
  \href{https://doi.org/10.1142/9789814678766_0005}{DOI}.

\bibitem{Marchesini:1987cf}
G.~Marchesini and B.~R. Webber, \emph{{Monte Carlo Simulation of General Hard
  Processes with Coherent QCD Radiation}},
  \href{https://doi.org/10.1016/0550-3213(88)90089-2}{\emph{Nucl. Phys.}
  {\bfseries B310} (1988) 461}.

\bibitem{Platzer:2009jq}
S.~Pl{\"a}tzer and S.~Gieseke, \emph{{Coherent Parton Showers with Local
  Recoils}}, \href{https://doi.org/10.1007/JHEP01(2011)024}{\emph{JHEP}
  {\bfseries 01} (2011) 024},
  [\href{https://arxiv.org/abs/0909.5593}{{\ttfamily 0909.5593}}].

\bibitem{Bothmann:2019yzt}
{\scshape Sherpa} collaboration, E.~Bothmann et~al., \emph{{Event Generation
  with Sherpa 2.2}},
  \href{https://doi.org/10.21468/SciPostPhys.7.3.034}{\emph{SciPost Phys.}
  {\bfseries 7} (2019) 034},
  [\href{https://arxiv.org/abs/1905.09127}{{\ttfamily 1905.09127}}].

\bibitem{Moore:1995ua}
G.~D. Moore and T.~Prokopec, \emph{{Bubble wall velocity in a first order
  electroweak phase transition}},
  \href{https://doi.org/10.1103/PhysRevLett.75.777}{\emph{Phys. Rev. Lett.}
  {\bfseries 75} (1995) 777--780},
  [\href{https://arxiv.org/abs/hep-ph/9503296}{{\ttfamily hep-ph/9503296}}].

\bibitem{Moore:1995si}
G.~D. Moore and T.~Prokopec, \emph{{How fast can the wall move? A Study of the
  electroweak phase transition dynamics}},
  \href{https://doi.org/10.1103/PhysRevD.52.7182}{\emph{Phys. Rev. D}
  {\bfseries 52} (1995) 7182--7204},
  [\href{https://arxiv.org/abs/hep-ph/9506475}{{\ttfamily hep-ph/9506475}}].

\bibitem{John:2000zq}
P.~John and M.~Schmidt, \emph{{Do stops slow down electroweak bubble walls?}},
  \href{https://doi.org/10.1016/S0550-3213(00)00768-9}{\emph{Nucl. Phys. B}
  {\bfseries 598} (2001) 291--305},
  [\href{https://arxiv.org/abs/hep-ph/0002050}{{\ttfamily hep-ph/0002050}}].

\bibitem{Konstandin:2014zta}
T.~Konstandin, G.~Nardini and I.~Rues, \emph{{From Boltzmann equations to
  steady wall velocities}},
  \href{https://doi.org/10.1088/1475-7516/2014/09/028}{\emph{JCAP} {\bfseries
  09} (2014) 028}, [\href{https://arxiv.org/abs/1407.3132}{{\ttfamily
  1407.3132}}].

\bibitem{Kozaczuk:2015owa}
J.~Kozaczuk, \emph{{Bubble Expansion and the Viability of Singlet-Driven
  Electroweak Baryogenesis}},
  \href{https://doi.org/10.1007/JHEP10(2015)135}{\emph{JHEP} {\bfseries 10}
  (2015) 135}, [\href{https://arxiv.org/abs/1506.04741}{{\ttfamily
  1506.04741}}].

\bibitem{Dorsch:2018pat}
G.~C. Dorsch, S.~J. Huber and T.~Konstandin, \emph{{Bubble wall velocities in
  the Standard Model and beyond}},
  \href{https://doi.org/10.1088/1475-7516/2018/12/034}{\emph{JCAP} {\bfseries
  12} (2018) 034}, [\href{https://arxiv.org/abs/1809.04907}{{\ttfamily
  1809.04907}}].

\bibitem{Ellis:2019oqb}
J.~Ellis, M.~Lewicki, J.~M. No and V.~Vaskonen, \emph{{Gravitational wave
  energy budget in strongly supercooled phase transitions}},
  \href{https://doi.org/10.1088/1475-7516/2019/06/024}{\emph{JCAP} {\bfseries
  06} (2019) 024}, [\href{https://arxiv.org/abs/1903.09642}{{\ttfamily
  1903.09642}}].

\bibitem{Hogan:1983zz}
C.~J. Hogan, \emph{{Magnetohydrodynamic Effects of a First-Order Cosmological
  Phase Transition}},
  \href{https://doi.org/10.1103/PhysRevLett.51.1488}{\emph{Phys. Rev. Lett.}
  {\bfseries 51} (1983) 1488--1491}.

\bibitem{Hindmarsh:2014aa}
M.~Hindmarsh, S.~J. Huber, K.~Rummukainen and D.~J. Weir, \emph{Gravitational
  waves from the sound of a first order phase transition},
  \href{https://doi.org/10.1103/PhysRevLett.112.041301}{\emph{Phys. Rev. Lett.}
  {\bfseries 112} (2014) 041301},
  [\href{https://arxiv.org/abs/1304.2433}{{\ttfamily 1304.2433}}].

\bibitem{Randall:2006py}
L.~Randall and G.~Servant, \emph{{Gravitational waves from warped spacetime}},
  \href{https://doi.org/10.1088/1126-6708/2007/05/054}{\emph{JHEP} {\bfseries
  05} (2007) 054}, [\href{https://arxiv.org/abs/hep-ph/0607158}{{\ttfamily
  hep-ph/0607158}}].

\bibitem{Konstandin:2011dr}
T.~Konstandin and G.~Servant, \emph{{Cosmological Consequences of Nearly
  Conformal Dynamics at the TeV scale}},
  \href{https://doi.org/10.1088/1475-7516/2011/12/009}{\emph{JCAP} {\bfseries
  12} (2011) 009}, [\href{https://arxiv.org/abs/1104.4791}{{\ttfamily
  1104.4791}}].

\bibitem{Konstandin:2011ds}
T.~Konstandin and G.~Servant, \emph{{Natural Cold Baryogenesis from Strongly
  Interacting Electroweak Symmetry Breaking}},
  \href{https://doi.org/10.1088/1475-7516/2011/07/024}{\emph{JCAP} {\bfseries
  07} (2011) 024}, [\href{https://arxiv.org/abs/1104.4793}{{\ttfamily
  1104.4793}}].

\bibitem{Jinno:2016knw}
R.~Jinno and M.~Takimoto, \emph{{Probing a classically conformal B-L model with
  gravitational waves}},
  \href{https://doi.org/10.1103/PhysRevD.95.015020}{\emph{Phys. Rev. D}
  {\bfseries 95} (2017) 015020},
  [\href{https://arxiv.org/abs/1604.05035}{{\ttfamily 1604.05035}}].

\bibitem{Iso:2017uuu}
S.~Iso, P.~D. Serpico and K.~Shimada, \emph{{QCD-Electroweak First-Order Phase
  Transition in a Supercooled Universe}},
  \href{https://doi.org/10.1103/PhysRevLett.119.141301}{\emph{Phys. Rev. Lett.}
  {\bfseries 119} (2017) 141301},
  [\href{https://arxiv.org/abs/1704.04955}{{\ttfamily 1704.04955}}].

\bibitem{vonHarling:2017yew}
B.~von Harling and G.~Servant, \emph{{QCD-induced Electroweak Phase
  Transition}}, \href{https://doi.org/10.1007/JHEP01(2018)159}{\emph{JHEP}
  {\bfseries 01} (2018) 159},
  [\href{https://arxiv.org/abs/1711.11554}{{\ttfamily 1711.11554}}].

\bibitem{Kobakhidze:2017mru}
A.~Kobakhidze, C.~Lagger, A.~Manning and J.~Yue, \emph{{Gravitational waves
  from a supercooled electroweak phase transition and their detection with
  pulsar timing arrays}},
  \href{https://doi.org/10.1140/epjc/s10052-017-5132-y}{\emph{Eur. Phys. J. C}
  {\bfseries 77} (2017) 570},
  [\href{https://arxiv.org/abs/1703.06552}{{\ttfamily 1703.06552}}].

\bibitem{Marzola:2017jzl}
L.~Marzola, A.~Racioppi and V.~Vaskonen, \emph{{Phase transition and
  gravitational wave phenomenology of scalar conformal extensions of the
  Standard Model}},
  \href{https://doi.org/10.1140/epjc/s10052-017-4996-1}{\emph{Eur. Phys. J. C}
  {\bfseries 77} (2017) 484},
  [\href{https://arxiv.org/abs/1704.01034}{{\ttfamily 1704.01034}}].

\bibitem{Prokopec:2018tnq}
T.~Prokopec, J.~Rezacek and B.~a. \'Swie\.zewska, \emph{{Gravitational waves
  from conformal symmetry breaking}},
  \href{https://doi.org/10.1088/1475-7516/2019/02/009}{\emph{JCAP} {\bfseries
  02} (2019) 009}, [\href{https://arxiv.org/abs/1809.11129}{{\ttfamily
  1809.11129}}].

\bibitem{Hambye:2018qjv}
T.~Hambye, A.~Strumia and D.~Teresi, \emph{{Super-cool Dark Matter}},
  \href{https://doi.org/10.1007/JHEP08(2018)188}{\emph{JHEP} {\bfseries 08}
  (2018) 188}, [\href{https://arxiv.org/abs/1805.01473}{{\ttfamily
  1805.01473}}].

\bibitem{Marzo:2018nov}
C.~Marzo, L.~Marzola and V.~Vaskonen, \emph{{Phase transition and vacuum
  stability in the classically conformal B--L model}},
  \href{https://doi.org/10.1140/epjc/s10052-019-7076-x}{\emph{Eur. Phys. J. C}
  {\bfseries 79} (2019) 601},
  [\href{https://arxiv.org/abs/1811.11169}{{\ttfamily 1811.11169}}].

\bibitem{Baratella:2018pxi}
P.~Baratella, A.~Pomarol and F.~Rompineve, \emph{{The Supercooled Universe}},
  \href{https://doi.org/10.1007/JHEP03(2019)100}{\emph{JHEP} {\bfseries 03}
  (2019) 100}, [\href{https://arxiv.org/abs/1812.06996}{{\ttfamily
  1812.06996}}].

\bibitem{Bruggisser:2018mrt}
S.~Bruggisser, B.~Von~Harling, O.~Matsedonskyi and G.~Servant,
  \emph{{Electroweak Phase Transition and Baryogenesis in Composite Higgs
  Models}}, \href{https://doi.org/10.1007/JHEP12(2018)099}{\emph{JHEP}
  {\bfseries 12} (2018) 099},
  [\href{https://arxiv.org/abs/1804.07314}{{\ttfamily 1804.07314}}].

\bibitem{Aoki:2019mlt}
M.~Aoki and J.~Kubo, \emph{{Gravitational waves from chiral phase transition in
  a conformally extended standard model}},
  \href{https://doi.org/10.1088/1475-7516/2020/04/001}{\emph{JCAP} {\bfseries
  04} (2020) 001}, [\href{https://arxiv.org/abs/1910.05025}{{\ttfamily
  1910.05025}}].

\bibitem{DelleRose:2019pgi}
L.~Delle~Rose, G.~Panico, M.~Redi and A.~Tesi, \emph{{Gravitational Waves from
  Supercool Axions}},
  \href{https://doi.org/10.1007/JHEP04(2020)025}{\emph{JHEP} {\bfseries 04}
  (2020) 025}, [\href{https://arxiv.org/abs/1912.06139}{{\ttfamily
  1912.06139}}].

\bibitem{Kuzmin:1985mm}
V.~A. Kuzmin, V.~A. Rubakov and M.~E. Shaposhnikov, \emph{{On the Anomalous
  Electroweak Baryon Number Nonconservation in the Early Universe}},
  \href{https://doi.org/10.1016/0370-2693(85)91028-7}{\emph{Phys. Lett.}
  {\bfseries B155} (1985) 36}.

\bibitem{Shaposhnikov:1986jp}
M.~Shaposhnikov, \emph{{Possible Appearance of the Baryon Asymmetry of the
  Universe in an Electroweak Theory}}, {\emph{JETP Lett.} {\bfseries 44} (1986)
  465--468}.

\bibitem{Morrissey:2012db}
D.~E. Morrissey and M.~J. Ramsey-Musolf, \emph{{Electroweak baryogenesis}},
  \href{https://doi.org/10.1088/1367-2630/14/12/125003}{\emph{New J. Phys.}
  {\bfseries 14} (2012) 125003},
  [\href{https://arxiv.org/abs/1206.2942}{{\ttfamily 1206.2942}}].

\bibitem{Cline:2020jre}
J.~M. Cline and K.~Kainulainen, \emph{{Electroweak baryogenesis at high bubble
  wall velocities}},
  \href{https://doi.org/10.1103/PhysRevD.101.063525}{\emph{Phys. Rev. D}
  {\bfseries 101} (2020) 063525},
  [\href{https://arxiv.org/abs/2001.00568}{{\ttfamily 2001.00568}}].

\bibitem{Joyce:1994zn}
M.~Joyce, T.~Prokopec and N.~Turok, \emph{{Nonlocal electroweak baryogenesis.
  Part 1: Thin wall regime}},
  \href{https://doi.org/10.1103/PhysRevD.53.2930}{\emph{Phys. Rev.} {\bfseries
  D53} (1996) 2930--2957},
  [\href{https://arxiv.org/abs/hep-ph/9410281}{{\ttfamily hep-ph/9410281}}].

\bibitem{Joyce:1994zt}
M.~Joyce, T.~Prokopec and N.~Turok, \emph{{Nonlocal electroweak baryogenesis.
  Part 2: The Classical regime}},
  \href{https://doi.org/10.1103/PhysRevD.53.2958}{\emph{Phys. Rev.} {\bfseries
  D53} (1996) 2958--2980},
  [\href{https://arxiv.org/abs/hep-ph/9410282}{{\ttfamily hep-ph/9410282}}].

\bibitem{Trodden:1998ym}
M.~Trodden, \emph{{Electroweak baryogenesis}},
  \href{https://doi.org/10.1103/RevModPhys.71.1463}{\emph{Rev.Mod.Phys.}
  {\bfseries 71} (1999) 1463--1500},
  [\href{https://arxiv.org/abs/hep-ph/9803479}{{\ttfamily hep-ph/9803479}}].

\bibitem{Katz:2016adq}
A.~Katz and A.~Riotto, \emph{{Baryogenesis and Gravitational Waves from Runaway
  Bubble Collisions}},
  \href{https://doi.org/10.1088/1475-7516/2016/11/011}{\emph{JCAP} {\bfseries
  11} (2016) 011}, [\href{https://arxiv.org/abs/1608.00583}{{\ttfamily
  1608.00583}}].

\bibitem{Grasso:2000wj}
D.~Grasso and H.~R. Rubinstein, \emph{{Magnetic fields in the early universe}},
  \href{https://doi.org/10.1016/S0370-1573(00)00110-1}{\emph{Phys. Rept.}
  {\bfseries 348} (2001) 163--266},
  [\href{https://arxiv.org/abs/astro-ph/0009061}{{\ttfamily
  astro-ph/0009061}}].

\bibitem{Hindmarsh:2017gnf}
M.~Hindmarsh, S.~J. Huber, K.~Rummukainen and D.~J. Weir, \emph{{Shape of the
  acoustic gravitational wave power spectrum from a first order phase
  transition}}, \href{https://doi.org/10.1103/PhysRevD.96.103520}{\emph{Phys.
  Rev. D} {\bfseries 96} (2017) 103520},
  [\href{https://arxiv.org/abs/1704.05871}{{\ttfamily 1704.05871}}].

\bibitem{Kahniashvili:2012uj}
T.~Kahniashvili, A.~G. Tevzadze, A.~Brandenburg and A.~Neronov,
  \emph{{Evolution of Primordial Magnetic Fields from Phase Transitions}},
  \href{https://doi.org/10.1103/PhysRevD.87.083007}{\emph{Phys. Rev. D}
  {\bfseries 87} (2013) 083007},
  [\href{https://arxiv.org/abs/1212.0596}{{\ttfamily 1212.0596}}].

\bibitem{Brandenburg:2017neh}
A.~Brandenburg, T.~Kahniashvili, S.~Mandal, A.~Roper~Pol, A.~G. Tevzadze and
  T.~Vachaspati, \emph{{Evolution of hydromagnetic turbulence from the
  electroweak phase transition}},
  \href{https://doi.org/10.1103/PhysRevD.96.123528}{\emph{Phys. Rev. D}
  {\bfseries 96} (2017) 123528},
  [\href{https://arxiv.org/abs/1711.03804}{{\ttfamily 1711.03804}}].

\bibitem{Durrer:2013pga}
R.~Durrer and A.~Neronov, \emph{{Cosmological Magnetic Fields: Their
  Generation, Evolution and Observation}},
  \href{https://doi.org/10.1007/s00159-013-0062-7}{\emph{Astron. Astrophys.
  Rev.} {\bfseries 21} (2013) 62},
  [\href{https://arxiv.org/abs/1303.7121}{{\ttfamily 1303.7121}}].

\bibitem{Ellis:2019tjf}
J.~Ellis, M.~Fairbairn, M.~Lewicki, V.~Vaskonen and A.~Wickens,
  \emph{{Intergalactic Magnetic Fields from First-Order Phase Transitions}},
  \href{https://doi.org/10.1088/1475-7516/2019/09/019}{\emph{JCAP} {\bfseries
  09} (2019) 019}, [\href{https://arxiv.org/abs/1907.04315}{{\ttfamily
  1907.04315}}].

\bibitem{Zhang:2019vsb}
Y.~Zhang, T.~Vachaspati and F.~Ferrer, \emph{{Magnetic field production at a
  first-order electroweak phase transition}},
  \href{https://doi.org/10.1103/PhysRevD.100.083006}{\emph{Phys. Rev. D}
  {\bfseries 100} (2019) 083006},
  [\href{https://arxiv.org/abs/1902.02751}{{\ttfamily 1902.02751}}].

\bibitem{Chen:2016wkt}
J.~Chen, T.~Han and B.~Tweedie, \emph{{Electroweak Splitting Functions and High
  Energy Showering}},
  \href{https://doi.org/10.1007/JHEP11(2017)093}{\emph{JHEP} {\bfseries 11}
  (2017) 093}, [\href{https://arxiv.org/abs/1611.00788}{{\ttfamily
  1611.00788}}].

\bibitem{Bauer:2017bnh}
C.~W. Bauer, N.~Ferland and B.~R. Webber, \emph{{Combining initial-state
  resummation with fixed-order calculations of electroweak corrections}},
  \href{https://doi.org/10.1007/JHEP04(2018)125}{\emph{JHEP} {\bfseries 04}
  (2018) 125}, [\href{https://arxiv.org/abs/1712.07147}{{\ttfamily
  1712.07147}}].

\bibitem{Kleiss:2020rcg}
R.~Kleiss and R.~Verheyen, \emph{{Collinear electroweak radiation in antenna
  parton showers}},
  \href{https://doi.org/10.1140/epjc/s10052-020-08510-w}{\emph{Eur. Phys. J. C}
  {\bfseries 80} (2020) 980},
  [\href{https://arxiv.org/abs/2002.09248}{{\ttfamily 2002.09248}}].

\bibitem{Field:1989uq}
R.~D. Field, \emph{{Applications of perturbative QCD}}.
\newblock Addison-Wesley, Redwood City, USA, 1989.

\bibitem{Ellis:1991qj}
R.~K. Ellis, W.~J. Stirling and B.~R. Webber, \emph{{QCD and collider
  physics}}, vol.~8.
\newblock Cambridge Monogr. Part. Phys. Nucl. Phys. Cosmol., 1~ed., 1996.

\bibitem{Catani:2002hc}
S.~Catani, S.~Dittmaier, M.~H. Seymour and Z.~Trocsanyi, \emph{{The dipole
  formalism for next-to-leading order QCD calculations with massive partons}},
  {\emph{Nucl. Phys.} {\bfseries B627} (2002) 189--265},
  [\href{https://arxiv.org/abs/hep-ph/0201036}{{\ttfamily hep-ph/0201036}}].

\bibitem{Schumann:2007mg}
S.~Schumann and F.~Krauss, \emph{{A parton shower algorithm based on
  Catani-Seymour dipole factorisation}}, {\emph{JHEP} {\bfseries 03} (2008)
  038}, [\href{https://arxiv.org/abs/0709.1027}{{\ttfamily 0709.1027}}].

\bibitem{Dittmaier:1999mb}
S.~Dittmaier, \emph{{A general approach to photon radiation off fermions}},
  \href{https://doi.org/10.1016/S0550-3213(99)00563-5}{\emph{Nucl. Phys.}
  {\bfseries B565} (2000) 69--122},
  [\href{https://arxiv.org/abs/hep-ph/9904440}{{\ttfamily hep-ph/9904440}}].

\bibitem{Byckling:1969sx}
E.~Byckling and K.~Kajantie, \emph{{N-particle phase space in terms of
  invariant momentum transfers}}, {\emph{Nucl. Phys.} {\bfseries B9} (1969)
  568--576}.

\bibitem{Byckling:1971vca}
E.~Byckling and K.~Kajantie, \emph{{Particle Kinematics}}.
\newblock University of Jyvaskyla, Jyvaskyla, Finland, 1971.

\bibitem{Catani:1996vz}
S.~Catani and M.~H. Seymour, \emph{{A general algorithm for calculating jet
  cross sections in NLO QCD}}, {\emph{Nucl. Phys.} {\bfseries B485} (1997)
  291--419}, [\href{https://arxiv.org/abs/hep-ph/9605323}{{\ttfamily
  hep-ph/9605323}}].

\end{thebibliography}\endgroup
\end{document}